\global\def\draftcontrol{0}

   \def\versionno{ ks casimir }

\catcode`\@=11

\expandafter\ifx\csname draftcontrol\endcsname\relax\global\def\draftcontrol{0}
\fi

{\count255=\time\divide\count255 by 60
\xdef\hourmin{\number\count255}
\multiply\count255 by-60\advance\count255 by\time
\xdef\hourmin{\hourmin:\ifnum\count255<10 0\fi\the\count255}}
\def\draftdate{\number\month/\number\day/\number\year\ \ \ \hourmin }

\newcommand\makepapertitle{\par
  \begingroup
    \renewcommand\thefootnote{\@fnsymbol\c@footnote}%
    \def\@makefnmark{\rlap{\@textsuperscript{\normalfont\@thefnmark}}}%
    \long\def\@makefntext##1{\parindent 1em\noindent
            \hb@xt@1.8em{%
                \hss\@textsuperscript{\normalfont\@thefnmark}}##1}%
     \newpage
     \global\@topnum\z@   
     \@makepapertitle
     \thispagestyle{empty}\@thanks
  \endgroup
  \setcounter{footnote}{0}%
  \global\let\thanks\relax
  \global\let\makepapertitle\relax
  \global\let\@makepapertitle\relax
  \global\let\@thanks\@empty
  \global\let\@author\@empty
  \global\let\@date\@empty
  \global\let\@title\@empty
  \global\let\title\relax
  \global\let\author\relax
  \global\let\date\relax
  \global\let\and\relax
  \def\version{\let\version\@version\@gobble}
}
\def\@makepapertitle{%
  \newpage
   \ifnum\draftcontrol=1 {}
   \version\versionno
   \vskip 3em%
   \else
   \hfill\hbox to 3cm {\parbox{4cm}{\@pubnum}\hss}%
   \vskip 3em%
   \fi
   \begin{center}%
   \let \footnote \thanks
     {\LARGE {\@title}}%
     \vskip 1.5em%
     {\normalsize
       \lineskip .5em%
       \begin{tabular}[t]{c}%
         \@author
       \end{tabular}\par}%
     \vskip 1.5em%
     {\@bstract}%
     \end{center}%
     \vskip 1.5em
     \@date%
   \par
}

\gdef\@pubnum{}
\def\pubnum#1{%
  \gdef\@pubnum{#1}}

\gdef\@bstract{}
\def\Abstract#1{%
  \gdef\@bstract{%
   \parbox{\textwidth-0pc}{%
   \centerline{\bf Abstract}\penalty1000%
\kern.2cm%
\noindent
\renewcommand\baselinestretch{1.0}%
{#1}}}
}

\def\ps@paper{\let\@mkboth\@gobbletwo%
     \ifnum\draftcontrol=1
    \def\@oddfoot{\hbox to \textwidth{\tiny \versionno \hfil\tiny\draftdate}%
    \hskip -\textwidth \hbox to \textwidth{\hfil\rm\thepage\hfil}}%
     \else\def\@oddfoot{\hbox to \textwidth{\hfil\rm\thepage\hfil}}
     \fi
     \let\@evenfoot\@oddfoot
}

\def\body{\clearpage
          \pagestyle{paper}
    }

\def\@version#1{\ifnum\draftcontrol=1
\typeout{}\typeout{#1}\typeout{}
\vskip3mm\centerline{\hbox{\fbox{\normalsize{\tt DRAFT -- #1 -- }
                   {\draftdate}}}}\vskip3mm
\fi}
\let\version\@version
\long\def\eqlabel#1{\ifnum\draftcontrol=1
                    \tag@false  
                    \tag*{(\theequation) \hbox to -0.2cm{\hspace{0cm}\small{#1}\hss}}
                    \refstepcounter{equation}
                    \edef\@currentlabel{\theequation}
                    \ltx@label{#1}          
                    \else
                    \label{#1}
                    \fi
                    }
\let\st@bibitem\@bibitem
\let\st@lbibitem\@lbibitem
\ifnum\draftcontrol=1
  \def\@bibitem#1{%
    \st@bibitem{#1}\a@@label{#1}\ignorespaces}
  \def\@lbibitem[#1]#2{%
    \st@lbibitem[#1]{#2}\a@@label{#2}\ignorespaces}
  \def\a@@label#1{%
    \gdef\a@lab{\smash{\normalfont\small#1}}
    \ifvmode
      \if@inlabel
        \global\setbox\@labels\hbox{%
          \llap{\a@lab\let\a@lab\relax
                \kern\@totalleftmargin\kern\marginparsep}%
          \box\@labels}%
      \fi
    \fi}
\fi

\documentclass[12pt,letterpaper]{article}

\usepackage{amsmath,amssymb,array,calc,epsfig,rotating,bm}
\usepackage[sort]{cite}
\usepackage{graphicx}
\usepackage{psfrag,verbatim}

\ifnum\draftcontrol=1
\fi

\renewcommand\baselinestretch{1.25}
\renewcommand\section{\@startsection {section}{1}{\z@}%
                                   {-3.5ex \@plus -1ex \@minus -.2ex}%
                                   {2.3ex \@plus.2ex}%
                                   {\normalfont\large\bfseries}}
\renewcommand\subsection{\@startsection{subsection}{2}{\z@}%
                                   {-3.25ex\@plus -1ex \@minus -.2ex}%
                                   {1.5ex \@plus .2ex}%
                                   {\normalfont\normalsize\bfseries}}
\renewcommand\subsubsection{\@startsection{subsubsection}{3}{\z@}%
                                   {-3.25ex\@plus -1ex \@minus -.2ex}%
                                   {1.5ex \@plus .2ex}%
                                   {\normalfont\normalsize\it}}
\renewcommand\paragraph{\@startsection{paragraph}{4}{\z@}%
                                   {-3.25ex\@plus -1ex \@minus -.2ex}%
                                   {1.5ex \@plus .2ex}%
                                   {\normalfont\normalsize\bf}}

\numberwithin{equation}{section}

\def\revise#1       {\raisebox{-0em}{\rule{3pt}{1em}}%
                     \marginpar{\raisebox{.5em}{\vrule width3pt\
                     \vrule width0pt height 0pt depth0.5em
                     \hbox to 0cm{\hspace{0cm}{%
                     \parbox[t]{4em}{\raggedright\footnotesize{#1}}}\hss}}}}

\newcommand\nxt[1]  {\\\fnxt#1}
\newcommand{\ie}{{\it i.e.,}\ }

\def\cala         {{\cal A}}

\def\cale         {{\cal E}}
\def\calf         {{\cal F}}
\def\calg         {{\cal G}}

\def\calk         {{\cal K}}
\def\call         {{\cal L}}
\def\calm         {{\cal M}}
\def\caln         {{\cal N}}
\def\calo         {{\cal O}}
\def\calp         {{\cal P}}
\def\calq         {{\cal Q}}

\def\zet          {{\mathbb Z}}

\def\del          {\partial}

\def\tr           {\mathop{\rm Tr}}

\def\sqr#1#2{{\vcenter{\vbox{\hrule height.#2pt
 \hbox{\vrule width.#2pt height#1pt \kern#1pt
 \vrule width.#2pt}\hrule height.#2pt}}}}
\def\square{%
  \mathop{\mathchoice{\sqr{12}{15}}{\sqr{9}{12}}{\sqr{6.3}{9}}{\sqr{4.5}{9}}}}

\def\a{\alpha}
\def\b{\beta}
\def\w{\omega}
\def\r{\rho}
\def\dd{\delta}
\def\e{\epsilon}
\def\c{\chi}

\def\g{\gamma}

\def\hh{\hat{h}}
\def\hf{\hat{f}}

\def\hK{\hat{K}}
\def\aa1{\phi}
\def\cc1{\psi}
\def\hh{\hat{h}}

\def\k{\kappa}
\def\dilog{\rm dilog}

\def\l{\lambda}

\def\Om{\Omega}
\def\om{\Omega}

\def\hr{\hat{\r}}
\def\hf{\hat{f}}
\def\hK{\hat{K}}

\def\vev#1{\langle #1 \rangle}
\def\pz{\hat{p}_0}
\def\kz{\hat{K}_0}
\def\ha{\hat{a}}
\def\hb{\hat{b}}
\def\hG{\hat{G}}
\def\csb{{\chi\rm{SB}}}

\catcode`\@=12

\begin{document}


\title{\bf  Quantum phase transitions in cascading gauge theory}
\pubnum{UWO-TH-11/8}

\date{August 30, 2011}

\author{
Alex Buchel\\[0.4cm]
\it Department of Applied Mathematics\\
\it University of Western Ontario\\
\it London, Ontario N6A 5B7, Canada\\[0.2cm]
\it Perimeter Institute for Theoretical Physics\\
\it Waterloo, Ontario N2J 2W9, Canada\\[0.2cm]
 }

\Abstract{
We study a ground state of $\caln=1$ supersymmetric $SU(K+P)\times
SU(K)$ cascading gauge theory of Klebanov et.al \cite{kt,ks} on
$R\times S^3$ at zero temperature. A radius of $S^3$ sets a
compactification scale $\mu$. An interplay between $\mu$ and the
strong coupling scale $\Lambda$ of the theory leads to an interesting
pattern of quantum phases of the system.  For $\mu\ge \mu_{\c{\rm
SB}}=1.240467(8) \Lambda$ the ground state of the theory is chirally
symmetric.  At $\mu=\mu_{\c{\rm SB}}$ the theory undergoes the
first-order transition to a phase with spontaneous breaking of the
chiral symmetry. We further demonstrate that the chirally symmetric
ground state of cascading gauge theory becomes perturbatively unstable
at scales below $\mu_c=0.950634(5) \mu_{\c{\rm SB}}$.  Finally, we
point out that for $\mu<1.486402(5)\Lambda$ the stress-energy tensor
of cascading gauge theory can source inflation of a closed Universe.
}

\makepapertitle

\body

\version\versionno
\tableofcontents

\section{Introduction and Summary}
Consider $\caln=1$ four-dimensional supersymmetric $SU(K+P)\times SU(K)$
gauge theory with two chiral superfields $A_1, A_2$ in the $(K+P,\overline{K})$
representation, and two fields $B_1, B_2$ in the $(\overline{K+P},K)$ in Minkowski space-time.
Perturbatively, this gauge theory has two gauge couplings $g_1, g_2$ associated with 
two gauge group factors,  and a quartic 
superpotential
\begin{equation}
W\sim \tr \left(A_i B_j A_kB_\ell\right)\e^{ik}\e^{j\ell}\,.
\end{equation}
When $P=0$ above theory flows in the infrared to a 
superconformal fixed point, commonly referred to as Klebanov-Witten (KW) 
theory \cite{kw}. At the IR fixed point KW gauge theory is 
strongly coupled --- the superconformal symmetry together with 
$SU(2)\times SU(2)\times U(1)$ global symmetry of the theory implies 
that anomalous dimensions of chiral superfields $\gamma(A_i)=\gamma(B_i)=-\frac 14$, \ie non-perturbatively large.

When $P\ne 0$, conformal invariance of the above $SU(K+P)\times SU(K)$
gauge theory is broken. It is useful to consider an effective description 
of this theory at energy scale $\mu$ with perturbative couplings
$g_i(\mu)\ll 1$. It is straightforward to evaluate NSVZ beta-functions for 
the gauge couplings. One finds that while the sum of the gauge couplings 
does not run
\begin{equation}
\frac{d}{d\ln\mu}\left(\frac{\pi}{g_s}\equiv \frac{4\pi}{g_1^2(\mu)}+\frac{4\pi}{g_2^2(\mu)}\right)=0\,,
\eqlabel{sum}
\end{equation}
the difference between the two couplings is  
\begin{equation}
\frac{4\pi}{g_2^2(\mu)}-\frac{4\pi}{g_1^2(\mu)}\sim P \ \left[3+2(1-\g_{ij})\right]\ \ln\frac{\mu}{\Lambda}\,,
\eqlabel{diff}
\end{equation}
where $\Lambda$  is the strong coupling scale of the theory and $\g_{ij}$ is an anomalous dimension of operators $\tr A_i B_j$.
Given \eqref{diff} and \eqref{sum} it is clear that the effective weakly coupled description of $SU(K+P)\times SU(K)$ gauge theory 
can be valid only in a finite-width energy band centered about $\mu$ scale. Indeed, extending effective description both to the UV 
and to the  IR one necessarily encounters strong coupling in one or the other gauge group factor. As  explained 
in \cite{ks}, to extend the theory past the strongly coupled region(s) one must perform a Seiberg duality \cite{sd}. 
Turns out, in this gauge theory, a Seiberg duality transformation is a self-similarity transformation of the effective description 
so that $K\to K-P$ as one flows to the IR, or $K\to K+P$ as the energy increases. Thus, extension of the effective 
$SU(K+P)\times SU(K)$ description to all energy scales involves and infinite sequence - a {\it cascade } - of Seiberg dualities
where the rank of the gauge group is not constant along RG flow, but changes with energy according to \cite{b,k,aby2} 
\begin{equation}
K=K(\mu)\sim 2 P^2 \ln \frac \mu\Lambda\,, 
\eqlabel{effk}
\end{equation}
at least as $\mu\gg \Lambda$.
To see \eqref{effk}, note that the rank changes by $\Delta K\sim P$ as $P\Delta\left(\ln\frac\mu\Lambda\right)\sim 1$.
Although there are infinitely many duality cascade steps in the UV, there is only a finite number of duality transformations as one 
flows to the IR (from a given scale $\mu$). The space of vacua of a generic cascading gauge 
theory was studied in details in 
\cite{dks}. In the simplest case, when $K(\mu)$ is an integer multiple of $P$, cascading gauge 
theory confines in the 
infrared with a spontaneous breaking of the chiral symmetry $U(1)\supset \zet_2$ \cite{ks}. 
Here, the full global symmetry of the ground state is $SU(2)\times SU(2)\times \zet_2$.  

Effective description of  cascading gauge theory in the UV suggests
that it must be ultimately defined as a theory with an infinite number
of degrees of freedom. If so, an immediate concern is whether such a
theory is renormalizable as a four dimensional quantum field
theory, \ie whether a definite prescription can be made for the
computation of all gauge invariant correlation functions in the
theory.  As was pointed out in \cite{ks}, whenever $g_s K(\mu)\gg 1$,
cascading gauge theory allows for a dual holographic
description \cite{juan,adscft} as type IIB supergravity on a warped
deformed conifold with fluxes. The duality is always valid in the UV
of  cascading gauge theory; if, in addition, $g_s P\gg 1$ the
holographic correspondence is valid in the IR as well. It was shown
in \cite{aby} that a cascading gauge theory {\it defined} by its
holographic dual as an RG flow of type IIB supergravity on a warped
deformed conifold with fluxes is holographically renormalizable as a
four dimensional quantum field theory.

In this paper\footnote{See \cite{bt} for a related early work.} 
we study the properties of the ground state of strongly coupled cascading gauge theory 
on $R\times S^3$ at zero temperature\footnote{Thermodynamics and 
the hydrodynamic transport of  cascading gauge theory plasma are
discussed in \cite{b,kbh1,kbh2,abk,bp1,ksbh,hyd1,
hyd2,hyd3}.}. 
The radius $f_0$ of the $S^3$ sets a compactification scale 
\begin{equation}
\mu\equiv \frac{1}{f_0}\,.
\eqlabel{defmu}
\end{equation} 
In the limit $\frac{\mu}{\Lambda}\to 0$ the ground state of the theory has a (spontaneously) broken
chiral symmetry \cite{ks}, while for $\frac{\mu}{\Lambda}\gg 1$ the chiral symmetry of  cascading gauge 
theory is expected to be restored 
\cite{bt}. Thus, one expects that there is a critical scale $\mu_{\c{\rm SB}}\sim \Lambda$ 
above which the ground state of the theory is chirally symmetric, while at $\mu< \mu_{\c{\rm SB}}$
the chiral symmetry of the ground state is spontaneously broken. We explicitly confirm a quantum phase transition
(QPT) of this type in cascading gauge theory.  Specifically, we compute the difference of the ground state 
energy densities in the symmetric $\cale^{s}$ and in the broken $\cale^{b}$ phases  and find that 
\begin{equation}
\Delta \cale\left(\frac{\mu}{\Lambda}\right)\ \equiv\  \cale^{b}-\cale^{s}\ \propto\ +(\mu- \mu_{\c{\rm SB}})
\,,\qquad \frac{|\mu-\mu_\csb|}{\Lambda}\ll 1\,.  
\eqlabel{deltaE}
\end{equation}
Since 
\begin{equation}
\frac{d \Delta \cale}{d\ln\mu}\  \bigg|_{\mu=\mu_{\c{\rm SB}}}\ne 0\,,
\eqlabel{order}
\end{equation}
the chiral symmetry breaking QPT in cascading gauge theory is of the first-order. 
We further study in details the spectrum of linearized chiral symmetry breaking ($\c\rm{SB}$) fluctuations in 
the symmetric phase of  cascading gauge theory and identify modes which become tachyonic for 
\begin{equation}
\mu\ <\ \mu_c=0.950634(5)\ \mu_{\c{\rm SB}}\,.
\eqlabel{mutachyon}
\end{equation} 

The rest of the paper is organized as follows. 
In the next section we recall the gravitational low-energy effective action 
realizing holographic dual to strongly coupled cascading gauge theory, and the effective action of the 
$\csb$ fluctuations about a chirally symmetric state of the theory \cite{ksbh}. 
In section 3 we present equations of motion and  the asymptotics of the gravitational dual to a 
chirally symmetric ground  state of cascading gauge theory. We discuss various scaling symmetries of the
relevant gravitational solution, explain the encoding of the physical parameters of cascading gauge theory in
their dual gravitational description, outline the numerical procedure for obtaining the gravitational solution,
 and compute the energy density $\cale^{s}$ and the 
pressure $\calp^{s}$ of this ground state as a function of\footnote{See \eqref{scalel} 
for a precise definition of $\Lambda$.} 
$\frac{\mu}{\Lambda}$. 
To leading order in $\delta\equiv \left(\ln\frac{\mu^2}{\Lambda^2P^2g_0}\right)^{-1}$ we find
\begin{equation}
\begin{split}
\cale^{s}=& \frac{\mu^4}{8\pi G_5}\ \frac{1}{32}\left(\frac{P^2g_0}{\dd}+P^2g_0\ln\frac{P^2g_0}{\dd}\right)^2
\biggl(1-2.272588(7)\ \dd+\calo(\dd^2)\ \biggr)\,,\\
\calp^{s}=&\frac{\mu^4}{8\pi G_5}\ \frac{1}{96}\left(\frac{P^2g_0}{\dd}+P^2g_0\ln\frac{P^2g_0}{\dd}\right)^2
\biggl(1+1.727411(3)\ \dd+\calo(\dd^2)\ \biggr)\,,
\end{split}
\eqlabel{highscale}
\end{equation}
where $g_0$ is the asymptotic values of the string coupling (see \eqref{guv}). 
Rather interestingly, we find that the  energy density $\cale^s$ of cascading gauge theory chirally symmetric phase 
is negative for $\mu<2.010798(8)\Lambda$, while the pressure  $\calp^s$ becomes negative for somewhat smaller compactification 
scales $\mu<1.375284(4) \Lambda$. For $\mu<1.486402(5)\Lambda$ the combination $(\cale^s+3\calp^s)$ becomes negative, which 
implies that cascading gauge theory compactified on sufficiently small $S^3$ would lead to 
an inflating closed Universe when coupled to 
four-dimensional Einstein gravity\footnote{Expanding $S^3$ would ultimately end  cascading gauge theory 
driven inflation. Further cosmological aspects 
of cascading gauge theory will be discussed elsewhere.}. 
In section 4 we study the spectrum of linearized $\csb$ fluctuations about a chirally symmetric ground state of 
cascading gauge theory on $S^3$. A mass of a generic $\csb$ state in the spectrum depends on the $S^3$ eigenvalue $L$ of 
its wavefunction, and  an integer $q$ which quantizes its radial wavefunction. For each pair  $\{L,q\}$ there are two branches 
in the spectrum associated with the non-analytic dependence of the mass on $\sqrt{\dd}$. This is evident from the 
(semi-)analytic analysis of the spectrum in the limit $\dd\to 0$. Specifically, we find that the mass-squared $\w^2$ 
of $\{L=0\,, q=1\}$
states is given by 
\begin{equation}
\frac{\w^2}{\mu^2}=9\mp 6\sqrt{2}\ \sqrt{\dd}+0.077172(8)\ \dd+\calo(\dd^{3/2})\,.
\eqlabel{massdelta}
\end{equation}
The lighter of the two states in \eqref{massdelta} eventually becomes tachyonic as $\dd$ (or equivalently $\frac \Lambda\mu$)
becomes sufficiently large:
\begin{equation}
\frac{\w^2}{\mu^2}\bigg|_{L=0,q=1}\ <\ 0\qquad {\rm if}\qquad \mu<\mu_c=1.179231(5)\ \Lambda\,.
\eqlabel{tachyonres}
\end{equation}
An interesting question is whether or not the final state associated with the condensation of $\csb$ 
tachyons below $\mu_c$ can be continuous 
connected to a chirally symmetric ground state in the limit $\mu\to \mu_c$ (from below). To address it, we mass-deform the 
cascading gauge theory at  $\mu=\mu_*< \mu_c$, thus explicitly breaking the chiral symmetry. We show that the $\csb$ condensates 
from the explicit breaking vanish as the mass-deformation parameter vanishes\footnote{This analysis are equivalent to 
the one in \cite{ksbh} where it was shown that $\csb$ tachyons of cascading gauge theory plasma do not condense 
to a homogeneous and isotropic state continuously connected to a chirally symmetric state of the plasma.}. 
In section 5 we construct a new state  of cascading gauge theory on $S^3$ with spontaneously broken chiral symmetry.
We begin with the supersymmetric state of cascading gauge theory on $R^{3,1}$  with $\csb$ \cite{ks} and ``continuously compactify'' 
$R^3$ to $S^3$ (see section \ref{kscompact} for details). During the "compactification'' process the chiral symmetry of the gauge theory 
remains spontaneously broken. Next we compare the energies of the chirally symmetric state of section \ref{symmetric}, 
$\cale^s$, and that of the newly
constructed state, $\cale^b$,
as a function of the compactification scale $\mu$. We show that the new state with spontaneously broken chiral symmetry 
is energetically favourable, \ie $\cale^s>\cale^b$, for
\begin{equation}
\mu<\mu_{\csb}\,,\qquad \mu_{\csb}=1.240467(8)\Lambda > \mu_c\,.
\end{equation}   
This quantum phase transition is of the first order, see \eqref{order}.
Notice that this transition occurs {\it prior to} (at higher compactification scales than) the tachyon condensation in the chirally symmetric 
phase. One possibility\footnote{In the context of cascading gauge theory plasma \cite{ksbh} 
this scenario was advocated by Ofer Aharony \cite{oa}.} 
is that the tachyon discussed in section \ref{tachyon} condenses into this new phase, which would also explain
the absence of the $\csb$ phase of cascading gauge theory continuously connected to a chirally symmetric phase as $\mu\to \mu_c$.
Another possibility\footnote{In the context of cascading gauge theory plasma this scenario was advocated in 
\cite{ksbh}.} is  that the end point of the tachyon condensation would produce a, yet undiscovered, state which 
is not $SO(4)$-invariant. Indeed, while we established the condensation of $SO(4)$-invariant ($L=0$) state in the 
spectrum of $\csb$ fluctuations, from figure \ref{figure5} is it likely that the $SO(4)$ non-invariant state ($L=3$)      
would condense as well, albeit at scales lower than $\mu_c$. 
The latter can also explain while an $SO(4)$ non-invariant state with $\csb$ (if it exists) is not continuously connected 
to the $SO(4)$-invariant chirally symmetric state at $\mu=\mu_c$. 
We hope to resolve these issues in the future work.

\section{Dual effective actions of cascading gauge theory}
Consider $SU(2)\times SU(2)\times \zet_2$ invariant states of cascading gauge theory on a 4-dimensional 
manifold $\calm_4\equiv \del\calm_5$. 
Effective gravitational action on a 5-dimensional manifold $\calm_5$ 
describing holographic dual of such states was derived in \cite{ksbh}:
\begin{equation}
\begin{split}
S_5\left[g_{\mu\nu},\Omega_i,h_i,\Phi\right]=& \frac{108}{16\pi G_5} 
\int_{\calm_5} {\rm vol}_{\calm_5}\ \Omega_1 \Omega_2^2\om_3^2\ 
\biggl\lbrace 
 R_{10}-\frac 12 \left(\nabla \Phi\right)^2\\
&-\frac 12 e^{-\Phi}\left(\frac{(h_1-h_3)^2}{2\om_1^2\om_2^2\om_3^2}+\frac{1}{\om_3^4}\left(\nabla h_1\right)^2
+\frac{1}{\om_2^4}\left(\nabla h_3\right)^2\right)
\\
&-\frac 12 e^{\Phi}\left(\frac{2}{\om_2^2\om_3^2}\left(\nabla h_2\right)^2
+\frac{1}{\om_1^2\om_2^4}\left(h_2-\frac P9\right)^2
+\frac{1}{\om_1^2\om_3^4} h_2^2\right)
\\
&-\frac {1}{2\Omega_1^2\Omega_2^4\om_3^4}\left(4{\om}_0+ h_2\left(h_3-h_1\right)+\frac 19 P h_1\right)^2
\biggr\rbrace\,,\\
\end{split}
\eqlabel{5action}
\end{equation}
where $\Omega_0$ is a constant, $R_{10}$ is given by
\begin{equation}
\begin{split}
R_{10}=R_5&+\left(\frac{1}{2\om_1^2}+\frac{2}{\om_2^2}+\frac{2}{\om_3^2}-\frac{\om_2^2}{4\om_1^2\om_3^2}
-\frac{\om_3^2}{4\om_1^2\om_2^2}-\frac{\om_1^2}{\om_2^2\om_3^2}\right)-2\Box \ln\left(\om_1\om_2^2\om_3^2\right)\\
&-\biggl\{\left(\nabla\ln\om_1\right)^2+2\left(\nabla\ln\om_2\right)^2
+2\left(\nabla\ln\om_3\right)^2+\left(\nabla\ln\left(\om_1\om_2^2\om_3^2\right)\right)^2\biggr\}\,,
\end{split}
\eqlabel{ric5}
\end{equation}
and $R_5$ is the five dimensional Ricci scalar of the metric 
\begin{equation}
ds_{5}^2 =g_{\mu\nu}(y) dy^{\mu}dy^{\nu}\,.
\eqlabel{5met}
\end{equation}
All the covariant derivatives $\nabla_\lambda$  are
with respect to the metric \eqref{5met}.
Finally, $G_5$ is the five dimensional effective gravitational constant  
\begin{equation}
G_5\equiv \frac{729}{4\pi^3}G_{10}\,,
\eqlabel{g5deff}
\end{equation}
where $G_{10}$ is a 10-dimensional gravitational constant of 
type IIB supergravity.

From \eqref{5action} we obtain the following equations of motion:
\begin{equation}
\begin{split}
0=&\Box\Phi+\nabla\Phi \nabla \ln \Om_1\Om_2^2\Om_3^2+\frac 12 e^{-\Phi}\left(\frac{(h_1-h_3)^2}{2\om_1^2\om_2^2\om_3^2}+\frac{1}{\om_3^4}\left(\nabla h_1\right)^2
+\frac{1}{\om_2^4}\left(\nabla h_3\right)^2\right)
\\
&-\frac 12 e^{\Phi}\left(\frac{2}{\om_2^2\om_3^2}\left(\nabla h_2\right)^2
+\frac{1}{\om_1^2\om_2^4}\left(h_2-\frac P9\right)^2
+\frac{1}{\om_1^2\om_3^4} h_2^2\right)\,,
\end{split}
\eqlabel{ea1}
\end{equation}
\begin{equation}
\begin{split}
0=&\Box h_1-\nabla\Phi\nabla h_1+\nabla h_1\nabla \ln\frac{\om_1\om_2^2}{\om_3^2}
+\frac{(h_3-h_1)\om_3^2}{2\om_1^2\om_2^2}
+\frac{\left(h_2-\frac P9\right)e^\Phi}{\om_1^2\om_2^4}
\biggl(4{\om}_0\\
&+ h_2\left(h_3-h_1\right)
+\frac 19 P h_1\biggr)\,,
\end{split}
\eqlabel{ea2}
\end{equation}
\begin{equation}
\begin{split}
0=&\Box h_2+\nabla\Phi\nabla h_2+\nabla h_2\nabla \ln{\om_1}
-\frac{h_2\om_2^2}{2\om_1^2\om_3^2}-\frac{\left(h_2-\frac P9\right)\om_3^2}{2\om_1^2\om_2^2}
+\frac{(h_1-h_3)e^{-\Phi}}{2\om_1^2\om_2^2\om_3^2}
\biggl(4{\om}_0\\
&+ h_2\left(h_3-h_1\right)+\frac 19 P h_1\biggr)\,,
\end{split}
\eqlabel{ea3}
\end{equation}
\begin{equation}
\begin{split}
0=&\Box h_3-\nabla\Phi\nabla h_3+\nabla h_3\nabla \ln\frac{\om_1\om_3^2}{\om_2^2}
+\frac{(h_1-h_3)\om_2^2}{2\om_1^2\om_3^2}
-\frac{h_2 e^\Phi}{\om_1^2\om_3^4}
\biggl(4{\om}_0
+ h_2\left(h_3-h_1\right)\\
&+\frac 19 P h_1\biggr)\,,
\end{split}
\eqlabel{ea4}
\end{equation}
\begin{equation}
\begin{split}
0=&\om_1^{-1}\Box\om_1+\nabla\ln\om_1\nabla\ln\om_2^2\om_3^2+\frac{(\om_2^2-\om_3^2)^2-4\om_1^4}
{4\om_1^2\om_2^2\om_3^2}-\frac{e^{-\Phi}}{8\om_3^4}\left(\nabla h_1\right)^2
-\frac{e^{\Phi}}{4\om_2^2\om_3^2}\left(\nabla h_2\right)^2
\\&-\frac{e^{-\Phi}}{8\om_2^4}\left(\nabla h_3\right)^2
+\frac{3(h_1-h_3)^2 e^{-\Phi}}{16\om_1^2\om_2^2\om_3^2}
+\frac{3h_2^2 e^{\Phi}}{8\om_1^2\om_3^4}
+\frac{3\left(h_2-\frac P9\right)^2 e^{\Phi}}{8\om_1^2\om_2^4}
+\frac {1}{4\Omega_1^2\Omega_2^4\om_3^4}
\biggl(4{\om}_0\\
&+ h_2\left(h_3-h_1\right)+\frac 19 P h_1\biggr)^2\,,
\end{split}
\eqlabel{ea5}
\end{equation}
\begin{equation}
\begin{split}
0=&\om_2^{-1}\Box\om_2+\nabla\ln\om_2\nabla\ln\om_1\om_2\om_3^2+\frac{(2\om_1^2-\om_3^2)^2
-4\om_1^2\om_3^2-\om_2^4}
{8\om_1^2\om_2^2\om_3^2}-\frac{e^{-\Phi}}{8\om_3^4}\left(\nabla h_1\right)^2\\
&+\frac{e^{\Phi}}{4\om_2^2\om_3^2}\left(\nabla h_2\right)^2
+\frac{3e^{-\Phi}}{8\om_2^4}\left(\nabla h_3\right)^2
+\frac{(h_1-h_3)^2 e^{-\Phi}}{16\om_1^2\om_2^2\om_3^2}
-\frac{h_2^2 e^{\Phi}}{8\om_1^2\om_3^4}
+\frac{3\left(h_2-\frac P9\right)^2 e^{\Phi}}{8\om_1^2\om_2^4}
\\
&+\frac {1}{4\Omega_1^2\Omega_2^4\om_3^4}
\biggl(4{\om}_0
+ h_2\left(h_3-h_1\right)+\frac 19 P h_1\biggr)^2\,,
\end{split}
\eqlabel{ea6}
\end{equation}
\begin{equation}
\begin{split}
0=&\om_3^{-1}\Box\om_3+\nabla\ln\om_3\nabla\ln\om_1\om_2^2\om_3+\frac{(2\om_1^2-\om_2^2)^2
-4\om_1^2\om_2^2-\om_3^4}
{8\om_1^2\om_2^2\om_3^2}+\frac{3e^{-\Phi}}{8\om_3^4}\left(\nabla h_1\right)^2\\
&+\frac{e^{\Phi}}{4\om_2^2\om_3^2}\left(\nabla h_2\right)^2
-\frac{e^{-\Phi}}{8\om_2^4}\left(\nabla h_3\right)^2
+\frac{(h_1-h_3)^2 e^{-\Phi}}{16\om_1^2\om_2^2\om_3^2}
+\frac{3h_2^2 e^{\Phi}}{8\om_1^2\om_3^4}
-\frac{\left(h_2-\frac P9\right)^2 e^{\Phi}}{8\om_1^2\om_2^4}
\\
&+\frac {1}{4\Omega_1^2\Omega_2^4\om_3^4}
\biggl(4{\om}_0
+ h_2\left(h_3-h_1\right)+\frac 19 P h_1\biggr)^2\,,
\end{split}
\eqlabel{ea7}
\end{equation}
\begin{equation}
\begin{split}
R_{5\mu\nu}=&\om_1^{-1}\nabla_\mu\nabla_\nu\om_1+2 \om_2^{-1}\nabla_\mu\nabla_\nu\om_2+
2 \om_3^{-1}\nabla_\mu\nabla_\nu\om_3+\frac 12\nabla_\mu\Phi\nabla_\nu\Phi
\\
&+\frac{e^{-\Phi}}{2\om_3^4}\nabla_\mu h_1\nabla_\nu h_1
+\frac{e^{\Phi}}{\om_2^2\om_3^2}\nabla_\mu h_2\nabla_\nu h_2
+\frac{e^{-\Phi}}{2\om_2^4}\nabla_\mu h_3\nabla_\nu h_3-\frac 18 
g_{\mu\nu}\biggl[\\
&
e^{-\Phi}\left(\frac{(h_1-h_3)^2}{2\om_1^2\om_2^2\om_3^2}+\frac{1}{\om_3^4}\left(\nabla h_1\right)^2
+\frac{1}{\om_2^4}\left(\nabla h_3\right)^2\right)
+e^{\Phi}\biggl(\frac{2}{\om_2^2\om_3^2}\left(\nabla h_2\right)^2\\
&
+\frac{1}{\om_1^2\om_2^4}\left(h_2-\frac P9\right)^2
+\frac{1}{\om_1^2\om_3^4} h_2^2\biggr)+
\frac {2}{\Omega_1^2\Omega_2^4\om_3^4}
\left(4{\om}_0+ h_2\left(h_3-h_1\right)+\frac 19 P h_1\right)^2
\biggr]\,.
\end{split}
\eqlabel{ea8}
\end{equation}
We explicitly verified that equations \eqref{ea1}-\eqref{ea8} are equivalent to type IIB 
supergravity equations of motion provided the uplift is given by:
\begin{equation}
ds_{10}^2 =g_{\mu\nu}(y) dy^{\mu}dy^{\nu}+\om_1^2(y) g_5^2
+\om_2^2(y) \left[g_3^2+g_4^2\right]+\om_3^2(y) \left[g_1^2+g_2^2\right],
\eqlabel{10dmetric}
\end{equation}
for the 10-dimensional Einstein frame metric, and  
\begin{equation}
\begin{split}
&B_2=h_1(y)\ g_1\wedge g_2+h_3(y)\ g_3\wedge g_4\,,\\
&F_3=\frac 19 P\ g_5\wedge g_3\wedge g_4+h_2(y)\ \left(g_1\wedge g_2-g_3\wedge g_4\right)\wedge g_5
\\
&\qquad +\left(g_1\wedge g_3+g_2\wedge g_4\right)\wedge d\left(h_2(y)\right)\,,\\
&F_5=\biggl(1+\star_{10}\biggr)\bigg(4{\om}_0+ h_2(y)\left(h_3(y)-h_1(y)\right)+\frac 19 P h_1(y)\bigg)\ 
g_5 \wedge
g_3\wedge g_4\wedge 
g_1\wedge g_2\,,\\
&\Phi= \Phi(y)\,,
\end{split}
\eqlabel{fdil}
\end{equation}
for the fluxes and the dilaton.
In \eqref{10dmetric}, \eqref{fdil} $g_i$'s are the following 1-forms on $T^{1,1}$
\begin{equation}
\begin{split}
&g_1=\frac{\a^1-\a^3}{\sqrt 2}\,,\qquad g_2=\frac{\a^2-\a^4}{\sqrt 2}\,,\\
&g_3=\frac{\a^1+\a^3}{\sqrt 2}\,,\qquad g_4=\frac{\a^2+\a^4}{\sqrt 2}\,,\\
&g_5=\a^5\,,
\end{split}
\eqlabel{3form1}
\end{equation}
where 
\begin{equation}
\begin{split}
&\a^1=-\sin\theta_1 d\phi_1\,,\qquad \a^2=d\theta_1\,,\\
&\a^3=\cos\psi\sin\theta_2 d\phi_2-\sin\psi d\theta_2\,,\\
&\a^4=\sin\psi\sin\theta_2 d\phi_2+\cos\psi d\theta_2\,,\\
&\a^5=d\psi+\cos\theta_1 d\phi_1+\cos\theta_2 d\phi_2\,.
\end{split}
\eqlabel{3form2}
\end{equation}

\subsection{$\csb$ fluctuations in $SU(2)\times SU(2)\times U(1)$ invariant states of cascading gauge 
theory}\label{csbfluctuations}

In what follows we will be interested in the spectrum of $\csb$ fluctuations about 
chirally-symmetric states of  cascading gauge theory. These chirally-symmetric 
states are described by the gravitational configurations of \eqref{5action} subject to constraints   
\begin{equation}
h_1=h_3\,,\qquad h_2=\frac{P}{18}\,,\qquad \om_2=\om_3\,.
\eqlabel{cinv}
\end{equation}

Introducing 
\begin{equation}
\begin{split}
h_1=&\frac 1P\left(\frac{K_1}{12}-36\Om_0\right)\,,\qquad h_2=\frac{P}{18}\ K_2\,,\qquad 
h_3=\frac 1P\left(\frac{K_3}{12}-36\Om_0\right)\,,\\
\Om_1=&\frac 13 f_c^{1/2} h^{1/4}\,,\qquad \Om_2=\frac {1}{\sqrt{6}} f_a^{1/2} h^{1/4}\,,\qquad 
\Om_3=\frac {1}{\sqrt{6}} f_b^{1/2} h^{1/4}\,,
\end{split}
\eqlabel{redef}
\end{equation}
and
\begin{equation}
\begin{split}
K_1=&K+\dd k_1\,,\qquad K_2=1+\dd k_2\,,\qquad K_3=K-\dd k_1\,,
\\
f_c=&f_2\,,\qquad f_a=f_3+\dd f\,,\qquad f_b=f_3-\dd f\,,
\end{split}
\eqlabel{deffl}
\end{equation}
we find the following effective action for the  linearized fluctuations $\{\dd f,\dd k_1,\dd k_2\}$
 \cite{ksbh}
\begin{equation}
S_{\c\rm{SB}}\bigg[\dd f,\dd k_1,\dd k_2\bigg]
=\frac{1}{16\pi G_5}\int_{\calm_5}\ {\rm vol}_{\calm_5}\ h^{5/4}f_2^{1/2}f_3^2
\biggl\{\call_1+\call_2+\call_3+\call_4+\call_5\biggr\}\,,
\eqlabel{flaction}
\end{equation}
\begin{equation}
\begin{split}
\call_1=&-\frac{(\dd f)^2}{f_3^2}\left(
-\frac{P^2 e^\Phi}{2 f_2 h^{3/2} f_3^2}-\frac{(\nabla K)^2}{8 f_3^2 h P^2 e^\Phi}- \frac{K^2}{2f_2 h^{5/2} f_3^4}
\right)\,,
\end{split}
\eqlabel{call1}
\end{equation}
\begin{equation}
\begin{split}
\call_2=&-\frac{9f_3^2-24 f_2 f_3+4f_2^2}{f_2h^{1/2}f_3^4}\ (\dd f)^2+2\square\frac{(\dd f)^2}{f_3^2}
-\left(\nabla \frac{\dd f}{f_3}\right)^2\\
&+2\nabla\left(\ln h^{1/4}f_3^{1/2}\right)\nabla 
\left(\frac{(\dd f)^2}{f_3^2}\right)+2\nabla\left(\ln f_2^{1/2} h^{5/4}f_3^2\right)\nabla\left(\frac{(\dd f)^2}
{f_3^2}\right)\,,
\end{split}
\eqlabel{call2}
\end{equation}
\begin{equation}
\begin{split}
\call_3=&-\frac {1}{2P^2 e^\Phi}\biggl(\frac {9}{2 f_2 h^{3/2}f_3^2}\ (\dd k_1)^2
+\frac {1}{2h f_3^4} \biggl(3(\nabla K)^2\ (\dd f)^2+f_3^2\ (\nabla\dd k_1)^2
\\
&+4f_3 \dd f\ \nabla K\nabla \dd k_1
\biggr)
\biggr)\,,
\end{split}
\eqlabel{call3}
\end{equation}
\begin{equation}
\begin{split}
\call_4=&\frac{P^2 e^\Phi}{2}\biggl(\frac {2}{9hf_3^2}\ (\nabla \dd k_2)^2
+\frac{2}{f_2h^{3/2}f_3^4}\left(3\ (\dd f)^2+4f_3\ \dd f\dd k_2+f_3^3\ (\dd k_2)^2 \right)
\biggr)\,,
\end{split}
\eqlabel{call4}
\end{equation}
\begin{equation}
\begin{split}
\call_5=&\frac{K}{f_2 h^{5/2} f_3^6}\ \left(f_3^2\ \dd k_1\dd k_2-K\ (\dd f)^2\right)\,.
\end{split}
\eqlabel{call5}
\end{equation}

\section{Chirally symmetric phase of cascading gauge theory on $S^3$}\label{symmetric}

We consider here $SU(2)\times SU(2)\times U(1)\times SO(4)$ (chirally-symmetric) states of the 
strongly coupled cascading gauge theory. 
We find it convenient to use  a radial coordinate introduced in \cite{aby}: 
\begin{equation}
ds_5^2=g_{\mu\nu}(y)dy^\mu dy^\nu=h^{-1/2}\r^{-2}\ \biggl(-dt^2+f_1^2\ \left(d S_3\right)^2\biggr)
+h^{1/2}\r^{-2}\ (d\r)^2\,,
\eqlabel{metricaby}
\end{equation}  
where $(dS_3)^2$ is the metric on a round $S^3$ of unit size, and $h=h(\r)$, $f_1=f_1(\r)$.
Furthermore, we use parametrization \eqref{redef} and denote\footnote{Recall that for the 
unbroken chiral symmetry we must set $K_2(\r)\equiv 1$.}
\begin{equation}
f_c=f_2\,,\qquad f_a=f_b=f_3\,,\qquad K_1=K_3=K\,,\qquad \Phi=\ln g\,,
\eqlabel{om12}
\end{equation}
with $f_i=f_i(\r)$, and $K=K(\r)$, $g=g(\r)$. 

Notice that parametrization \eqref{metricaby} is not unique --- the diffeomorphisms 
of the type 
\begin{equation}
\left( \begin{array}{c}
\r  \\
h  \\
f_1\\
f_2\\
f_3\\
K\\
g   \end{array} \right)\
\Longrightarrow \left( \begin{array}{c}
\hr  \\
\hh  \\
\hf_1\\
\hf_2\\
\hf_3\\
\hK\\
\hat{g}   \end{array} \right)
=
\left( \begin{array}{c}
{\r}/{(1+\a\ \r)}  \\
(1+\a\ \r)^4\ h \\
f_1\\
(1+\a\ \r)^{-2}\ f_2\\
(1+\a\ \r)^{-2}\ f_3\\
K\\
{g}   \end{array} \right)\,,\qquad \a={\rm const}\,,
\eqlabel{leftover}
\end{equation}
preserve the general form of the metric. We can completely fix \eqref{leftover}, \ie
parameter $\a$ in \eqref{leftover},
requiring that for a geodesically complete $\calm_5$  the radial coordinate $\rho$
extends as 
\begin{equation}
\rho\in [0,+\infty)\,.
\eqlabel{extend}
\end{equation}

\subsection{Equations of motion}

For a background ansatz \eqref{metricaby}, \eqref{om12}, 
the  equations of motion obtained from \eqref{ea1}-\eqref{ea8} take form
\begin{equation}
\begin{split}
0=&f_1''-\frac{(f_1')^2}{f_1}+\left(\frac{6}{\r}- \frac{f_2'}{f_2}- \frac{3h'}{2h}-4 \frac{f_3'}{f_3}\right) f_1'
+\frac{f_1 (2 f_3-f_3' \r) f_2'}{f_3 \r f_2}-\frac{3 f_1 (f_3')^2}{2f_3^2}+ \frac{8f_1 f_3'}{f_3 \r}
\\
&+\frac{f_1 (h')^2}{4h^2}+\frac{2 f_1 h'}{h \r}+\frac{f_1 (K')^2}{8g h f_3^2 P^2}+\frac{f_1 (g')^2}{4g^2}-
\frac{f_1 K^2}{4f_2 f_3^4 h^2 \r^2}-\frac{f_1 P^2 g}{2f_2 f_3^2 h \r^2}+\frac{h}{f_1}
\\&+\frac{2 f_1 (6 f_3-f_2-3 f_3^2)}{f_3^2 \r^2}\,,
\end{split}
\eqlabel{eq1}
\end{equation}
\begin{equation}
\begin{split}
0=&f_2''+\frac{3 f_2 (f_1')^2}{f_1^2}- \left( \frac{15f_2}{f_1 \r}-\frac{9 f_2'}{2f_1}
-\frac{6 f_3' f_2}{f_3 f_1}-\frac{3h' f_2}{2h f_1}\right) f_1'
-\frac{(f_2')^2}{2f_2}-\frac{3 (2 f_3-f_3' \r) f_2'}{f_3 \r}\\
&+\frac{3f_2 (f_3')^2}{2f_3^2}
-\frac{12 f_2 f_3'}{f_3 \r}-\frac{f_2 (h')^2}{4h^2}
-\frac{2 f_2 h'}{h \r}-\frac{3f_2 (K')^2}{8g h f_3^2 P^2}-\frac{f_2 (g')^2}{4g^2}
+\frac{K^2}{4f_3^4 h^2 \r^2}\\
&+ \frac{3P^2 g}{2f_3^2 h \r^2}
-\frac{3 f_2 h}{f_1^2}+\frac{2 f_2 (7 f_3^2-3 f_2-6 f_3)}{f_3^2 \r^2}\,,
\end{split}
\eqlabel{eq2}
\end{equation}
\begin{equation}
\begin{split}
0=&f_3''+\frac{3 f_3 (f_1')^2}{f_1^2}- \left(\frac{15 f_3}{f_1 \r}-\frac{3f_2' f_3}{2f_1 f_2}-\frac{9 f_3'}{f_1}-\frac{3h' f_3}{2h f_1}\right) f_1'
-\frac{3(2 f_3-f_3' \r) f_2'}{2\r f_2}
+\frac{5(f_3')^2}{2f_3}\\
&-\frac{15 f_3'}{\r}-\frac{ f_3 (h')^2}{4h^2}-\frac{2 f_3 h'}{h \r}-\frac{(K')^2}{8g h f_3 P^2}-\frac{f_3 (g')^2}{4g^2}
+ \frac{K^2}{4f_2 f_3^3 h^2 \r^2}+\frac{P^2 g}{2f_3 h f_2 \r^2}-\frac{3 f_3 h}{f_1^2}\\
&-\frac{2 (12 f_3-3 f_2-7 f_3^2)}{f_3 \r^2}\,,
\end{split}
\eqlabel{eq3}
\end{equation}
\begin{equation}
\begin{split}
0=&h''-\frac{9 h (f_1')^2}{f_1^2}+\left(\frac{39 h}{\r f_1}-\frac{3h'}{2f_1}-\frac{18 f_3' h}{f_3 f_1}
-\frac{9 f_2' h}{2f_1 f_2}\right) f_1'+\left(\frac{8 h}{\r f_2}-\frac{3 f_3' h}{f_3 f_2}+\frac{h'}{2f_2}\right) f_2'
\\
&-\frac{9h (f_3')^2}{2f_3^2}+\frac{2 (16 h+h' \r) f_3'}{f_3 \r}-\frac{(h')^2}{4h}+\frac{3 h'}{\r}
+ \frac{5(K')^2}{8g f_3^2 P^2}+\frac{3h (g')^2}{4g^2}+\frac{K^2}{4f_2 f_3^4 h \r^2}
\\
&-\frac{P^2 g}{2f_3^2 f_2 \r^2}+\frac{9 h^2}{f_1^2}+\frac{2 h (18 f_3-3 f_2-17 f_3^2)}{f_3^2 \r^2}\,,
\end{split}
\eqlabel{eq4}
\end{equation}
\begin{equation}
\begin{split}
0=&K''+\left(\frac{3 f_1'}{f_1}- \frac{g'}{g}+\frac{f_2'}{2f_2}-\frac{ h'}{h}-\frac 3\r\right) K'-\frac{2 g P^2 K}{f_2 f_3^2 h 
\r^2}\,,
\end{split}
\eqlabel{eq5}
\end{equation}
\begin{equation}
\begin{split}
0=&g''-\frac{(g')^2}{g}+\left(\frac{3 f_1'}{f_1}+\frac{f_2'}{2f_2}+\frac{2 f_3'}{f_3}-\frac 3\r\right) g'+\frac{(K')^2}{4f_3^2 h P^2}
-\frac{g^2 P^2}{f_2 f_3^2 h \r^2}\,,
\end{split}
\eqlabel{eq6}
\end{equation}
Additionally we have the first order constraint
\begin{equation}
\begin{split}
0=&(K')^2+\frac{2 P^2 h f_3^2 (g')^2}{g}-\frac{24 P^2 g h f_3^2 (f_1')^2}{f_1^2}+12 P^2 g f_3 \biggl(\frac{6 f_3 h}{f_1 r}
-\frac{f_2' f_3 h}{f_1 f_2}-\frac{h' f_3}{f_1}\\
&-\frac{4 f_3' h}{f_1}\biggr) f_1'
+\frac{8 P^2 g f_3 h (2 f_3-f_3' r) f_2'}{r f_2}-12 P^2 g h (f_3')^2+\frac{64 P^2 g f_3 h f_3'}{r}+\frac{2 g P^2 f_3^2 (h')^2}{h}
\\
&+\frac{16 g P^2 f_3^2 h'}{r}-2 g P^2 \biggl(\frac{K^2}{f_2 h f_3^2 r^2}-\frac{12 h^2 f_3^2}{f_1^2}-\frac{48 h f_3}{r^2}+\frac{8 f_2 h}{r^2}
+\frac{2 g P^2}{f_2 r^2}+\frac{24 h f_3^2}{r^2}\biggr)\,.
\end{split}
\eqlabel{eq7}
\end{equation}
We explicitly verified that the constraint \eqref{eq7} is consistent with 
\eqref{eq1}-\eqref{eq6}.

\subsection{UV asymptotics}\label{uvcond}
The general UV (as $\r\to 0$) asymptotic solution of \eqref{eq1}-\eqref{eq7}  describing the symmetric phase of cascading 
gauge theory takes form
\begin{equation}
f_1=f_0 \left(1+\left(-\frac {K_0}{8}-\frac{1}{16} P^2 g_0+\frac 14 P^2 g_0 \ln \r\right)\frac{\r^2}{f_0^2}
+\sum_{n=3}^\infty\sum_{k} f_{n,k}\ \frac{\r^{n}}{f_0^{n}}\ln^k \r\right)\,,
\eqlabel{f1uv}
\end{equation}
\begin{equation}
f_2=1-\a_{1,0}\frac{\r}{f_0}
+\left(\frac{ K_0}{4}+\frac{ \a_{1,0}^2}{4}+\frac38 P^2 g_0-\frac12 P^2 g_0 
\ln \r\right)\frac{\r^2}{f_0^2} +\sum_{n=3}^\infty\sum_{k} a_{n,k}\ \frac{\r^{n}}{f_0^{n}}\ln^k \r\,,
\eqlabel{f2uv}
\end{equation}
\begin{equation}
f_3=1-\a_{1,0}\frac{\r}{f_0}+ \left(\frac{ K_0}{4}+\frac{ \a_{1,0}^2}{4}
+\frac{5}{16} P^2 g_0-\frac12 P^2 g_0 \ln \r\right)\frac{\r^2}{f_0^2}+\sum_{n=3}^\infty\sum_{k} b_{n,k}\ 
\frac{\r^{n}}{f_0^{n}}\ln^k \r\,,
\eqlabel{f3uv}
\end{equation}
\begin{equation}
\begin{split}
h=&\frac18 P^2 g_0+\frac14 K_0-\frac12 P^2 g_0 \ln \r+\a_{1,0}\left(\frac 12 K_0-P^2 g_0\ln\r\right)
\frac{\r}{f_0}+\biggl(
\frac{23}{288}P^4g_0^2-\frac18K_0^2\\
&-\frac16 P^2 g_0K_0+\frac{\a_{1,0}^2}{8}(5 K_0-2 P^2 g_0)
+\frac16 P^2g_0\left(3K_0+2P^2g_0-\frac{15}{2}\a_{1,0}^2
\right)\ln \r\\&
-\frac12 P^4g_0^2\ln^2\r
\biggr)\frac{\r^2}{f_0^2}
+\sum_{n=3}^\infty\sum_{k} h_{n,k}\ 
\frac{\r^{n}}{f_0^{n}}\ln^k \r\,,
\end{split}
\eqlabel{huv}
\end{equation}
\begin{equation}
\begin{split}
K=&K_0-2 P^2 g_0 \ln \r-P^2 g_0 \a_{10} \frac{\r}{f_0}
+ \left(\frac14 P^2 g_0 (K_0+3 P^2 g_0-\a_{1,0}^2)-\frac 12 P^4 g_0^2 \ln \r\right)\frac{\r^2}{f_0^2}
\\
&+\sum_{n=3}^\infty\sum_{k} K_{n,k}\ \frac{\r^{n}}{f_0^{n}}\ln^k \r
\,,
\end{split}
\eqlabel{kuv}
\end{equation}
\begin{equation}
g=g_0 \left(1-\frac14 P^2 g_0\ \frac{\r^2}{f_0^2}+\sum_{n=3}^\infty\sum_{k} g_{n,k}\ \frac{\r^{n}}{f_0^{n}}\ln^k \r\right)\,.
\eqlabel{guv}
\end{equation}
It is characterized by 9 parameters:
\begin{equation}
\{K_0\,,\ f_0\,,\ g_0\,,\  \a_{1,0}\,,\ b_{4,0}\,,\ a_{4,0}\,,\ a_{6,0}\,,\ a_{8,0}\,,\ g_{4,0}\}\,.
\eqlabel{uvpar}
\end{equation}
In what follows we developed the UV expansion to order $\calo(\r^{12})$ inclusive.

\subsection{IR asymptotics}\label{ircond}
We use a radial coordinate $\r$ that extends to infinity, see \eqref{extend}. Introducing 
\begin{equation}
y\equiv \frac 1\r\,,\qquad h^h\equiv y^{-4}\ h\,,\qquad f_1^h\equiv y^{-1}\ f_1\,,\qquad  f^h_{2,3}\equiv y^2\ f_{2,3}\,,
\eqlabel{horfunc}
\end{equation}
the general IR (as $y\to 0$) asymptotic solution of  \eqref{eq1}-\eqref{eq7} describing the symmetric phase of cascading 
gauge theory takes form
\begin{equation}
\begin{split}
f_1^h=h_0^h-\frac{g^h_0 P^2 (h_0^h)^2 (f_{3,0}^h)^2+2 (K_0^h)^2-8 f_{2,0}^h (h_0^h)^4 (f_{3,0}^h)^2 (f_{2,0}^h-6 f_{3,0}^h)}
{144(f_{3,0}^h)^4 (h_0^h)^3 f_{2,0}^h}  
y^2+\sum_{n=2} f_{1,n}^h y^{2n}\,,
\end{split}
\eqlabel{f1hy}
\end{equation}
\begin{equation}
f_2^h=f_{2,0}^h-\frac{g^h_0 P^2-8 (h_0^h)^2 (f_{2,0}^h)^2}{8(f_{3,0}^h)^2 (h_0^h)^2}  y^2+\sum_{n=2} f_{2,n}^h y^{2n}\,,
\eqlabel{f2hy}
\end{equation}
\begin{equation}
f_3^h=f_{3,0}^h+\frac{3 f_{3,0}^h-f_{2,0}^h}{2f_{3,0}^h} y^2+\sum_{n=2} f_{3,n}^h y^{2n}\,,
\eqlabel{f3hy}
\end{equation}
\begin{equation}
h^h=(h_0^h)^2-\frac{(K_0^h)^2+g^h_0 P^2 (h_0^h)^2 (f_{3,0}^h)^2}{8(h_0^h)^2 (f_{3,0}^h)^4 f_{2,0}^h}  y^2+\sum_{n=2} h_{n}^h y^{2n}\,,
\eqlabel{hhy}
\end{equation}
\begin{equation}
K=K_0^h+\frac{g^h_0 P^2 K_0^h}{4(f_{3,0}^h)^2 (h_0^h)^2 f_{2,0}^h}  y^2+\sum_{n=2} K_{n}^h y^{2n}\,,
\eqlabel{khy}
\end{equation}
\begin{equation}
g=g^h_0+\frac{(g^h_0)^2 P^2}{8(f_{3,0}^h)^2 (h_0^h)^2f_{2,0}^h}   y^2+\sum_{n=2} g_{n}^h y^{2n}\,.
\eqlabel{ghy}
\end{equation}
It is characterized by 5 parameters:
\begin{equation}
\{K_0^h\,,\ h_0^h\,,\ g_0^h\,,\  f_{2,0}^h\,,\ f_{3,0}^h\}\,.
\eqlabel{irpar}
\end{equation}
In what follows we developed the IR expansion to order $\calo(y^{12})$ inclusive.

\subsection{Symmetries}\label{symmetries}
The background geometry \eqref{metricaby}, \eqref{om12} enjoys 4 distinct scaling symmetries. 
We now discuss these symmetries and exhibit their action on the asymptotic parameters \eqref{uvpar}.
\nxt First, we have:
\begin{equation}
\begin{split}
&P\to \l\ P\,,\ g\to \frac 1\l\ g\,,\qquad \{\r,f_i,h,K\}\to \{\r,f_i,h,K\}\,,\qquad 
\{y,f_i^h,h^h\}\to \{y,f_i,h^h\}\,,
\end{split}
\eqlabel{scale1}
\end{equation} 
which acts on the asymptotic parameters as 
\begin{equation}
\begin{split}
&g_0\to \frac 1\l\ g_0\,,\\
&\{K_0\,,
f_0\,,  \a_{1,0}\,, b_{4,0}\,, a_{4,0}\,, a_{6,0}\,, a_{8,0}\,, g_{4,0}\}
\to \{K_0\,, f_0\,,  \a_{1,0}\,, b_{4,0}\,, a_{4,0}\,, a_{6,0}\,, a_{8,0}\,, g_{4,0}\}\,,
\end{split}
\eqlabel{action1}
\end{equation}
and
\begin{equation}
\begin{split}
\{K_0^h\,,\ h_0^h\,,\ g_0^h\,,\ f_{2,0}^h\,,\ f_{3,0}^h\}\to \{K_0^h\,,\ h_0^h\,,\  \l^{-1} g_0^h\,,\ f_{2,0}^h\,,\ f_{3,0}^h\}\,.
\end{split}
\eqlabel{action1h}
\end{equation}
We can use the exact symmetry \eqref{scale1} to set 
\begin{equation}
g_0=1\,.
\eqlabel{res1}
\end{equation}
\nxt Second, we have:
\begin{equation}
\begin{split}
&P\to \l\ P\,,\ \r\to \frac 1\l\ \r\,,\ h\to \l^2\ h\,,\ K\to \l^2 K\,,\qquad \{f_i,g\}\to\{f_i, g\}\,,
\\
&\{y,f_1^h,f_2^h,f_3^h,h^h\}\to \{\l y,\l^{-1} f_1^h,\l^2 f_2^h,\l^2 f_3^h,\l^{-2} h^h\}\,,
\end{split}
\eqlabel{scale2}
\end{equation}
which acts on the asymptotic parameters as 
\begin{equation}
\begin{split}
\{g_0\,,\ f_0\}\to \{g_0\,,\ f_0\} \,,
\end{split}
\eqlabel{action21}
\end{equation}
\begin{equation}
\begin{split}
\a_{1,0}\to \l\a_{1,0}\,,
\end{split}
\eqlabel{action21h}
\end{equation}
\begin{equation}
\begin{split}
K_0\to \l^2 \biggl(K_0-2P^2 g_0\ \ln\l \biggr)\,,
\end{split}
\eqlabel{action22}
\end{equation}
\begin{equation}
\begin{split}
b_{4,0}&\to \l^4\biggl(b_{4,0}-\left(\frac{5}{32} P^4 g_0^2+\frac{1}{12} P^2 g_0 K_0\right) \ln\l
+\frac{1}{12} P^4 g_0^2 \ln^2\l
\biggr)\,,
\end{split}
\eqlabel{action23}
\end{equation}
\begin{equation}
\begin{split}
g_{4,0}&\to \l^4\biggl(g_{4,0}+\left(3 (a_{4,0}-b_{4,0})-\frac{3}{64} P^2 g_0 K_0
-\frac{3}{64} P^4 g_0^2\right) \ln\l
\biggr)\,,
\end{split}
\eqlabel{action25}
\end{equation}
\begin{equation}
\begin{split}
a_{4,0}&\to \l^4\biggl(a_{4,0}-\left(\frac{1}{12} P^2 g_0 K_0+\frac{3}{16} P^4 g_0^2\right) 
\ln\l+\frac{1}{12} P^4 g_0^2 \ln^2\l
\biggr)\,,
\end{split}
\eqlabel{action26}
\end{equation}
\begin{equation}
\begin{split}
&a_{6,0}\to \l^6\biggl(a_{6,0}+\biggl(\frac{107}{80} a_{4,0} P^2 g_0-\frac{1}{640} P^2 g_0 K_0^2
-\frac{629}{76800} P^4 g_0^2 K_0
-\frac{101}{80} P^2 g_0 b_{4,0}\\
&-\frac{1}{10} P^2 g_0 g_{4,0}-\frac{1}{10} K_0 b_{4,0}+\frac{11959}{768000} P^6 g_0^3
+\frac{1}{10} a_{4,0} K_0-\frac{97P^2g_0+120K_0}{1920}P^2g_0\a_{1,0}^2\biggr) \ln\l\\
&+\left(\frac14 P^2 g_0 b_{4,0}-\frac14 a_{4,0} P^2 g_0-\frac{623}{38400} P^6 g_0^3
+\frac{1}{1280} P^4 g_0^2 K_0+\frac{1}{16}\a_{1,0}^2 P^4 g_0^2\right) \ln^2\l\\
&+\frac{1}{480} P^6 g_0^3 \ln^3\l
\biggr)\,,
\end{split}
\eqlabel{action27}
\end{equation}
\begin{equation}
\begin{split}
&a_{8,0}\to \l^8\biggl(a_{8,0}+\frac{1}{P^2 g_0 (70 K_0-141 P^2 g_0)} \biggl
(18 K_0^2 (b_{4,0}-a_{4,0})^2+(\frac{5}{16} K_0^3 (b_{4,0}-a_{4,0})\\
&-\frac{5706}{35} (b_{4,0}-a_{4,0})^2 K_0+24 K_0 g_{4,0} (b_{4,0}-a_{4,0})
+\frac{35}{2}a_{1,0}^2K_0^2(a_{4,0}-b_{4,0})) P^2 g_0+(2 a_{4,0}^2\\
&-36 a_{4,0} (b_{4,0}-a_{4,0})
+12 g_{4,0} (b_{4,0}-a_{4,0})
-\frac{23}{48} a_{4,0} K_0^2+\frac{3437}{480} K_0^2 (b_{4,0}-a_{4,0})\\
&+\frac{47193}{70} (b_{4,0}-a_{4,0})^2
-140 a_{8,0}+\frac98 K_0^2 g_{4,0}-8 g_{4,0}^2+\frac{17}{2048} K_0^4
+\frac{2731}{16} a_{4,0} K_0 \a_{1,0}^2\\
&+350 \a_{1,0}^2 a_{6,0}-\frac{875}{4} \a_{1,0}^4 a_{4,0}-\frac{35}{128}
 \a_{1,0}^2 K_0^3-\frac{35}{2} 
\a_{1,0}^2 K_0 g_{4,0}-\frac{175}{96} \a_{1,0}^4 K_0^2\\
&-\frac{2521}{16} K_0 b_{4,0}
\a_{1,0}^2) P^4 g_0^2
+(\frac{2051699}{100800} K_0 (b_{4,0}-a_{4,0})
-\frac{927}{560} a_{4,0} K_0+\frac{2063}{32256} K_0^3\\
&+\frac{1049}{1680} K_0 g_{4,0}-\frac{185}{3} a_{6,0}
-\frac{575}{4} a_{4,0} \a_{1,0}^2-\frac{2533}{3072} \a_{1,0}^2 K_0^2+\frac{33}{2} g_{4,0} \a_{1,0}^2-\frac{2135}{192}
 \a_{1,0}^6\\
&+\frac{2945}{192} \a_{1,0}^4 K_0+\frac{1405}{8} b_{4,0} \a_{1,0}^2
) P^6 g_0^3
+(\frac{1000999}{43200} a_{4,0}
-\frac{13889}{15120} g_{4,0}+\frac{30969307}{270950400} K_0^2\\
&-\frac{3931199}{151200}
 b_{4,0}+\frac{8993}{288}\a_{1,0}^4-\frac{995093}{645120}\a_{1,0}^2K_0) P^8 g_0^4
+(\frac{2541334849}{4741632000} K_0\\
&-\frac{2003273}{403200}\a_{1,0}^2) P^{10} g_0^5
+\frac{6274690897}{14224896000} P^{12} g_0^6\biggr) \ln\l
+\biggl(-\frac{36}{35} (b_{4,0}-a_{4,0})^2\\
&-\frac{3}{80} (K_0-\frac{50}{3}\a_{1,0}^2) (b_{4,0}-a_{4,0}) P^2 g_0
+(\frac{389}{1200} a_{4,0}
-\frac{1}{2016} K_0^2
-\frac{3}{140} g_{4,0}-\frac{219}{700} b_{4,0}\\
&+\frac{1}{512}\a_{1,0}^2K_0+\frac{5}{192}
\a_{1,0}^4) P^4 g_0^2-(\frac{3103}{7526400} K_0 +\frac{3961}{107520}\a_{1,0}^2)P^6 g_0^3
+\frac{19094567}{2370816000} P^8 g_0^4\biggr) \\
&\times\ln^2\l+\biggl(\frac{1}{40} (b_{4,0}-a_{4,0}) P^4 g_0^2+(\frac{47}{161280} K_0+\frac{1}{192}\a_{1,0}^2) P^6 g_0^3
-\frac{37889}{11289600}
 P^8 g_0^4\biggr) 
\ln^3\l\\
&+\frac{1}{20160} P^8 g_0^4 \ln^4\l
\biggr)\,,
\end{split}
\eqlabel{action28}
\end{equation}
and 
\begin{equation}
\begin{split}
&\{K_0^h\,,\ h_0^h\,,\ g_0^h\,,\ f_{2,0}^h\,,\ f_{3,0}^h\}\to \{\l^2 K_0^h\,,\ \l^{-1} h_0^h\,,\ g_0^h\,,\ \l^2 f_{2,0}^h\,,\ \l^2 f_{3,0}^h\}\,.
\end{split}
\eqlabel{action2h}
\end{equation}
We can use the exact symmetry \eqref{scale2} to relate different sets of $\{K_0,P\}$.
For the study of perturbative in $P^2/K_0$ expansion we find it convenient to set $K_0=1$ and 
vary $P^2$. To access the infrared properties of the theory we set $P=1$ and vary $K_0$.
Notice that the two approaches connect at $\{K_0=1,P=1\}$.
\nxt Third, we have:
\begin{equation}
\begin{split}
&\r\to \l\ \r\,,\ f_1\to \l\ f_1\,,\qquad \{P\,,\ f_2\,,\ f_3\,, h\,, K\,, g\}\to 
 \{P\,,\ f_2\,,\ f_3\,, h\,, K\,, g\}\,,\\
&\{y,f_1^h,f_2^h,f_3^h,h^h\}\to \{\l^{-1} y,\l^2 f_1^h,\l^{-2}f_2^h,\l^{-2}f_3^h,\l^4 h^h\}\,,
\end{split}
\eqlabel{scale3} 
\end{equation}
provided we rescale the four-dimensional metric component $G_{tt}=-1\to -\l^2$.
This scaling symmetry acts on the asymptotic parameters as 
\begin{equation}
\begin{split}
\{g_0\,,\ \a_{1,0}\,,\ f_0\}\to \{g_0\,,\ \a_{1,0}\,,\ \l\ f_0 \}\,,
\end{split}
\eqlabel{action31}
\end{equation}
\begin{equation}
\begin{split}
K_{0}\to K_0+2 P^2 g_0 \ln\l\,,
\end{split}
\eqlabel{action312}
\end{equation}
\begin{equation}
\begin{split}
b_{4,0}\to b_{4,0}+\left(\frac{5}{32} P^4 g_0^2+\frac{1}{12} P^2 g_0 K_0\right) \ln\l+\frac{1}{12} P^4 g_0^2 \ln^2\l\,,
\end{split}
\eqlabel{action32}
\end{equation}
\begin{equation}
\begin{split}
g_{4,0}\to g_{4,0}+\left(3 (b_{4,0}-a_{4,0})+\frac{3}{64} P^2 g_0 K_0+\frac{3}{64} P^4 g_0^2\right) \ln\l\,,
\end{split}
\eqlabel{action33}
\end{equation}
\begin{equation}
\begin{split}
a_{4,0}\to a_{4,0}+\left(\frac{1}{12} P^2 g_0 K_0+\frac{3}{16} P^4 g_0^2\right) \ln\l+\frac{1}{12} P^4 g_0^2 \ln^2\l\,,
\end{split}
\eqlabel{action34}
\end{equation}
\begin{equation}
\begin{split}
&a_{6,0}\to a_{6,0}+\biggl(\frac{1}{10} K_0 (b_{4,0}-a_{4,0})+(-\frac{107}{80} a_{4,0}+\frac{1}{10} g_{4,0}
+\frac{1}{640} K_0^2+\frac{101}{80} b_{4,0}\\
&+\frac{1}{16}a_{1,0}^2K_0) P^2 g_0
+(\frac{629}{76800} K_0+\frac{97}{1920}\a_{1,0}^2) P^4 g_0^2-\frac{11959}{768000} P^6 g_0^3\biggr) \ln\l\\
&+\biggl(
\frac 14 (b_{4,0}-a_{4,0}) P^2 g_0+(\frac{1}{16}\a_{1,0}^2+\frac{1}{1280}K_0) P^4 g_0^2
-\frac{623}{38400} P^6 g_0^3\biggr) \ln^2\l\\
&-\frac{1}{480} P^6 g_0^3 \ln^3\l\,,
\end{split}
\eqlabel{action35}
\end{equation}
\begin{equation}
\begin{split}
&a_{8,0}\to a_{8,0}+\frac{1}{P^2 g_0 (70 K_0-141 P^2 g_0)} \biggl(-18 K_0^2 (b_{4,0}-a_{4,0})^2
+(\frac{5706}{35} K_0 (b_{4,0}-a_{4,0})^2\\
&-24 K_0 g_{4,0} (b_{4,0}-a_{4,0})-\frac{5}{16} K_0^3 (b_{4,0}-a_{4,0})
-\frac{35}{2}\a_{1,0}^2K_0^2(a_{4,0}-b_{4,0})) P^2 g_0+(8 g_{4,0}^2\\
&-12 g_{4,0} (b_{4,0}-a_{4,0})
-\frac{17}{2048} K_0^4-2 a_{4,0}^2
+36 a_{4,0} (b_{4,0}-a_{4,0})-\frac{47193}{70} (b_{4,0}-a_{4,0})^2\\
&+\frac{23}{48} K_0^2 a_{4,0}
-\frac{3437}{480} K_0^2 (b_{4,0}-a_{4,0})-\frac 98 K_0^2 g_{4,0}+140 a_{8,0}
-350 \a_{1,0}^2 a_{6,0}+\frac{2521}{16} K_0 b_{4,0} \a_{1,0}^2\\
&+\frac{875}{4} \a_{1,0}^4 a_{4,0}-\frac{2731}{16}
 a_{4,0} K_0 \a_{1,0}^2+\frac{35}{128} \a_{1,0}^2 K_0^3+\frac{175}{96} \a_{1,0}^4 K_0^2+\frac{35}{2} \a_{1,0}^2K_0 
g_{4,0}
) P^4 g_0^2
\\
&+(\frac{927}{560} a_{4,0} K_0
-\frac{2051699}{100800} K_0 (b_{4,0}-a_{4,0})+\frac{185}{3} a_{6,0}-\frac{1049}{1680} K_0 g_{4,0}
-\frac{2063}{32256} K_0^3
\\
&+\frac{2533}{3072}\a_{1,0}^2K_0^2-\frac{2945}{192}\a_{1,0}^4K_0-\frac{1405}{8}
b_{4,0}\a_{1,0}^2-\frac{33}{2}g_{4,0}\a_{1,0}^2+\frac{2135}{192}\a_{1,0}^6
\\
&+\frac{575}{4}a_{4,0}\a_{1,0}^2
) P^6 g_0^3
+(\frac{13889}{15120} g_{4,0}
-\frac{1000999}{43200} a_{4,0}
-\frac{30969307}{270950400} K_0^2+\frac{3931199}{151200} b_{4,0}\\
&+\frac{995093}{645120}\a_{1,0}^2K_0
-\frac{8993}{288}\a_{1,0}^4) 
P^8 g_0^4+(\frac{2003273}{403200}\a_{1,0}^2-\frac{2541334849}{4741632000} K_0) P^{10} g_0^5
\\
&-\frac{6274690897}{14224896000} P^{12} g_0^6\biggr) \ln\l
+\biggl(-\frac{36}{35} (b_{4,0}-a_{4,0})^2-\frac{3}{80} 
(K_0-\frac{50}{3}\a_{1,0}^2) (b_{4,0}-a_{4,0}) P^2 g_0
\\
&+(-\frac{1}{2016} K_0^2+\frac{389}{1200} a_{4,0}
-\frac{3}{140} g_{4,0}-\frac{219}{700} b_{4,0}+\frac{5}{192}\a_{1,0}^4+\frac{1}{512}\a_{1,0}^2K_0) P^4 g_0^2
\\&-(\frac{3103}{7526400}K_0+\frac{3961}{107520}\a_{1,0}^2) P^6 g_0^3
+\frac{19094567}{2370816000} P^8 g_0^4\biggr) \ln^2\l
+\biggl(\frac{1}{40} (a_{4,0}-b_{4,0}) P^4 g_0^2\\
&-(\frac{47}{161280}K_0 +\frac{1}{192}\a_{1,0}^2)
P^6 g_0^3
+\frac{37889}{11289600} P^8 g_0^4\biggr) \ln^3\l
+\frac{1}{20160} P^8 g_0^4 \ln^4\l\,,
\end{split}
\eqlabel{action36}
\end{equation}
and 
\begin{equation}
\begin{split}
&\{K_0^h\,,\ h_0^h\,,\ g_0^h\,,\ f_{2,0}^h\,,\ f_{3,0}^h\}\to \{K_0^h\,,\ \l^{2} h_0^h\,,\ g_0^h\,,\ \l^{-2} f_{2,0}^h\,,\ \l^{-2} f_{3,0}^h\}\,.
\end{split}
\eqlabel{action3h}
\end{equation}
We can use the exact symmetry \eqref{scale3} 
to set
\begin{equation}
f_0=1\,.
\eqlabel{res3f0}
\end{equation}
\nxt Forth, we have residual diffeomorphisms \eqref{leftover} of the metric parametrization \eqref{metricaby}.
The latter transformations act on asymptotic parameters as 
\begin{equation}
\begin{split}
\{g_0\,,\  f_0\,,\ K_0\}\to \{g_0\,,\ f_0\,, K_0 \}\,,
\end{split}
\eqlabel{action41}
\end{equation}
\begin{equation}
\begin{split}
\a_{1,0}\to \a_{1,0}+2 \a f_0 \,,
\end{split}
\eqlabel{action412}
\end{equation}
\begin{equation}
\begin{split}
a_{4,0}\to a_{4,0}-\frac 14 P^2 g_0 \a f_0 (\a_{1,0}+\a f_0)\,,
\end{split}
\eqlabel{action42}
\end{equation}
\begin{equation}
\begin{split}
b_{4,0}\to  b_{4,0}-\frac 14 P^2 g_0 \a f_0 (\a_{1,0}+\a f_0)\,,
\end{split}
\eqlabel{action43}
\end{equation}
\begin{equation}
\begin{split}
g_{4,0}\to  g_{4,0}-\frac 34 P^2 g_0 \a f_0 (\a_{1,0}+\a f_0)\,,
\end{split}
\eqlabel{action44}
\end{equation}
\begin{equation}
\begin{split}
&a_{6,0}\to a_{6,0}+3 a_{4,0} \a f_0 (\a f_0+\a_{1,0})-\frac{\a f_0}{24}  (\a f_0+\a_{1,0}) (5 K_0-3 \a_{1,0}^2+3 \a_{1,0} \a f_0\\
&+3 \a^2 f_0^2) P^2 g_0-\frac{37}{96} \a f_0 (\a f_0+\a_{1,0}) P^4g_0^2\,,
\end{split}
\eqlabel{action45}
\end{equation}
\begin{equation}
\begin{split}
&a_{8,0}\to a_{8,0}+\frac{\a f_0}{20}  (\a f_0+\a_{1,0}) (9 K_0 a_{4,0}-100 a_{4,0} \a_{1,0}^2+100 a_{4,0} 
\a_{1,0} \a f_0+100 a_{4,0} \a^2 f_0^2
\\
&-9 K_0 b_{4,0}+200 a_{6,0})
-\frac{\a f_0}{11520}  (\a f_0+\a_{1,0}) (2880 \a_{1,0}^4-2880 \a_{1,0}^3 \a f_0-1920 \a_{1,0}^2 \a^2 f_0^2\\
&-2920 \a_{1,0}^2 K_0+1920 \a^3 f_0^3 \a_{1,0}+6160 \a f_0 K_0 \a_{1,0}+960 \a^4 f_0^4+6160 \a^2 f_0^2 K_0+5184 g_{4,0}
\\&+62568 b_{4,0}-66456 a_{4,0}+81 K_0^2) P^2g_0-\frac{\a f_0}{460800}  (\a f_0+\a_{1,0}) (16623 K_0
-251240 \a_{1,0}^2\\
&+327200 \a_{1,0} \a f_0+327200 \a^2 f_0^2) P^4g_0^2+\frac{82711}{1536000} \a f_0 (\a f_0+\a_{1,0}) P^6g_0^3\,,
\end{split}
\eqlabel{action46}
\end{equation}
and 
\begin{equation}
\begin{split}
\{K_0^h\,,\ h_0^h\,,\ g_0^h\,,\ f_{2,0}^h\,,\ f_{3,0}^h\}\to \{K_0^h\,,\  h_0^h\,,\ g_0^h\,,\  f_{2,0}^h\,,\  f_{3,0}^h\}\,.
\end{split}
\eqlabel{action4h}
\end{equation}
As mentioned earlier, the diffeomorphisms \eqref{leftover} can be completely fixed requiring that 
\begin{equation}
\lim_{\r\to +\infty} f_1(\rho)=0\,,
\eqlabel{geocoms}
\end{equation}
\ie in the holographic dual to the symmetric phase of  cascading gauge theory the manifold $\calm_5$ geodesically completes 
in the interior with smooth shrinking of $S^3$ (see \eqref{metricaby}) as $\r\to +\infty$.

\subsection{Keeping the physical parameters fixed}
Cascading gauge theory on $S^3$ has two dimensionfull physical parameters: the strong coupling 
scale $\Lambda$ and the scale $\mu\equiv \frac {1}{f_0}$, set by the size of the
3-sphere, and a dimensionless physical parameters $P^2\,, g_0$.  
Recall that a symmetry transformation \eqref{scale3} rescales $\mu$, and a symmetry transformation \eqref{scale2} rescales $P$ 
and affects $K_0$, while 
leaving the combination 
\begin{equation}
\frac{K_0}{P^2g_0}-2 \ln f_0+
\ln P^2 g_0\ =\ {\rm invariant}\ \equiv -2 \ln\Lambda + 2\ln\mu = \ln\frac{\mu^2}{\Lambda^2}
\eqlabel{scalel}
\end{equation}
invariant. The latter invariant defines the strong coupling scale $\Lambda$ of  cascading gauge theory. 
In particular, using the symmetry choices \eqref{res1} and \eqref{res3f0} we identify 
\begin{equation}
\frac{K_0}{P^2}= \ln \frac{\mu^2}{\Lambda^2 P^2 }\equiv \frac 1\dd \,.
\eqlabel{physical}
\end{equation}
Notice that \eqref{physical} is not invariant under the symmetry transformation \eqref{scale2}. This is because 
such transformation modifies $P^2 g_0$, and thus changes the theory;  \eqref{physical} is invariant under the residual 
diffeomorphisms \eqref{leftover}.

As defined in \eqref{physical}, a new dimensionless parameter $\dd$ is small when the IR cutoff set by the 
$S^3$ is much higher than the strong coupling scale $\Lambda$ (and thus  cascading gauge theory is close to be conformal).
In section \ref{pertubative} we develop perturbative expansion in $\dd$.

\subsection{Numerical procedure}\label{numerical}
Although we would like to have an analytic control over the gravitational solution dual to a symmetric 
phase of cascading gauge theory, the relevant equations 
for $\{f_1$, $f_2$, $f_3$, $h$, $K$, $g\}$ \eqref{eq1}-\eqref{eq7} are rather complicated. 
Thus, we have to resort to numerical analysis. Recall that various scaling symmetries of the background equations 
of motion allowed us to set  (see \eqref{res1} and \eqref{res3f0})
\begin{equation}
\lim_{\r\to 0} g\equiv g_0=1\,,\qquad \lim_{\r\to 0} f_1\equiv f_0=1\,.
\eqlabel{2symm}
\end{equation}
While the metric parametrization \eqref{metricaby} has  residual diffeomorphisms \eqref{leftover}, the latter are fixed 
once we insist on the IR asymptotics at $y\equiv \frac 1\r\to 0$ (see \eqref{geocoms}). Finally, a scaling symmetry 
\eqref{scale2} relates different pairs $\{K_0,P\}$ so that only the ratio $\frac{K_0}{P^2}\equiv \frac 1\dd$ is physically meaningful 
(see \eqref{physical}). In the end, for a fixed $\dd$, the gravitational solution is characterized by 6 parameters in the UV 
and 5 parameters in the IR:
\begin{equation}
\begin{split}
&{\rm UV}:\qquad \{\a_{1,0}\,,\ b_{4,0}\,,\ a_{4,0}\,,\ a_{6,0}\,,\ a_{8,0}\,,\ g_{4,0}\}\,,
\\
&{\rm IR}:\qquad \{K_0^h\,,\ h_0^h\,,\ g_0^h\,,\ f_{2,0}^h\,,\ f_{3,0}^h\}\,.
\end{split}
\eqlabel{uvirfinal}
\end{equation} 
Notice that $6+5=11$ is precisely the number of integration constants needed to specify a solution to \eqref{eq1}-\eqref{eq7} ---
we have 6 second order differential equations and a single first order differential constraint: $2\times 6-1=11$.

In practice, we replace the second-order differential equation for $f_2$ \eqref{eq2} with the constraint equation \eqref{eq7},
which we use to  algebraically eliminate $f_2'$ from \eqref{eq1}, \eqref{eq3}-\eqref{eq6}. The solution is found using the 
``shooting'' method as detailed in \cite{abk}. 
 
Finding a ``shooting'' solution in 11-dimensional parameter space  \eqref{uvirfinal} is quite challenging. Thus, we start with 
(leading) analytic results for $\dd\ll 1$ (see section \ref{pertubative}) and construct numerical solution 
for $(K_0=1, P^2)$ slowly incrementing $P^2$ from zero to one. Starting with the solution at $K_0=P^2=1$ 
we slowly decrease $K_0$ while keeping $P^2=1$. 

\subsection{Symmetric phase of cascading gauge theory at $\frac{\mu}{\Lambda}\gg 1$}
\label{pertubative}

In this section we describe perturbative solution in $\dd\ll 1$ \eqref{physical} to 
\eqref{eq1}-\eqref{eq7}.
Such gravitational backgrounds describe cascading gauge theory compactified on small $S^3$, \ie  the cutoff $\mu$ set by the 
compactification scale  is well above the strong coupling scale $\Lambda$ of cascading gauge theory. 

In the limit $\dd\to 0$ (or equivalently $P\to 0$) 
the gravitational background is simply that of the Klebanov-Witten model \cite{kw}:
\begin{equation}
\begin{split}
\dd=0: \qquad &f_1^{(0)}=\frac{2}{\sqrt{4+\hK_0 \r^2}}\,,\qquad f_2^{(0)}=f_3^{(0)}=1+\frac{\hK_0}{4}\r^2\,,
\qquad h^{(0)}=\frac{4 \hK_0}{(4+\hK_0\r^2)^2}\,,\\
&K^{(0)}=\hK_0\,,\qquad g^{(0)}=1\,,
\end{split}
\eqlabel{kw}
\end{equation}
where $\hK_0$ is a constant.
Perturbatively, we find
\begin{equation}
\begin{split}
&f_i(\r)=f_i^{(0)}\times\  \sum_{j=0}^\infty \left(\frac{P^2}{\hK_0}\right)^j\ f_{i,j}(\r^2 \hK_0)\,,\qquad 
h(\r)=h^{(0)}\times\ \sum_{j=0}^\infty \left(\frac{P^2}{\hK_0}\right)^j\ h_{j}(\r^2 \hK_0)\,,\\
&K(\r)=\hK^{(0)}\times\ \sum_{j=0}^\infty \left(\frac{P^2}{\hK_0}\right)^j\ K_{j}(\r^2 \hK_0)\,,\qquad
g(\r)=g^{(0)}\times\ \sum_{j=0}^\infty \left(\frac{P^2}{\hK_0}\right)^j\ g_{j}(\r^2 \hK_0)\,.
\end{split}
\eqlabel{pertgen}
\end{equation}
Apart from technical complexity, there is no obstacle of developing perturbative 
solution to any order in $\frac{P^2}{\hK_0}$. For our purposes it is sufficient to do so to order $\calo\left(\frac{P^2}
{\hK_0}\right)$.
Notice that explicit $\r$ dependence enters only in combination $\r \sqrt{\hK_0}$, thus, we can set $\hK_0=1$
and reinstall explicit $\hK_0$ dependence when necessary.

Substituting \eqref{pertgen} in \eqref{eq1}-\eqref{eq7} we find to  order $\calo(\dd)$
the following equations
\begin{equation}
\begin{split}
0=&f_{1,1}''-\frac{4 (\r^2+3)}{\r (4+\r^2)} f_{1,1}'-\frac{\r}{2(4+\r^2)} f_{2,1}'-\frac{2 \r}{4+\r^2} f_{3,1}'
-\frac{2 (h_{1}-2 f_{1,1})}{4+\r^2}\,,
\end{split}
\eqlabel{pereq1}
\end{equation}
\begin{equation}
\begin{split}
0=&f_{2,1}''+\frac{12}{\r} f_{1,1}'+\frac{2 \r}{4+\r^2} f_{2,1}'+\frac{12}{\r} f_{3,1}'+\frac{3 \r^2+16}{\r (4+\r^2)}
 h_{1}'-\frac{2}{(4+\r^2) \r^2} (3 \r^2 (2 f_{1,1}- h_{1})\\
&-64 f_{3,1}+8 f_{2,1}+32 K_{1}-32 h_{1})\,,
\end{split}
\eqlabel{pereq2}
\end{equation}
\begin{equation}
\begin{split}
0=&f_{3,1}''-\frac{\r^2+28}{\r (4+\r^2)} f_{3,1}'-\frac{4}{\r (4+\r^2)} f_{2,1}'
-\frac{24}{\r (4+\r^2)} f_{1,1}'+\frac{16 (f_{3,1}+f_{2,1})}{(4+\r^2) \r^2}\,,
\end{split}
\eqlabel{pereq3}
\end{equation}
\begin{equation}
\begin{split}
0=&h_{1}''-\frac{12 (\r^2+2)}{\r (4+\r^2)} f_{1,1}'-\frac{3 \r^2+8}{\r (4+\r^2)} f_{2,1}'
-\frac{4 (3 \r^2+8)}{\r (4+\r^2)} f_{3,1}'-\frac{4 (7+\r^2)}{\r (4+\r^2)} h_{1}'\\
&-\frac{2}{(4+\r^2) \r^2}
 (3 \r^2 (h_{1}-2 f_{1,1})-16+40 f_{2,1}+96 h_{1}+160 f_{3,1}-96 K_{1}))\,,
\end{split}
\eqlabel{pereq4}
\end{equation}
\begin{equation}
\begin{split}
0=&K_{1}''-\frac{\r^2+12}{\r (4+\r^2)} K_{1}'-\frac{32}{(4+\r^2) \r^2}\,,
\end{split}
\eqlabel{pereq5}
\end{equation}
\begin{equation}
\begin{split}
0=&g_{1}''-\frac{3 \r^2+16}{\r (4+\r^2)} h_{1}'-\frac{12 (\r^2+6)}{\r (4+\r^2)} f_{1,1}'
-\frac{3 \r^2+16}{\r (4+\r^2)} f_{2,1}'-\frac{4 (3 \r^2+16)}{\r (4+\r^2)} f_{3,1}'
-\frac{\r^2+12}{\r (4+\r^2)} g_{1}'\\
&-\frac{2}{(4+\r^2) \r^2} (3 \r^2 (h_{1}-2 f_{1,1})
+32 h_{1}+32 f_{3,1}+8 f_{2,1}-32 K_{1})\,,
\end{split}
\eqlabel{pereq6}
\end{equation}
along with the first order constraint
\begin{equation}
\begin{split}
0=&(K_1')^2-\frac{12 (\r^2+6)}{\r (4+\r^2)} f_{1,1}'-\frac{3 \r^2+16}{\r (4+\r^2)} 
(f_{2,1}'+4 f_{3,1}'+h_{1}')+\frac{2}{(4+\r^2) \r^2} 
(3 \r^2 (2f_{1,1}- h_{1})\\
&+8-32 h_{1}-32 f_{3,1}+32 K_{1}-8 f_{2,1})\,.
\end{split}
\eqlabel{pereqc}
\end{equation}

Above equations should be solved with $\calo(\dd)$ UV and the IR boundary conditions 
prescribed in sections \ref{uvcond} and \ref{ircond}. 
Solving \eqref{pereq5} we find
\begin{equation}
K_1=\ln\left(\frac{1}{\r^2}+\frac 14\right)\,.
\eqlabel{k1sol}
\end{equation} 
Next, we can use the constraint \eqref{pereqc} to decouple the equation for $g_1$:
\begin{equation}
0=g_1''-\frac{\r^2+12}{\r(4+\r^2)} g_1'-\frac{16}{(4+\r^2)^2}\,.
\eqlabel{decg1}
\end{equation}
We find 
\begin{equation}
g_1=\frac 14\ln\left(\frac{4}{\r^2}+1\right) \left(\r^2-2\ln\left(\frac {4}{\r^2}
+1\right)\right)-{\rm dilog}\left(\frac {4}{\r^2}+1\right)-\frac{\pi^2}{6}\,.
\eqlabel{g1sol}
\end{equation}
The remaining equations (for $\{f_{i,1},h_1\}$) we solved numerically. Parametrizing 
the asymptotics as follows
\nxt UV, \ie $\r\to 0$, (the independent coefficients being $\{f_{1,1,4}, f_{2,1,1},  f_{2,1,6}, f_{2,1,8}\}$):
\begin{equation}
\begin{split}
&f_{1,1}=\left(\frac{1}{16}+\frac14 \ln\r\right) \r^2
+\frac18 f_{2,1,1} \r^3+\left(f_{1,1,4}-\frac{1}{48} \ln\r\right) \r^4-\frac{1}{32} f_{2,1,1} \r^5
-\biggl(\frac14 f_{1,1,4}\\
&+\frac{31}{9216}-\frac{1}{192} \ln\r\biggr) \r^6+\frac{1}{128} f_{2,1,1} \r^7
+\left(\frac{1069}{921600}+\frac{1}{16} f_{1,1,4}-\frac{1}{768} \ln\r\right) \r^8
+\calo(\r^9)\,,
\end{split}
\eqlabel{f11uv}
\end{equation}
\begin{equation}
\begin{split}
&f_{2,1}=f_{2,1,1} \r+\left(\frac38-\frac12 \ln\r\right) \r^2-\frac14 f_{2,1,1} \r^3
+\left(-\frac{5}{144}-2 f_{1,1,4}+\frac{1}{24} \ln\r\right) \r^4\\
&+\frac{1}{16} f_{2,1,1} \r^5+\left(
f_{2,1,6}-\frac{1}{80} \ln\r\right) \r^6-\frac{1}{64} f_{2,1,1} \r^7+\biggr(f_{2,1,8}+\frac{11}{3360} \ln\r\biggr) \r^8
+\calo(\r^9)\,,
\end{split}
\eqlabel{f21uv}
\end{equation}
\begin{equation}
\begin{split}
&f_{3,1}=f_{2,1,1} \r+\left(\frac{5}{16}-\frac12 \ln\r\right) \r^2-\frac14 f_{2,1,1} \r^3+\left(-\frac{1}{72}-2 f_{1,1,4}
+\frac{1}{24} \ln\r\right) \r^4\\
&+\frac{1}{16} f_{2,1,1} \r^5+\biggl(\frac58 f_{1,1,4}-\frac14 f_{2,1,6}
+\frac{1123}{92160}-\frac{19}{1920} \ln\r\biggr) \r^6-\frac{1}{64} f_{2,1,1} \r^7
+\biggl(-\frac{3}{16} f_{1,1,4}\\
&+\frac38 f_{2,1,6}+f_{2,1,8}-\frac{1319}{307200}+\frac{67}{26880} \ln\r\biggr) \r^8
+\calo(\r^9)\,,
\end{split}
\eqlabel{f31uv}
\end{equation}
\begin{equation}
\begin{split}
&h_{1}=\frac12-2 \ln\r-2 f_{2,1,1} \r-\left(\frac{5}{12}-\ln\r\right) \r^2+\frac12 f_{2,1,1} \r^3
+\left(\frac{11}{1152}+4 f_{1,1,4}-\frac{1}{12} \ln\r\right) \r^4\\
&-\frac18 f_{2,1,1} \r^5
+\biggl(-f_{1,1,4}-\frac{113}{7200}+\frac{1}{48} \ln\r\biggr) \r^6+\frac{1}{32} f_{2,1,1} \r^7
+\biggl(\frac{11}{32} f_{1,1,4}-\frac98 f_{2,1,6}\\
&-\frac{15}{4} f_{2,1,8}+\frac{92753}{12902400}-\frac{289}{53760} \ln\r
\biggr) \r^8
+\calo(\r^9)\,;
\end{split}
\eqlabel{h1uv}
\end{equation}
\nxt IR, \ie $y=\frac 1\r\to 0$, (the independent coefficients being $\{f_{1,1}^h, f_{2,1}^h, f_{3,1}^h\}$: 
\begin{equation}
\begin{split}
f_{i,1}=f_{i,1}^h+\calo(y^2)\,,\qquad h_{1}=2 f_{1,1}^h+\calo(y^2)\,,
\end{split}
\eqlabel{fhhir}
\end{equation}
we find 
\begin{equation}
\begin{split}
&f_{1,1,4} = -0.020200(6)\,,\ f_{2,1,1} = -0.606789(8)\,,\ f_{2,1,6} = 0.005345(5)\,,\\
&f_{2,1,8} = -0.001621(4)\,,\
f_{1,1}^h = -0.274253(9)\,,\
f_{2,1}^h = -0.031228(4)\,,\\
& f_{3,1}^h = -0.241360(3)\,.
\end{split}
\eqlabel{numres}
\end{equation}

\begin{figure}[t]
\begin{center}
\psfrag{a10}{{$\a_{1,0}$}}
\psfrag{b40}{{$b_{4,0}$}}
\psfrag{a40}{{$a_{4,0}$}}
\psfrag{dd}{{$\dd$}}
\includegraphics[width=2in]{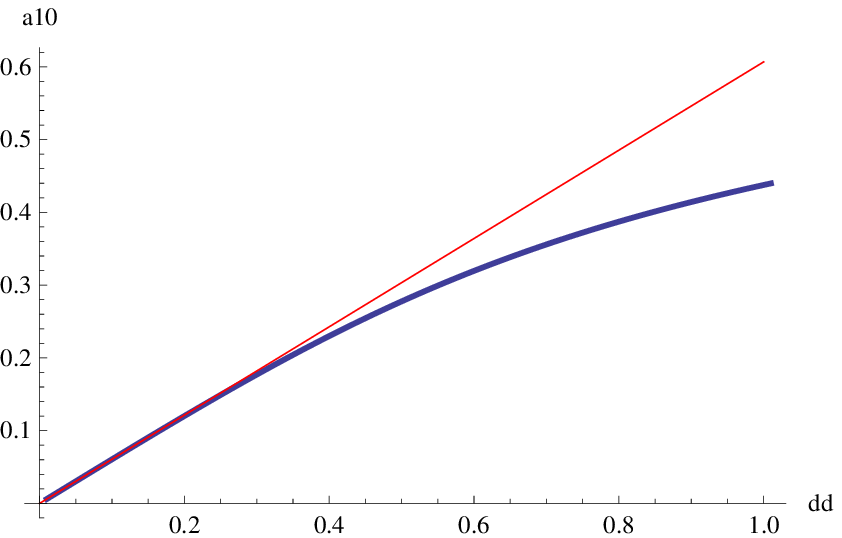}
\includegraphics[width=2in]{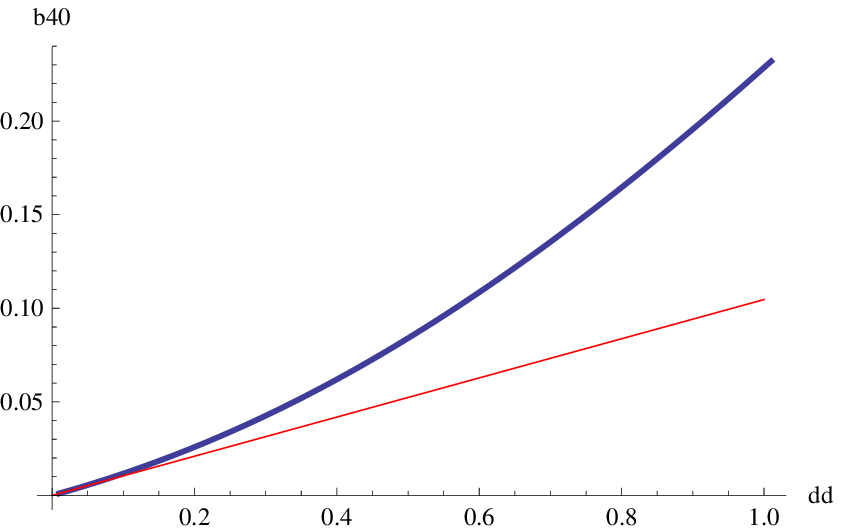}
\includegraphics[width=2in]{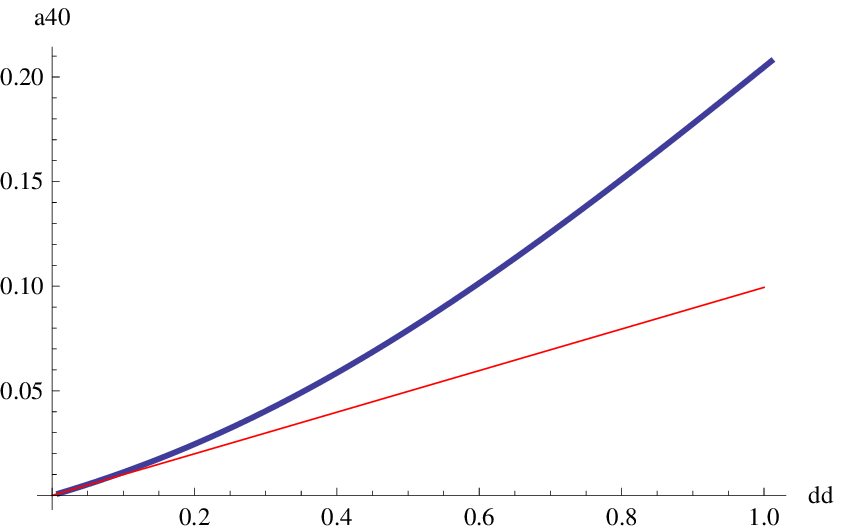}
\end{center}
  \caption{(Colour online) Comparison of values of  UV  parameters $\{\a_{1,0},b_{4,0},a_{4,0}\}$ 
(see \eqref{uvirfinal})
in the range $\dd\in[0,1]$ (blue curves) with their perturbative predictions \eqref{matching} (red curves).
} \label{figure1}
\end{figure}

\begin{figure}[t]
\begin{center}
\psfrag{a60}{{$a_{6,0}$}}
\psfrag{a80}{{$a_{8,0}$}}
\psfrag{g40}{{$g_{4,0}$}}
\psfrag{dd}{{$\dd$}}
\includegraphics[width=2in]{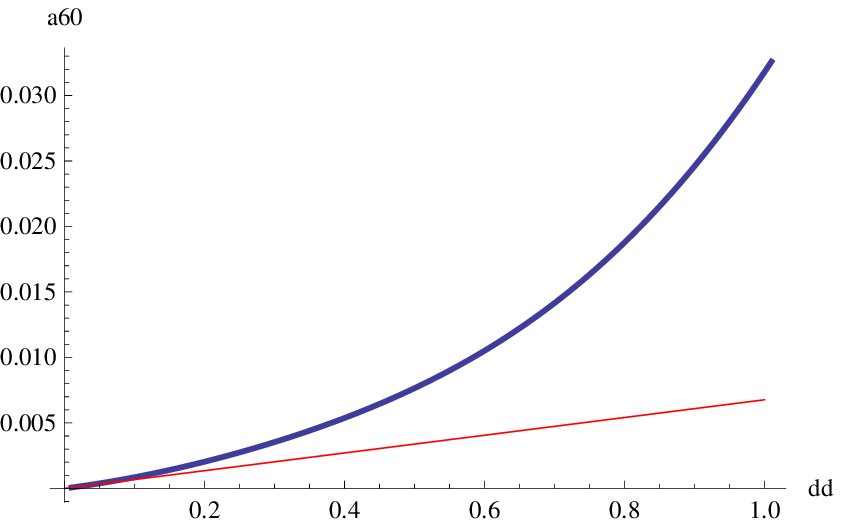}
\includegraphics[width=2in]{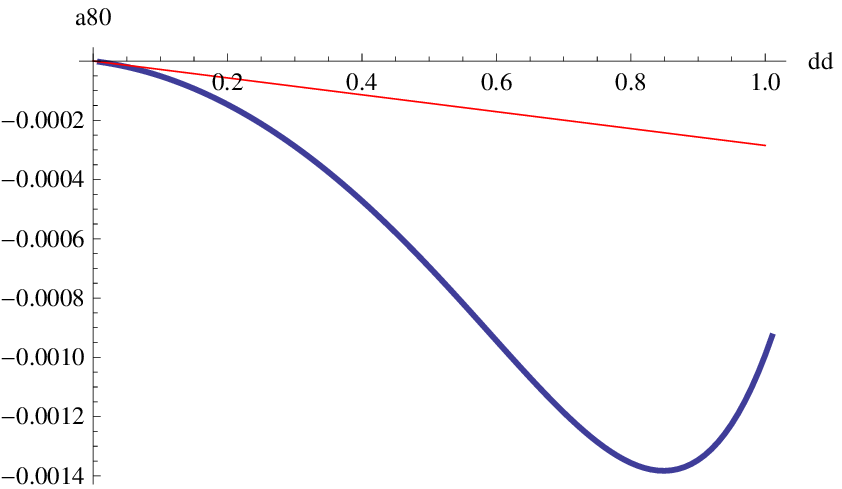}
\includegraphics[width=2in]{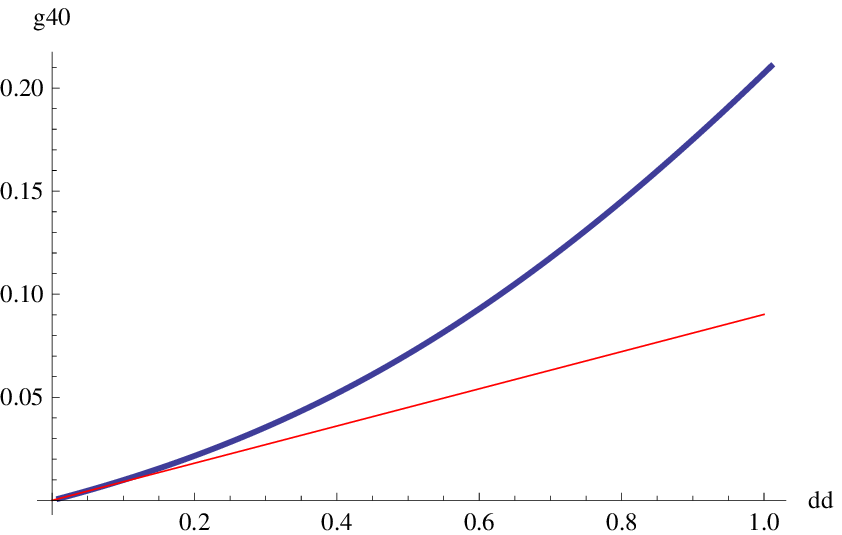}
\end{center}
  \caption{(Colour online) Comparison of values of  UV  parameters $\{a_{6,0},a_{8,0},g_{4,0}\}$ 
(see \eqref{uvirfinal})
in the range $\dd\in[0,1]$ (blue curves) with their perturbative predictions \eqref{matching} (red curves).
} \label{figure2}
\end{figure}

We can now identify the leading $\calo(\dd)$ 
values of general UV and IR parameters (see \eqref{uvirfinal}):
\begin{equation}
\begin{split}
&\a_{1,0}=- f_{2,1,1}\ \dd \,,\
b_{4,0}=\left(\frac{37}{576}-2 f_{1,1,4}\right)\dd\,,
\ a_{4,0}=\left(\frac{17}{288}-2 f_{1,1,4}\right)\dd\,,\\
&a_{6,0}=\left(f_{2,1,6}-\frac 12 f_{1,1,4}-\frac{5}{576}\right)\,,
\ a_{8,0}=\left(f_{2,1,8}+\frac 14 f_{2,1,6}\right)\dd\,,\ g_{4,0}
=\left(\frac{3}{64}+\frac{1}{16}\ln 2\right)\dd\,,\\
&K_0^h=1-2\dd\ \ln 2\,,\ h_0^h=2+2\dd\ f_{1,1}^h\,,\ 
g_0^h=1+\left(1-\frac{\pi^2}{6}\right)\dd\,,\\
&f_{2,0}^h=\frac 14+\frac 
14\dd\ f_{2,1}^h\,,\  f_{3,0}^h=\frac 14+\frac 
14\dd\ f_{3,1}^h\,,
\end{split}
\eqlabel{matching}
\end{equation}
where we set $K_0=\hK_0=1$. General relation between $\{K_0,\hK_0\}$ can be obtained while acting with the symmetry 
\eqref{scale2}:
\begin{equation}
K_0=\hK_0-2 P^2 \ln \hK_0+\calo(P^4)\,.
\eqlabel{k0hk0}
\end{equation}

Figures \ref{figure1} and \ref{figure2} compare the values of general UV  parameters $\a_{1,0}$,
$ b_{4,0}$, $a_{4,0}$, $a_{6,0}$, $a_{8,0}$,
$g_{4,0}$  (see \eqref{uvirfinal})
in the range $\dd\in[0,1]$ (blue curves) with their perturbative predictions \eqref{matching} (red curves).

\subsection{Stress-energy tensor}
Holographic renormalization of  cascading gauge theory was discussed in details in \cite{aby}.
For a general curved boundary background $\calm_4$ with 
\begin{equation}
ds_{\calm_4}^2=G_{ij}^{(0)}\ dx^i dx^j\,,
\eqlabel{m4}
\end{equation} 
the one point correlation function of the boundary stress-energy tensor $\vev{T_{ij}}$ takes form  
\begin{equation}
\begin{split}
8\pi G_5 \vev{T_{ij}}=&G_{ij}^{(0)} 
\biggl(R_{ab}R^{ab} \left(\frac{1921}{276480} \pz^2 P^4-\frac{1}{512} \kz^2+\frac{1}{96} \kz  P^2 \pz
\right)\\
&\qquad -R^2 \left(\frac{1}{4608} \kz^2+\frac{337}{51840} \pz^2 P^4+\frac{175}{27648} \kz  P^2 \pz\right)\\
&\qquad +R\left(\frac{1}{16}\kz \ha^{(2,0)}+\frac{1}{128}P^2\pz \ha^{(2,0)}+\frac{5}{256}P^2\pz \ha^{(2,1)}\right)\\
&\qquad +\square R \left(\frac{391}{82944} \pz^2 P^4-\frac{53}{23040} \kz^2+\frac{323}{46080} \kz  P^2 \pz
\right)\biggr)\\
&+R_{aijb} R^{ab} \left(\frac{17}{8640} \pz^2 P^4-\frac{1}{32} \kz^2+\frac{7}{192} \kz  P^2 \pz\right)\\
&-R_i^{\ a} R_{aj} \left( \frac{1}{64} \kz^2+\frac{1}{256} \pz^2 P^4+\frac{1}{64} \kz  P^2 \pz\right)\\
&+R R_{ij} \left(\frac{1691}{103680} \pz^2 P^4-\frac{1}{576} \kz^2+\frac{13}{432} \kz  P^2 \pz\right)\\
&-R_{ij}\left(\frac{1}{16}P^2\pz \ha^{(2,1)}+\frac 14 \kz \ha^{(2,0)}\right) \\
&-\nabla_i\nabla_j R \left( \frac{2773}{207360} \pz^2 P^4
+\frac{5}{3456} \kz  P^2 \pz+\frac{7}{1152} \kz^2\right)\\
&+\square R_{ij} \left(-\frac{17}{17280} \pz^2 P^4-\frac{7}{384} \kz  P^2 \pz+\frac{1}{64} \kz^2
\right)\\
&-\nabla_i\nabla_j \ha^{(2,0)}\left(\frac{1}{16}P^2\pz+\frac{1}{16}\kz\right)
+\nabla_i\nabla_j \ha^{(2,1)}\left(\frac{7}{128}P^2\pz+\frac{3}{64}\kz\right)\\
&+2 \hG_{ij}^{(4,0)}-\frac 12 G_{ij}^{(0)} \hG_a^{(4,0)a}+\frac 32 G_{ij}^{(0)} \left(\hb^{(4,0)}-\ha^{(4,0)}\right)
\\
&+T_{ij}^{ambiguity}\,,
\end{split}
\eqlabel{tijfinal}
\end{equation}
where 
\begin{equation}
\begin{split}
T_{ij}^{ambiguity}=&\left(\frac 12\pz^2 P^4 \kappa_3+\frac 12  
\pz P^2 \kappa_2 \kz +\frac 12 \kappa_1 \kz^2\right)\times \\
&\biggl(-2 \nabla_i\nabla_j R+6 \square R_{ij}-12 R_{aijb} R^{ab}-3 G_{ij}^{(0)} R_{ab}R^{ab}+R^2 G_{ij}^{(0)}
-4 R R_{ij}\\
&-\square R G_{ij}^{(0)}\bigg)\,.
\end{split}
\eqlabel{ambiguity}
\end{equation}
We use $\ \hat{}\ $ to indicated asymptotic parameters used in \cite{aby}. Notice that the asymptotic expansions in 
\cite{aby} are done in $\hat{\a}_{1,0}=0$ radial gauge.
All the derivatives in \eqref{tijfinal}
are with respect to the boundary metric \eqref{m4}; $R_{aijb}$, $R_{ab}$ and $R$ are the various Riemann tensors 
constructed from \eqref{m4}. $T_{ij}^{ambiguity}$, parametrized by $\kappa_i$, indicates ambiguities in renormalization 
prescription discussed in \cite{aby} due to defining cascading gauge theory on general manifold $\calm_4$. 
In a special case 
\begin{equation}
\calm_4\ =\ R\times S^3\,,
\eqlabel{special}
\end{equation}  
the one-point correlation function of the stress-energy tensor $\vev{T_{ij}}$ is actually 
ambiguity-free\footnote{There are no ambiguities in vevs of dimension-4 operators 
$\vev{\calo_{p_0}}$ and $\vev{\calo_{K_0}}$ as well --- see \cite{aby} for details.}. 

In the asymptotic UV parametrization \eqref{f1uv}-\eqref{guv} we have 
\begin{equation}
\begin{split}
&\kz=K_0\,,\qquad \pz=g_0\,,\qquad G_{ij}^{(0)}dx^idx^j=-dt^2+f_0^2 (dS_3)^2\,,\\
&\ha^{(2,0)}=\frac{1}{f_0^{2}}\left(\frac 14 K_0+\frac38 P^2 g_0\right)\,,\qquad \ha^{(2,1)}=-\frac{1}{2f_0^2}P^2 g_0\,,\\
&\ha^{(4,0)}=\frac{1}{f_0^4}\left(a_{4,0}+\frac{1}{16}P^2 g_0 \a_{1,0}^2\right)\,,\qquad \hb^{(4,0)}=\frac{1}{f_0^4}\left(b_{4,0}
+\frac{1}{16}P^2 g_0 \a_{1,0}^2\right)\,,\\ 
&\hG^{(4,0)}_{ij}dx^idx^j=
\frac{1}{f_0^2}\left(\frac{7}{576}P^4g_0^2+\frac{1}{16}K_0^2-b_{4,0}-\frac{1}{16}P^2g_0\a_{1,0}^2+\frac{55}{576}P^2g_0K_0\right) 
(dS_3)^2\,.
\end{split}
\eqlabel{abyparameters}
\end{equation}

Since 
\begin{equation}
\vev{T^{ij}}=\cale^s\ \dd^i_0 \dd^j_0+ \calp^s \left(G^{(0)ij}+\dd^i_0 \dd^j_0\right)\,,
\eqlabel{casdef}
\end{equation}
for the casimir energy density $\cale^s$ and the casimir pressure\footnote{We use the superscript $\ ^s$ to 
indicate the {\it symmetric}
phase of  cascading gauge theory.}, using \eqref{abyparameters}
we find from \eqref{tijfinal}
\begin{equation}
\begin{split}
\cale^s=&\frac{1}{8\pi G_5}\ \frac{1}{f_0^4}\left(\frac{403}{1920}P^4g_0^2+\frac{1}{32}K_0^2+\frac{3}{32}K_0P^2g_0-3b_{4,0}
+\frac32 a_{4,0}-\frac{3}{32}P^2 g_0\a_{1,0}^2\right)\,,\\
\calp^s=&\frac{1}{8\pi G_5}\ \frac{1}{f_0^4}\left(
\frac{283}{5760}P^4g_0^2+\frac{1}{96}K_0^2+\frac{1}{16}K_0P^2g_0+b_{4,0}-\frac{3}{2}a_{4,0}-\frac{1}{32}P^2 g_0\a_{1,0}^2
\right)\,.
\end{split}
\eqlabel{epc}
\end{equation}

It is instructive to understand the transformation of $\{\cale^c\,, \calp^c\}$ 
under the scaling symmetries \eqref{scale1}, \eqref{scale2} and \eqref{scale3}:
\nxt Under \eqref{scale1}, $\cale^c$ and $\calp^c$ are invariant. 
\nxt Under \eqref{scale2}
\begin{equation}
\{\cale^s\,,\calp^s\}\to \l^4\ \{\cale^s\,, \calp^s\} \,.
\eqlabel{scale2cas}
\end{equation} 
It is easy to understand the origin of \eqref{scale2cas}: 
the transformation \eqref{scale2} rescales the five-dimensional effective
gravitational action (or equivalently $G_5^{-1}$) by $\l^4$.
\nxt Under \eqref{scale3} 
\begin{equation}
\{\cale^s\,,\calp^s\}\to \l^{-4}\ \{\cale^s\,, \calp^s\} \,,
\eqlabel{scale3cas}
\end{equation} 
precisely as expected, given that $f_0\to \l f_0$.
\nxt As expected, under \eqref{leftover}, $\cale^s$ and $\calp^s$ are invariant.

\begin{figure}[t]
\begin{center}
\psfrag{lnmuls}{{$\ln\frac{\mu^2}{\Lambda^2}$}}
\psfrag{en}{{$\frac{8\pi G_5}{P^4 g_0^2}\times \frac{\cale^s}{\mu^4}$}}
\psfrag{pr}{{$\frac{8\pi G_5}{P^4 g_0^2}\times \frac{\calp^s}{\mu^4}$}}
\psfrag{dd}{{$\dd$}}
\includegraphics[width=3in]{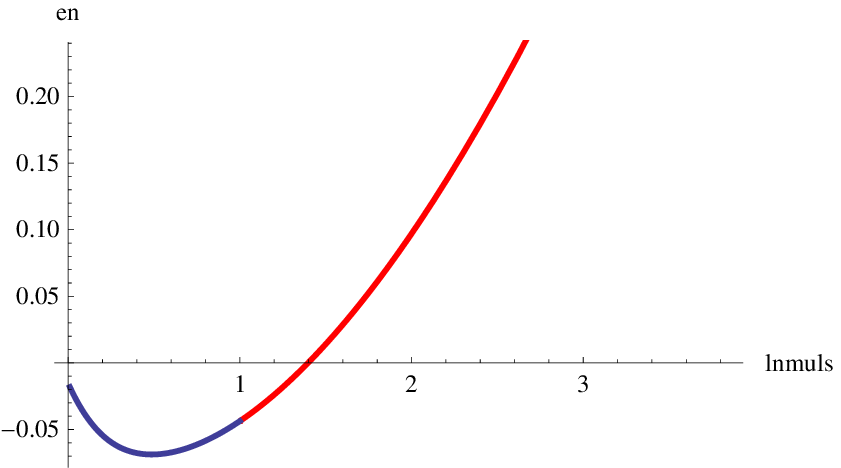}
\includegraphics[width=3in]{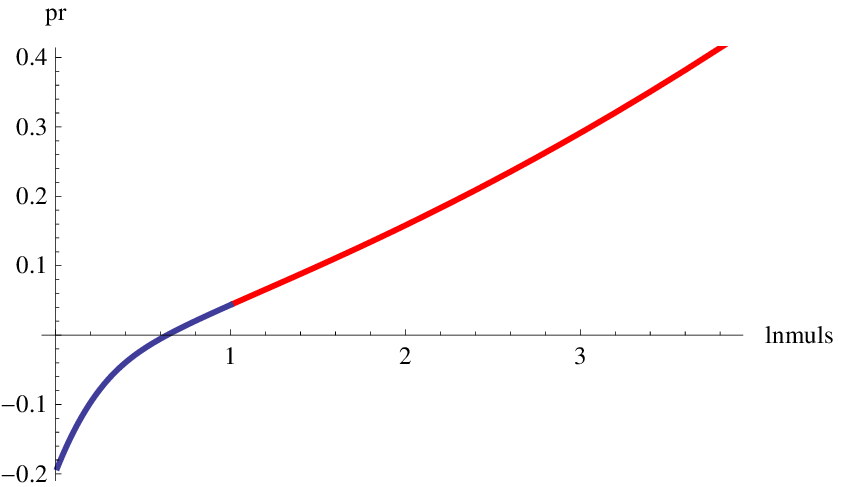}
\end{center}
  \caption{(Colour online) The energy density $\cale^s$ and the pressure $\calp^s$ of the chirally symmetric 
phase of cascading gauge theory compactified on $S^3$ of radius $\frac 1\mu$ as a function of $\ln\frac{\mu^2}{\Lambda^2}$.
The red curves are obtained from numerical solutions with $K_0=1$ and varying $P^2$, while the blue curves are obtained 
with $P^2=1$ and varying $K_0$  (see section \ref{numerical} for more details).  
} \label{figure3}
\end{figure}

Using \eqref{matching}, to leading order in $\dd$ (see \eqref{physical})
we have 
\begin{equation}
\begin{split}
\cale^s=&\frac{1}{8\pi G_5}\ \frac{1}{f_0^4}\ \frac{1}{32}(K_0-2P^2 g_0 \ln f_0+P^2g_0 \ln K_0)^2\ \left(1+\left(
-\frac13+96 f_{1,1,4}
\right)\dd+\calo(\dd^2)\right)\\
=&\frac{\mu^4}{8\pi G_5}\ \frac{1}{32}\left(\frac{P^2g_0}{\dd}+P^2g_0\ln\frac{P^2g_0}{\dd}\right)^2
\biggl(1-2.272588(7)\ \dd+\calo(\dd^2)\ \biggr)\,,
\\
\calp^s=&\frac{1}{8\pi G_5}\ \frac{1}{f_0^4}\ \frac{1}{96}(K_0-2P^2 g_0 \ln f_0+P^2g_0 \ln K_0)^2\ \left(1+\left(
\frac{11}{3}+96 f_{1,1,4}
\right)\dd+\calo(\dd^2)\right)\\
=&\frac{\mu^4}{8\pi G_5}\ \frac{1}{96}\left(\frac{P^2g_0}{\dd}+P^2g_0\ln\frac{P^2g_0}{\dd}\right)^2
\biggl(1+1.727411(3)\ \dd+\calo(\dd^2)\ \biggr)\,.
\end{split}
\eqlabel{PEperturbative}
\end{equation}

The results for the energy density $\cale^s$ and the pressure $\calp^s$ for general $\ln\frac{\mu^2}{\Lambda^2}$ are presented in 
figure \ref{figure3}. The red curves are obtained from 
numerical solutions with $K_0=1$ and varying $P^2$, while the blue curves are obtained 
with $P^2=1$ and varying $K_0$  (see section \ref{numerical} for more details). Notice that 
\begin{equation}
\begin{split}
&\frac{\cale^s}{\mu^4}\ <\ 0\,,\qquad {\rm once}\qquad \ln\frac{\mu^2}{\Lambda^2}\ <\ 1.397064(1)\,,\\
&\frac{\calp^s}{\mu^4}\ <\ 0\,,\qquad {\rm once}\qquad \ln\frac{\mu^2}{\Lambda^2}\ <\ 0.637321(1) \,. 
\end{split}
\eqlabel{negative}
\end{equation}
Likewise, we find 
\begin{equation}
\frac{\cale^s+3 \calp^s}{\mu^4}\ <\ 0\,, \qquad {\rm once}\qquad \ln\frac{\mu^2}{\Lambda^2}\ <\ 0.792717(5)\,,
\eqlabel{inflation}
\end{equation}
implying that cascading gauge theory compactified on sufficiently small $S^3$ would result in a closed 
inflationary Universe when coupled to four-dimensional Einstein gravity.

\section{$\csb$ fluctuations about chirally symmetric phase of $S^3$ compactified cascading gauge theory }
In previous sections we studied chirally symmetric phase of cascading gauge theory on $S^3$. Such a 
phase is expected to describe the ground state of the theory once the $S^3$ compactification scale $\mu$ 
sufficiently exceeds the strong coupling scale $\Lambda$ of the theory \cite{bt}.
On the other hand, in the $S^3$ decompactification limit, \ie $\frac{\mu}{\Lambda}\to 0$,
we expect the chiral symmetry to be spontaneously broken. In this section we identify $\csb$ fluctuations 
that become tachyonic once $\mu < \mu_c$, see \eqref{mucres}. 

\subsection{Equations of motion and boundary conditions for $\csb$ fluctuations} \label{fleombc}
Effective action for $\csb$ fluctuations about $SU(2)\times SU(2)\times U(1)$ symmetric 
states was summarized in section \ref{csbfluctuations}. Specializing to chirally symmetric states of cascading 
gauge theory on $S^3$ \eqref{metricaby}, and introducing 
\begin{equation}
\dd f=\calf(\r)\ e^{-i\w t}\ \Omega_L(S^3)\,,\qquad  \dd k_{1,2}=\calk_{1,2}(\r)\ e^{-i\w t}\ \Omega_L(S^3)\,,
\eqlabel{fluctwave}
\end{equation} 
where $\Omega_L(S^3)$ are $S^3$ Laplace-Beltrami operator eigenfunctions with eigenvalues $L=\ell (\ell+2)$ for 
integer $\ell$ 
\begin{equation}
\Delta_{S^3}\ \Omega_L(S^3)=-L\ \Omega_L(S^3)=-\ell (\ell+2)\ \Omega_L(S^3)\,,
\eqlabel{harmonics}
\end{equation}
we find the following equations of motion
\begin{equation}
\begin{split}
0=&\calf''+ \left(\frac{f_2'}{2f_2}-\frac3\r+2 \frac{f_3'}{f_3}+3 \frac{f_1'}{f_1}\right) \calf'
-\frac{K'}{2h f_3^2 g P^2} \calk_1'+h \left(\w^2-\frac{L}{f_1^2}\right) \calf
-\frac{2 g P^2}{h \r^2 f_2 f_3^2} \calk_2\\
&- \left(\frac{2 g P^2}{h \r^2 f_2 f_3^2}+\frac{(K')^2}{2h g P^2 f_3^2}+\frac{9}{\r^2 f_2}-\frac{12}{\r^2 f_3}\right) \calf\,,
\end{split}
\eqlabel{ffl}
\end{equation}
\begin{equation}
\begin{split}
0=&\calk_1''+ \left(3 \frac{f_1'}{f_1}- \frac{h'}{h}-\frac3\r+\frac{f_2'}{3f_2}- \frac{g'}{g}\right) \calk_1'
+2 K' \calf'+h \left(\w^2-\frac{L}{f_1^2}\right) \calk_1
+\frac{2 g P^2K}{ h \r^2 f_2 f_3^2} \calk_2\\
&-\frac{9}{\r^2 f_2} \calk_1+\frac{4  g P^2 K}{h \r^2 f_2 f_3^2}\calf\,,
\end{split}
\eqlabel{k1fl}
\end{equation}
\begin{equation}
\begin{split}
0=&\calk_2''+ \left(3 \frac{f_1'}{f_1}- \frac{h'}{h}-\frac3\r+ \frac{g'}{g}+\frac{f_2'}{2f_2}\right) \calk_2'
+h \left(\w^2-\frac{L}{f_1^2}\right) \calk_2+\frac{ 9K}{2h \r^2 g f_2 P^2 f_3^2} \calk_1\\
&-\frac{9}{\r^2 f_2} \calk_2-\frac{18}{\r^2 f_2} \calf\,.
\end{split}
\eqlabel{k2fl}
\end{equation}
In the UV (as $\r\to 0$) only the normalizable modes of $\{\calf, \calk_{1,2}\}$ can be nonzero; thus the 
asymptotic solution to \eqref{ffl}-\eqref{k2fl} is given by\footnote{For the numerics we developed expansions to order $\calo(\r^{10})$ inclusive.} 
\begin{equation}
\calf=\calf_{3,0}\ \r^3+\frac 32 \a_{1,0} \calf_{3,0}\ \r^4
+\sum_{n=5}^\infty\sum_k
\calf_{n,k}\ \r^n \ln^k\r\,,
\eqlabel{ffluv}
\end{equation}
\begin{equation}
\begin{split}
\calk_1=&P^2g_0\left(\calk_{1,3,0}+2\calf_{3,0}\ln\r\right)\r^3
+P^2g_0\a_{1,0}\left(\frac32\calk_{1,3,0}+\calf_{3,0}
+3\calf_{3,0}\ln\r\right)\r^4\\
&+\sum_{n=5}^\infty\sum_k
\calk_{1,n,k}\ \r^n \ln^k\r\,,
\end{split}
\eqlabel{k1fluv}
\end{equation}
\begin{equation}
\begin{split}
\calk_2=&\left(\frac 32\calk_{1,3,0}-\calf_{3,0}+3\calf_{3,0}\ln\r
\right)\r^3+\frac {9\a_{1,0}}{4}\left(\calk_{1,3,0}+2\calf_{3,0}\ln\r
\right)\r^4\\
&+\sum_{n=5}^\infty\sum_k
\calk_{2,n,k}\ \r^n \ln^k\r\,.
\end{split}
\eqlabel{k2fluv}
\end{equation}
It is characterized by 4 parameters:
\begin{equation}
\{\w^2\,, \calf_{3,0}\,, \calf_{7,0}\,, \calk_{1,3,0}\}\,.
\eqlabel{uvfl}
\end{equation}
In the IR (as $y\equiv\frac 1\r\to 0$ ) the non-singular asymptotic solution to \eqref{ffl}-\eqref{k2fl} is given 
by\footnote{For the numerics we developed expansions to order $\calo(y^{10})$ inclusive.}
\begin{equation}
\begin{split}
\calf=&y^{\sqrt{1+L}-1}\times \left(\calf_0^h+\sum_{n=1}^\infty \calf_n^h\ y^n\right)\,,\\
\calk_{1}=&y^{\sqrt{1+L}-1}\times \left(\calk_{1,0}^h+\sum_{n=1}^\infty \calk_{1,n}^h\ y^n\right)\,,\\
\calk_{2}=&y^{\sqrt{1+L}-1}\times \left(\calk_{2,0}^h+\sum_{n=1}^\infty \calk_{2,n}^h\ y^n\right)\,.
\end{split}
\eqlabel{irfluct}
\end{equation}
Since equations of motion for $\{\calf,\calk_{1,2}\}$ are homogeneous, without loss of generality we can set 
\begin{equation}
\calk_{2,0}^h=1\,.
\eqlabel{irbc}
\end{equation}
As a result, the asymptotic expansion \eqref{irfluct} is characterized by 2 additional parameters:
\begin{equation}
\{\calf_0^h,\calk_{1,0}^h\}\,.
\eqlabel{irfl} 
\end{equation}
Given \eqref{uvfl} and \eqref{irfl},
notice that $4+2=6$ is precisely the number of integration constants needed to specify 
a solution to \eqref{ffl}-\eqref{k2fl} for a given chirally symmetric state of cascading gauge theory on $S^3$ and  for a 
fixed $L$.

\subsection{Spectrum of $\csb$ fluctuations for $\ln\frac{\mu}{\Lambda}\gg 1$}\label{fluctpert}
We begin exploring the spectrum of $\csb$ fluctuations in the regime when the $S^3$ compactification scale $\mu$ 
is much larger than the strong couping scale $\Lambda$ of cascading gauge theory. Following the perturbative expansion 
of the background in section \ref{pertubative}, we find that perturbative in $\frac{P}{\sqrt{\hK_0}}$ solution to \eqref{ffl}-\eqref{k1fl}
takes form 
\begin{equation}
\begin{split}
&\calf=\sum_{n=1}^\infty \left(\frac{P}{\sqrt{\hK_0}}\right)^n\ \calf_{(n)}(\r^2\hK_0)\,,\qquad  
\calk_1=\hK_0\ \sum_{n=2}^\infty \left(\frac{P}{\sqrt{\hK_0}}\right)^n\ 
\calk_{1,(n)}(\r^2\hK_0)
\,,\\ 
&\calk_2=\sum_{n=0}^\infty \left(\frac{P}{\sqrt{\hK_0}}\right)^n\ \calk_{2,(n)}(\r^2\hK_0)\,,\qquad 
\frac{\w^2}{\mu^2}=\sum_{n=0} \left(\frac{P}{\sqrt{\hK_0}}\right)^n\ M_{(n)}\,.
\end{split}
\eqlabel{flpert}
\end{equation}    

\subsubsection{The leading order}

Introducing 
\begin{equation}
x\equiv \r^2 \hK_0\,,
\eqlabel{flcoord}
\end{equation}
to leading order in $\frac{P}{\sqrt{\hK_0}}$ we find:
\begin{equation}
\begin{split}
0=&\calf_{(1)}''-\frac{4}{x (4+x)} \calf_{(1)}'+\frac{12 x-x^2 L-4 x L+4 x M_{(0)}+48}{4(4+x)^2 x^2} \calf_{(1)}\,,
\end{split}
\eqlabel{fllead1}
\end{equation}
\begin{equation}
\begin{split}
0=&\calk_{1,(2)}''-\frac{4}{x (4+x)} \calk_{1,(2)}'+ \frac{4 x M_{(0)}-x^2 L-36 x-4 x L-144}{4(4+x)^2 x^2}
 \calk_{1,(2)}\\
&+\frac{8}{(4+x) x^2}\calk_{2,(0)}\,,
\end{split}
\eqlabel{fllead2}
\end{equation}
\begin{equation}
\begin{split}
0=&\calk_{2,(0)}''-\frac{4}{x (4+x)} \calk_{2,(0)}'+\frac{4 x M_{(0)}-x^2 L-36 x-4 x L-144}{4(4+x)^2 x^2} 
\calk_{2,(0)}\\
&+\frac{18}{(4+x)x^2}\calk_{1,(2)}\,.
\end{split}
\eqlabel{fllead3}
\end{equation}
From now on restrict discussion to the lowest $\ell=0$ (correspondingly $L=0$) harmonic\footnote{Extension of the 
analysis to higher $\ell$ is straightforward.}. Solving \eqref{fllead1}, subject to boundary conditions \eqref{ffluv} and 
\eqref{irfluct} we find
\begin{equation}
\begin{split}
\calf_{(1)}^{[q]}=&\cala_{q}\ \left(\frac{x}{4+x}\right)^{3/2} (4+x)\left(1+\frac14 x\right)^{-q}\
_2F_1 \left([-q, -q+1], [-2q], 1+\frac14 x\right)\,,\\
M_{(0)}^{[q]}=&(2q+1)^2\,,
\end{split}
\eqlabel{leadF}
\end{equation}
where an integer $q=1,2,\cdots$ labels different states in the spectrum of $\csb$ fluctuations, and $\cala_{q}$ is 
a normalization constant. From now on we restrict discussion to the lowest\footnote{Extension of the 
analysis to higher $q$-states is straightforward.} $q=1$ state in the spectrum of 
$\csb$ fluctuations: 
\begin{equation}
\calf_{(1)}\equiv \calf_{(1)}^{[1]}=\cala_1\ \left(\frac{x}{4+x}\right)^{3/2}\,,\qquad M_{(0)}\equiv M_{(0)}^{[1]}=9\,.
\eqlabel{lowest}
\end{equation}
Next, from \eqref{fllead2} and \eqref{fllead3} we find that the equation for  
\begin{equation}
\delta\calk\equiv \calk_{2,(0)}-\frac 32 \calk_{1,(2)}\,,
\end{equation}    
decouples (we used the value of $M_{(0)}$ as in \eqref{lowest}):
\begin{equation}
0=\delta\calk''-\frac{4}{x(4+x)}\ \delta\calk'-\frac{12(7+x)}{(4+x)^2 x^2}\ \delta\calk\,.
\eqlabel{dkequa}
\end{equation}
The only solution of \eqref{dkequa} consistent with the boundary conditions \eqref{k1fluv}, \eqref{k2fluv} and 
\eqref{irfluct} is 
\begin{equation}
\delta\calk= 0 \qquad \Longrightarrow\qquad  \calk_{2,(0)}=\frac 32 \calk_{1,(2)}\,.
\eqlabel{resdk}
\end{equation}
Given \eqref{resdk}, the equation for $\calk_{1,(2)}$ \eqref{fllead2}, subject to the boundary conditions \eqref{k1fluv}
and \eqref{irfluct}, can be solved analytically\footnote{The normalization constant is determined 
from the IR normalization of $\calk_{2,(0)}$, see \eqref{irbc}. }:
\begin{equation}
\calk_{1,(2)}=\frac 23 \left(\frac{x}{4+x}\right)^{3/2}\,.
\eqlabel{calk20res}
\end{equation}

For completeness, we summarize the leading order results for the $\csb$ fluctuations
\begin{equation}
\begin{split}
&\calf_{(1)}=\cala_1\ \left(\frac{x}{4+x}\right)^{3/2}\,,\ \ \calk_{1,(2)}=\frac 23 \left(\frac{x}{4+x}\right)^{3/2}\,,
\ \  \calk_{2,(0)}=\left(\frac{x}{4+x}\right)^{3/2}\,,\\
&M_{(0)}=9\,.
\end{split}
\eqlabel{summarylead} 
\end{equation}
Notice that $\cala_1$ is not fixed at this stage --- it will be fixed at the next subleading order\footnote{We find that this 
pattern continues at subleading orders.}.

\subsubsection{The subleading order}

To leading order in $\frac{P}{\sqrt{\hK_0}}$,  the 
$L=0$ (see \eqref{harmonics}) and the lowest $q=1$ (see \eqref{leadF}) state of the  linearized $\csb$ 
fluctuations is presented in \eqref{summarylead}. 
At the subleading order this state is described by the following equations 
\begin{equation}
\begin{split}
0=&\calf_{(2)}''-\frac{4}{x(x+4)} \calf_{(2)}'+\frac{12 (1+x)}{(x+4)^2 x^2} \calf_{(2)}+
\frac{32}{x (x+4)^3} \left(\frac{x}{x+4}\right)^{1/2}\\
&+\frac{M_{(1)} \cala_1 x-32-8 x}{(x+4)^2 x^2}
\left(\frac{x}{x+4}\right)^{3/2} \,,
\end{split}
\eqlabel{flsublead1}
\end{equation}
\begin{equation}
\begin{split}
0=&\calk_{1,(3)}''-\frac{4}{x (x+4)} \calk_{1,(3)}'-\frac{36}{(x+4)^2 x^2} \calk_{1,(3)}+\frac{8}{(x+4) x^2} \calk_{2,(1)}
\\
&-\frac{48 \cala_1}{x (x+4)^3} \left(\frac{x}{x+4}\right)^{1/2}
+ \frac{2(24 \cala_1 x+M_{(1)} x+96 \cala_1)}{3(x+4)^2 x^2} \left(\frac{x}{x+4}\right)^{3/2}\,,
\end{split}
\eqlabel{flsublead2}
\end{equation}
\begin{equation}
\begin{split}
0=&\calk_{2,(1)}''-\frac{4}{x (x+4)} \calk_{2,(1)}'-\frac{36}{(x+4)^2 x^2} \calk_{2,(1)}
+\frac{18}{(x+4) x^2} \calk_{1,(3)}\\
&+ \frac{M_{(1)} x-18 \cala_1 x-72 \cala_1}{(x+4)^2 x^2} \left(\frac{x}{x+4}\right)^{3/2}\,.
\end{split}
\eqlabel{flsublead3}
\end{equation}
Solving \eqref{flsublead1} subject to the boundary conditions \eqref{ffluv} and \eqref{irfluct}
we find
\begin{equation}
\cala_1=\frac{8}{M_{(1)}}\,,\qquad \calf_{(2)}=\cala_2\ \left(\frac{x}{x+4}\right)^{3/2}\,,
\eqlabel{sublowest}
\end{equation}
where $\cala_2$ is a (new) normalization constant.
Introducing 
\begin{equation}
\calk_{1,(3)}
=\left(\frac{x}{x+4}\right)^{1/2}\left(\calg_{1}+\calg_2\right)\,,\qquad 
\calk_{2,(1)}
=\frac 32\left(\frac{x}{x+4}\right)^{1/2}\left(\calg_{1}-\calg_2\right)\,,
\eqlabel{decsublead}
\end{equation}
the equations of motion for $\{\calg_1,\calg_2\}$ decouples:
\begin{equation}
\begin{split}
0=&\calg_1''+\frac{8}{x (x+4)^2} \calg_1+ \frac{2(M_{(1)}^2 x+24 x-192)}{3x (x+4)^3 M_{(1)}}\,,
\end{split}
\eqlabel{resg1}
\end{equation}
\begin{equation}
\begin{split}
0=&\calg_2''-\frac{16 (6+x)}{(x+4)^2 x^2} \calg_2+\frac{16 (7 x+16)}{x (x+4)^3 M_{(1)}}\,.
\end{split}
\eqlabel{resg2}
\end{equation}
Imposing the UV boundary conditions \eqref{k1fluv} and \eqref{k2fluv}, as well as regularity 
in the IR, we find solving \eqref{resg1} 
\begin{equation}
M_{(1)}=\mp 6\sqrt{2}\,,\qquad \calg_1=\frac{x\beta_2}{4+x}\pm
\frac{2\sqrt{2}x}{3(4+x)} \ln\left(1+\frac4x\right)\,,
\eqlabel{solveg1}
\end{equation}
where $\b_2$ is (so far) an arbitrary constant; we further find solving
\eqref{resg2}
\begin{equation}
\begin{split}
\calg_2=&\pm\biggl(\frac{4\sqrt{2} (4+x)^2}{x^2} \biggl(
{\dilog}\left(1+\frac{x}{4} \right)+\ln(4+x) \ln x-\frac12 \ln^2(4+x)+2 \ln^2 2\\
&-2 \ln 2  
\ln x\biggr)+\frac{\sqrt{2}(480 x+768+88 x^2+3 x^3)}{12x (4+x)}
\ln\left(1+\frac 4x\right)
\\&+\frac{\sqrt{2}(576+47 x^2+324 x)}{9x (4+x)}\biggr)\,.
\end{split}
\eqlabel{solveg2}
\end{equation} 
The $\pm$ signs in \eqref{solveg1} and \eqref{solveg2} are correlated.
To fix $\b_2$ we use the normalization condition \eqref{irbc}, which at the subleading 
order considered here implies that 
\begin{equation}
\lim_{x\to +\infty}\calk_{2,(1)}=0\qquad \Longrightarrow\qquad \b_2=\mp \frac{2\sqrt{2}(3 \pi^2-28)}{9}\,.
\eqlabel{fixb2}
\end{equation}

For completeness, we summarize the subleading order results for the $\csb$ fluctuations
\begin{equation}
\begin{split}
&\cala_1=\mp\frac{2\sqrt{2}}{3}\,,\qquad M_{(1)}=\mp 6\sqrt{2}\,,\qquad 
\calf_{(2)}=\cala_2\ \left(\frac{x}{x+4}\right)^{3/2}\,,\\
&\calk_{1,(3)}
=\left(\frac{x}{x+4}\right)^{1/2}\left(\calg_{1}+\calg_2\right)\,,\qquad 
\calk_{2,(1)}
=\frac 32\left(\frac{x}{x+4}\right)^{1/2}\left(\calg_{1}-\calg_2\right)\,,\\
&\calg_1=\mp \left(\frac{2\sqrt{2}(3 \pi^2-28)}{9}\frac{x}{4+x}-
\frac{2\sqrt{2}x}{3(4+x)} \ln\left(1+\frac4x\right)\right)\,,\\
&\calg_2=\pm\biggl(\frac{4\sqrt{2} (4+x)^2}{x^2} \biggl(
{\dilog}\left(1+\frac{x}{4} \right)+\ln(4+x) \ln x-\frac12 \ln^2(4+x)+2 \ln^2 2\\
&-2 \ln 2  
\ln x\biggr)+\frac{\sqrt{2}(480 x+768+88 x^2+3 x^3)}{12x (4+x)}
\ln\left(1+\frac 4x\right)
\\&+\frac{\sqrt{2}(576+47 x^2+324 x)}{9x (4+x)}\biggr)\,.
\end{split}
\eqlabel{sumsublead}
\end{equation}
Notice that the leading and the first subleading correction to the $L=0$, $q=1$ 
state in the 
spectrum of $\csb$ fluctuations (\eqref{summarylead} and \eqref{sumsublead}) 
is determined up to a constant $\cala_2$ --- the latter constant will be fixed 
at the second subleading order.

\subsubsection{The subsubleading order}

To leading and the first subleading order in $\frac{P}{\sqrt{\hK_0}}$,  the 
$L=0$ (see \eqref{harmonics}) and the lowest $q=1$ (see \eqref{leadF}) state of the  linearized $\csb$ 
fluctuations is presented in \eqref{summarylead} and \eqref{sumsublead} correspondingly. 
At the second subleading order this state is described by the following equations 
\begin{equation}
\begin{split}
&0=\calf_{(3)}''-\frac{4}{x (4+x)} \calf_{(3)}'+\frac{12(1+ x)}{(4+x)^2 x^2} \calf_{(3)}+\frac{8}{x (4+x)} \calk_{1,(3)}'
-\frac{24 \sqrt x}{(4+x)^{5/2} M_{(1)}} h_{1}'\\
&+\frac{144 \sqrt x}{M_{(1)} (3 x+16) (4+x)^{3/2}} f_{1,1}'+\frac{768-2\ln 2}{(4+x)^{5/2} \sqrt x M_{(1)} (3 x+16)}
 \ln\left(1+\frac4x\right)
\\
&+\frac{144 \sqrt x}{(4+x)^{5/2} M_{(1)} (3 x+16)} f_{1,1}
+\frac{24 (9 x+40)}{(3 x+16) (4+x)^{5/2} M_{(1)} \sqrt x } f_{2,1}\\
&-\frac{288 (x+8)}{(3 x+16) (4+x)^{5/2} M_{(1)} \sqrt x } f_{3,1}
+\frac{48 (2 x-64+3 x^2)}{\sqrt x M_{(1)} (4+x)^{7/2} (3 x+16)} h_{1}-\frac{8}{(4+x) x^2} \calk_{2,(1)}\\
&+\frac{3 M_{(1)}^2 \cala_2 x^2+16 M_{(1)}^2 \cala_2 x-8192-2368 x-192 x^2+128 x M_{(2)}+24 x^2 M_{(2)}}{\sqrt x (4+x)^{7/2} (3 x+16) M_{(1)}}\,,
\end{split}
\eqlabel{flssublead1}
\end{equation}
\begin{equation}
\begin{split}
&0=\calk_{1,(4)}''-\frac{4}{(4+x) x} \calk_{1,(4)}'-\frac{36}{(4+x)^2 x^2} \calk_{1,(4)}+\frac{8}{x^2 (4+x)} \calk_{2,(2)}
+\frac{M_{(1)}}{x (4+x)^2} \calk_{1,(3)}\\
&+\frac{12 \sqrt x}{(3 x+16) (4+x)^{3/2}} f_{1,1}'-\frac{8 \sqrt x}{(4+x)^{5/2}} f_{3,1}'
-\frac{6 \sqrt{x}}{(4+x)^{5/2}} h_{1}'-\frac{48 (x+8)\ln 2}{(3 x+16) \sqrt x (4+x)^{5/2}}\\
&-\frac{3 x^3-768-288 x-8 x^2}{(4+x)^{7/2} \sqrt x (3 x+16)}\ln\left(1+\frac4x\right)
+\frac{2}{3(4+x)^{7/2} \sqrt x  (3 x+16)} \biggl(18 f_{1,1} x^2-18 h_{1} x^2\\
&+18 x^2+72 \cala_2 x^2-72 f_{3,1} x^2-9 f_{2,1} x^2+3 x^2 M_{(2)}+36 g_{1} x^2+16 x M_{(2)}+120 x\\
&-324 h_{1} x
-108 f_{2,1} x+72 f_{1,1} x+456 \cala_2 x+336 g_{1} x-768 f_{3,1} x+384 \cala_2-1152 h_{1}\\
&-288 f_{2,1}
+768 g_{1}-1920 f_{3,1}\biggr)\,,
\end{split}
\eqlabel{flssublead2}
\end{equation}
\begin{equation}
\begin{split}
&0=\calk_{2,(2)}''-\frac{4}{(4+x) x} \calk_{2,(2)}'-\frac{36}{(4+x)^2 x^2} \calk_{2,(2)}+\frac{18}{x^2 (4+x} \calk_{1,(4)})
+\frac{M_{(1)}}{x (4+x)^2} \calk_{2,(1)}\\
&+\frac{18 \sqrt x}{(3 x+16) (4+x)^{3/2}} f_{1,1}'-\frac{12 \sqrt x}{(4+x)^{5/2}} f_{3,1}'-\frac{9 \sqrt x}{(4+x)^{5/2}} h_{1}'
-\frac{72 (x+8)\ln 2}{(3 x+16) \sqrt x  (4+x)^{5/2}}\\
&+\frac{3(768+288 x+40 x^2+3 x^3)}{2(4+x)^{7/2} \sqrt x  (3 x+16)} \ln\left(1+\frac4x\right)
+\frac{1}{(4+x)^{7/2} \sqrt x (3 x+16)}\biggl(18 f_{1,1} x^2\\
&-1152 \cala_2-54 \cala_2 x^2-324 h_{1} x-768 f_{3,1} x-108 f_{2,1} x+16 x M_{(2)}-336 g_{1} x-504 \cala_2 x\\
&+72 f_{1,1} x-18 h_{1} x^2-72 x-18 x^2-288 f_{2,1}-1152 h_{1}-1920 f_{3,1}-768 g_{1}+3 x^2 M_{(2)}\\
&-72 f_{3,1} x^2-9 f_{2,1} x^2-36 g_{1} x^2\biggr)\,,
\end{split}
\eqlabel{flssublead3}
\end{equation}
where in order to avoid unnecessary complications of the formulas we did not substitute the explicit 
subleading results for $M_{(1)}$, $\calk_{1,(3)}$ and $\calk_{2,(1)}$, see
\eqref{sumsublead}. 

We have to resort to numerics to solve \eqref{flssublead1}-\eqref{flssublead3}. 
Notice that all the equations are coupled --- at least via an undetermined so far constant $\cala_2$. 
To begin, notice that if 
 $\calf_{(3)}$ is a solution to \eqref{flssublead1}-\eqref{flssublead3} subject to the boundary conditions 
\eqref{ffluv}-\eqref{k2fluv} and \eqref{irfluct}, \eqref{irbc}, so is the combination
\begin{equation}
\calf_{(3)}+\cala_3\ \left(\frac{x}{x+4}\right)^{3/2}\,,
\eqlabel{zeromode3}
\end{equation}
for an arbitrary constant $\cala_3$. This constant plays the role of $\cala_1$ in \eqref{summarylead} and the role 
of $\cala_2$ in \eqref{sumsublead} --- it is fixed at the third subleading order in the perturbative expansion \eqref{flpert}.
Notice that the shift \eqref{zeromode3} adjusts the coefficients of $x^{3/2}$ in the UV expansion \eqref{ffluv}. Thus, up to a 
zero mode \eqref{zeromode3}, a particular solution to \eqref{flssublead1}-\eqref{flssublead3} has the 
following UV asymptotics 
\begin{equation}
\begin{split}
\calf_{(3)}=-\frac{3 f_{2,1,1}}{2 M_{(1)}}\ x^2+\sum_{n=5}^{\infty}\sum_{k=0}^1\ \calf_{(3),n,k}\ x^{n/2}\ln^k x\,,
\end{split}
\eqlabel{uv31}
\end{equation}
\begin{equation}
\begin{split}
\calk_{1,(4)}=&\left(\frac{1}{12}\cala_2+\frac 23 \calk_{2,(2),3,0}+\frac 18\cala_2 \ln x\right)x^{3/2}-\frac 18 f_{2,1,1}\ x^2
\\&+\sum_{n=5}^{\infty}\sum_{k=0}^1\ \calk_{1,(4),n,k}\ x^{n/2}\ln^k x\,,
\end{split}
\eqlabel{uv32}
\end{equation}
\begin{equation}
\begin{split}
\calk_{2,(2)}=\left(\calk_{2,(2),3,0}+\frac {3}{16}\cala_2 \ln x\right)x^{3/2}-\frac {3}{16} f_{2,1,1}\ x^2
+\sum_{n=5}^{\infty}\sum_{k=0}^1\ \calk_{2,(2),n,k}\ x^{n/2}\ln^k x\,.
\end{split}
\eqlabel{uv33}
\end{equation}
Asymptotics \eqref{uv31}-\eqref{uv33} are completely determined by 4 parameters:
\begin{equation}
\left\{\cala_2\,,\ M_{(2)}\,,\ \calk_{2,(2),3,0}\,,\ \calk_{1,(4),7,0}\right\}\,.
\eqlabel{uv3par}
\end{equation}
In the IR (as $y\equiv \frac 1x\to 0$) the asymptotic solution to \eqref{flssublead1}-\eqref{flssublead3}
is given by
\begin{equation}
\begin{split}
\calf_{(3)}=&\calf_{(3)}^h-\frac{1}{2M_{(1)}}\biggl(72 f_{2,1}^h-96f_{3,1}^h+12 \calf_{(3)}^h M_{(1)}
+144 f_{1,1}^h+M_{(1)}^2\cala_2+8M_{(2)}-64\biggr)y\\
&+\sum_{n=2}^\infty \calf_{(3),n}^h\ y^n\,,
\end{split}
\eqlabel{ir31}
\end{equation}
\begin{equation}
\begin{split}
\calk_{1,(4)}=&\calk_{1,(4)}^h+\biggl(8\ln 2-\frac13M_{2}+8f_{3,1}^h+f_{2,1}^h-8\cala_2+\frac{212}{3}-\frac{22}{3}\pi^2
+2f_{1,1}^h\biggr)y
\\
&+\sum_{n=2}^\infty \calk_{1,(4),n}^h\ y^n\,,
\end{split}
\eqlabel{ir32}
\end{equation}
\begin{equation}
\begin{split}
\calk_{2,(2)}=&\biggl(12\ln 2-9 \calk_{1,(4)}^h+3 f_{1,1}^h+6-\pi^2+9\cala_2+\frac 32 f_{2,1}^h-\frac12 M_{(2)}
+12f_{3,1}^h\biggr)y
\\&+\sum_{n=2}^\infty \calk_{2,(2),n}^h\ y^n\,.
\end{split}
\eqlabel{ir33}
\end{equation}
It is completely determined by 2 additional parameters: 
\begin{equation}
\left\{\calf_{(3)}^h\,,\ \calk_{1,(4)}^h\right\}\,.
\eqlabel{ir2par}
\end{equation}
Numerically, we find:
\begin{equation}
\begin{split}
&\cala_2=-0.831952(7)\,,\ M_{(2)}=0.077172(8)\,,\ \calk_{2,(2),3,0}=-0.000540(7)\,,\\
&\calk_{1,(4),7,0}=0.000371(6)\,,\ \calf_{(3)}^h=3.116692(8)\,,\ \calk_{1,(4)}^h=-0.954388(2)\,.
\end{split}
\eqlabel{order2results}
\end{equation}

To summarize, from \eqref{summarylead}, \eqref{sumsublead} and \eqref{order2results}, 
the mass-squared $\w^2$ of $L=0$ (see \eqref{harmonics}), $q=1$ (see \eqref{leadF}) 
state of linearized $\csb$ fluctuations about chirally symmetric state of cascading gauge theory with strong
coupling scale $\Lambda$, compactified  on $S^3$ 
of radius $\frac 1\mu$ is given by 
\begin{equation}
\frac{\w^2}{\mu^2}=9\mp 6\sqrt{2}\ \sqrt{\dd}+0.077172(8)\ \dd+\calo(\dd^{3/2})\,,\qquad \dd=\left(\ln\frac{\mu^2}{\Lambda^2P^2g_0}\right)^{-1}\,.
\eqlabel{w2pert}
\end{equation}

\subsection{Spectrum of $\csb$ fluctuations for general $\frac{\mu}{\Lambda}$}\label{tachyon}

\begin{figure}[t]
\begin{center}
\psfrag{w2}{{$\frac{\w^2}{\mu^2}$}}
\psfrag{dd}{{$\dd$}}
\includegraphics[width=4in]{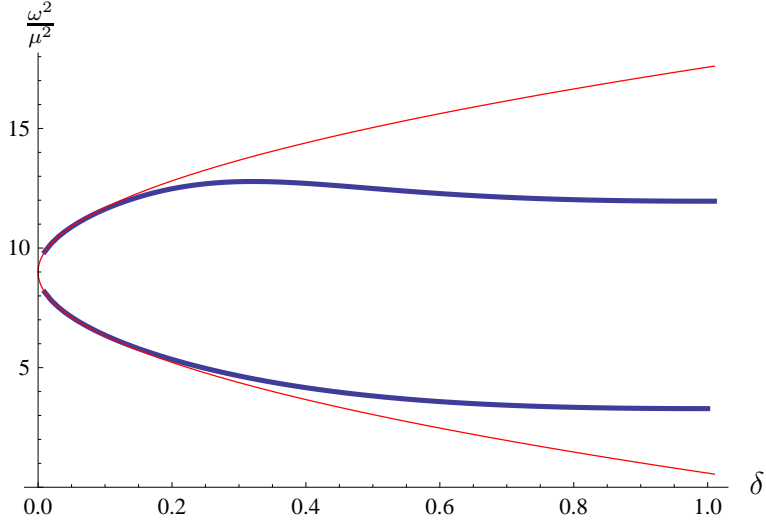}
\end{center}
  \caption{(Colour online) Comparison of the mass-squared $\w^2$ of the $L=0$, $q=1$ $\csb$ states 
 of cascading gauge theory on $S^3$ as a function of $\dd=(\ln({\mu^2}/{(\Lambda^2P^2g_0)})^{-1}$ 
in the range $\dd\in[0,1]$ (blue curves) with  perturbative predictions \eqref{w2pert} (red curves).
} \label{figure4}
\end{figure}

\begin{figure}[t]
\begin{center}
\psfrag{lnmuls}{{$\ln\frac{\mu^2}{\Lambda^2}$}}
\psfrag{w2}{{$\frac{\w^2}{\mu^2}$}}
\includegraphics[width=3in]{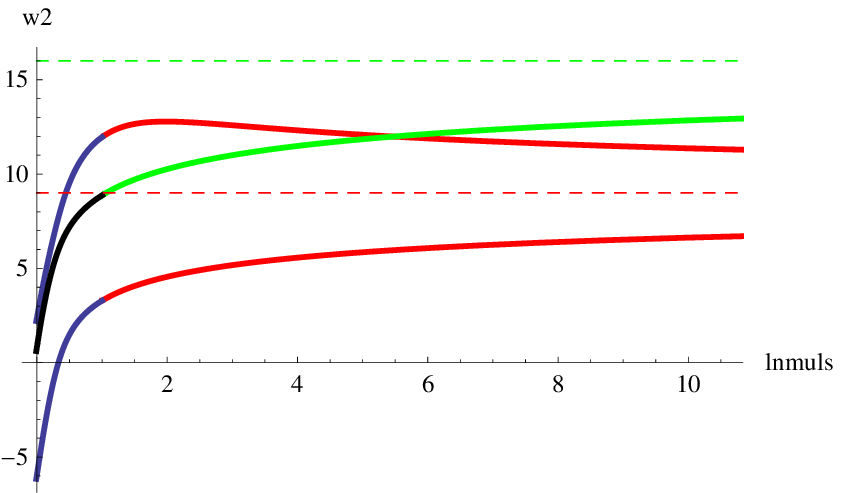}
\includegraphics[width=3in]{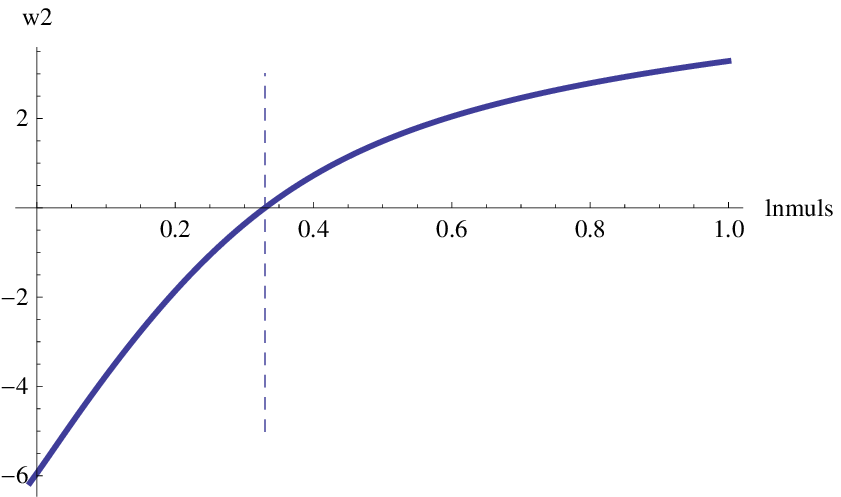}
\end{center}
  \caption{(Colour online) The spectrum of $L=0$, $q=1$ (blue and red curves) and of $L=3$, $q=1$ (black and green curves)  
states of $\csb$ fluctuations of cascading gauge theory on $S^3$ as a function of $\ln\frac{\mu^2}{\Lambda^2}$.
The red and green curves are obtained from numerical solutions with $K_0=1$ and varying $P^2$, while the 
blue and black curves are obtained with $P^2=1$ and varying $K_0$  (see section \ref{numerical} for more details).
The horizontal dashed lines represent the asymptotic mass-squared of the states with $L=0$ (red curve) and 
$L=3$ (green curve). The vertical blue dashed line represent the critical value $\mu_c$ (see \eqref{mucres}) 
below which  the lighter 
$L=0$, $q=1$ state becomes tachyonic.   
} \label{figure5}
\end{figure}

In this section we present results for the mass of the linearized $\csb$ fluctuations about the 
chirally symmetric state of cascading gauge theory on $S^3$ for general values of 
the $S^3$ compactification scale $\mu$ and the strong coupling scale $\Lambda$ of  
cascading gauge theory. The equations of motion and the boundary conditions 
for the $\csb$ fluctuations are presented in section \ref{fleombc}.
A mass of a  generic state in the $\csb$ spectrum depends on the $S^3$ eigenvalue $L$ (see \eqref{harmonics})
as well as on an integer $q=1,2,\cdots$ (see \eqref{leadF}) which quantizes its radial wavefunction \eqref{fluctwave}.
For each value $\{L,q\}$ there are two branches in the spectrum arising from non-analytic dependence of a 
mass on $\sqrt{P^2}$. 
The mass of $L=0$, $q=1$ state was computed perturbatively in $\dd=(\ln({\mu^2}/{(\Lambda^2P^2g_0)})^{-1}$ in 
section \ref{fluctpert}, see \eqref{w2pert} for the final expression. 

Figure \ref{figure4} compares the mass-squared $\frac{\w^2}{\mu^2}$ of $L=0$, $q=1$ $\csb$ states in the range 
$\dd\in[0,1]$ (blue curves) with perturbative predictions \eqref{w2pert} (red curves).

Figure \ref{figure5} presents the spectrum of $L=0$, $q=1$ (blue and red curves) and of $L=3$, $q=1$ (black and green curves)  
states of $\csb$ fluctuations of cascading gauge theory on $S^3$ as a function of $\ln\frac{\mu^2}{\Lambda^2}$.
The red and green curves are obtained from numerical solutions with $K_0=1$ and varying $P^2$, while the 
blue and black curves are obtained with $P^2=1$ and varying $K_0$  (see section \ref{numerical} for more details).
The horizontal dashed lines represent the asymptotic mass-squared of the $q=1$ states with $L=0$ (red curve) and 
$L=3$ (green curve)
\begin{equation}
\lim_{{\mu}/{\Lambda}\to \infty}\ \frac{\w^2}{\mu^2}\bigg|_{L=0,q=1}=9\,,\qquad \lim_{{\mu}/{\Lambda}\to \infty}\ \frac{\w^2}{\mu^2}
\bigg|_{L=3,q=1}=16\,.
\eqlabel{massl0}
\end{equation}
 The vertical blue dashed line represent the critical value $\mu_c$
\begin{equation}
\mu_c=1.179231(5)\ \Lambda\,,
\eqlabel{mucres}
\end{equation}
such that  for $\mu<\mu_c$  the lighter 
$L=0$, $q=1$ $\csb$ state becomes tachyonic.   

\subsection{The end point of $\csb$ tachyon condensation}\label{endpoint}

\begin{figure}[t]
\begin{center}
\psfrag{fa70}{{$f_{7,0}$}}
\psfrag{k130}{{$k_{1,3,0}$}}
\psfrag{fa10}{{$f_{1,0}$}}
\psfrag{fa30}{{$f_{3,0}$}}
\includegraphics[width=2in]{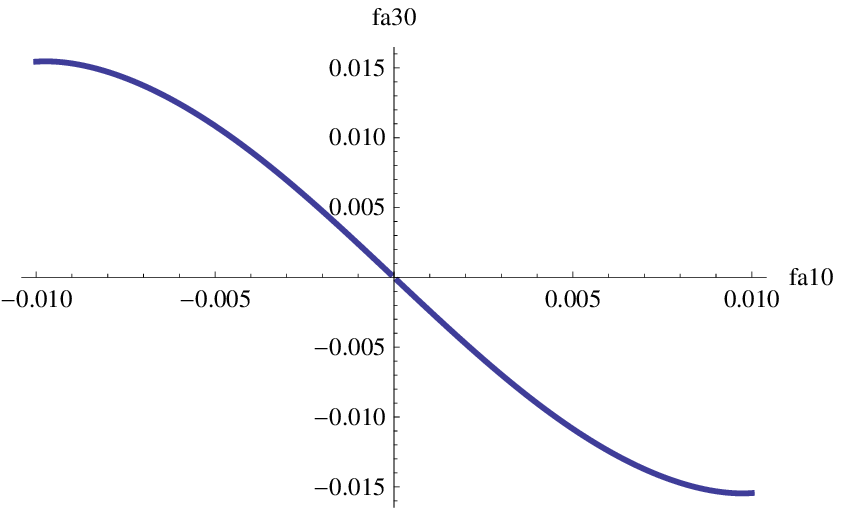}
\includegraphics[width=2in]{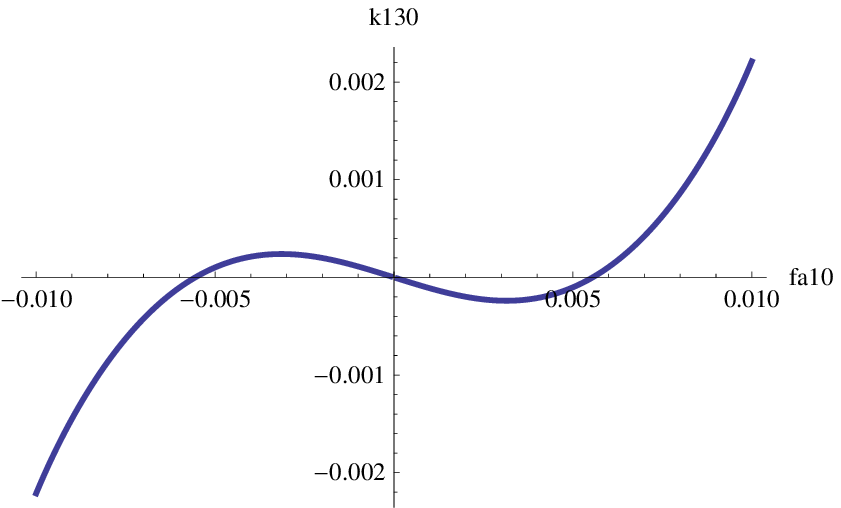}
\includegraphics[width=2in]{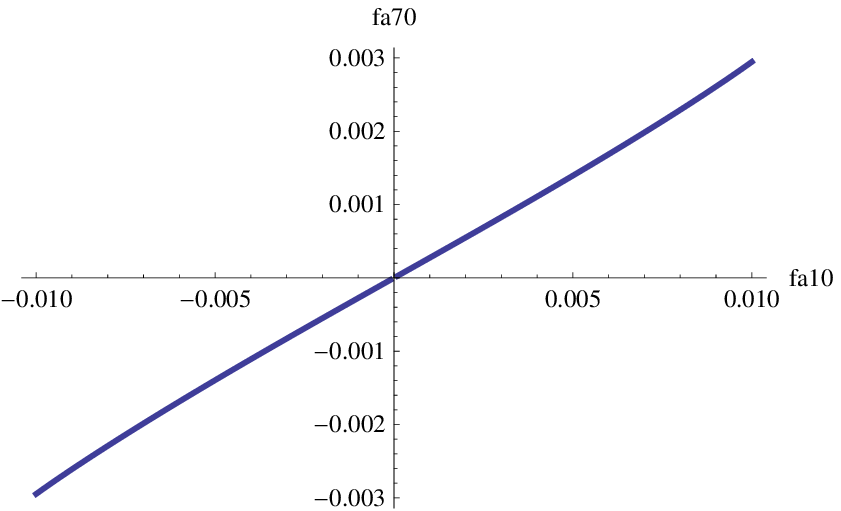}
\end{center}
  \caption{Expectation values of dimension-3 operators $\calo_{3,i}\propto \{f_{3,0},k_{1,3,0}\}$
and a dimension-7 operator $\calo_7\propto f_{7,0}$ of cascading gauge theory on $S^3$ with compactification 
scale $\mu_*<\mu_c$ as a function of gaugino mass deformation parameter $f_{1,0}\propto \frac{m}{\Lambda}$. 
} \label{figure6}
\end{figure}

In section \ref{tachyon} we identified the state in the spectrum of $\csb$ fluctuations of 
cascading gauge theory on $S^3$ which becomes tachyonic once the $S^3$ compactification scale $\mu$ 
becomes sufficiently low, see \eqref{mucres}. Notice that the mass of this state vanishes as $\mu\to \mu_c$,
see figure \ref{figure5}.
An interesting question is then whether the condensation of this tachyon signals a second-order 
(spontaneous) chiral symmetry breaking phase transition at $\mu=\mu_c$. We address this question 
following the analysis similar to the one in \cite{ksbh}. Notice that  $\csb$ fluctuations 
about a chirally symmetric state are associated
with the expectation values of two dimension-3 operators $\calo_{3,i}\,,\ \ i=1,2$ 
(corresponding to the normalizable mode coefficients 
$\{\calf_{3,0}\,, \calk_{1,3,0}\}$ in \eqref{uvfl}) and that of 
a single dimension-7 operator $\calo_7$ (corresponding to the normalizable 
mode coefficient $\calf_{7,0}$ in \eqref{uvfl}). In a chirally symmetric state, \ie for $\mu>\mu_c$ (prior to 
the tachyon condensation), these vevs vanish. If the end point of the $\csb$ tachyon condensation is continuously 
(via a second-order phase transition) connected to a chirally symmetric state we expect that there is a new phase of cascading 
gauge theory on $S^3$ at $\mu<\mu_c$, such that 
\begin{equation}
\{\calo_{3,i}\ne 0\,,\ \calo_7\ne 0\}\to \{0,0\}\qquad {\rm as}\qquad \mu_c-\mu\to 0_+\,.
\eqlabel{csbpahse}
\end{equation} 
As in \cite{ksbh}, to access this new state, we deform\footnote{We consider only 
$SO(4)$-invariant states.} cascading gauge theory on $S^3$ with the compactification scale
$\mu=\mu_*<\mu_c$, in practice we choose
\begin{equation}
\mu_*=0.960921(1)\ \mu_c\qquad \Longleftrightarrow\qquad \{K_0=0.25\,, P^2=1\}\,,
\eqlabel{mustardef}
\end{equation} 
by giving an explicit mass\footnote{Generically, there are two independent mass deformations 
of this type, see \cite{ksbh}.} $m$ to 
gauginous ($\caln=1$ fermionic superpartners of $SU(K+P)\times SU(K)$ gauge bosons).
Once $m\ne 0$, such a deformation explicitly breaks chiral symmetry and 
generates the nonzero vevs for $\{\calo_{3,i}\,, \calo_7\}$. Without establishing the precise holographic 
dictionary\footnote{This can be done as in \cite{ksbh}.}, it is clear that 
\begin{equation}
\frac{m}{\Lambda}\propto f_{1,0}\,,\qquad \calo_{3,i}\propto \{f_{3,0}\,, k_{1,3,0}\}\,,\qquad 
\calo_7\propto f_{7,0}\,,
\eqlabel{holrel}
\end{equation}
where, at the linearized level, $f_{1,0}$ corresponds to the UV non-normalizable coefficient in 
$\calf$ (it would modify 
the asymptotic expansion  \eqref{ffluv} with a leading term $f_{1,0}\ \r$ ), and  $\{f_{3,0}\,, f_{7,0}\,, k_{1,3,0}\}$
correspond to the normalizable coefficients  $\{\calf_{3,0}\,, \calf_{7,0}\,, \calk_{1,3,0}\}$
in the fluctuations  $\{\calf,\calk_1,\calk_2\}$, see \eqref{ffluv}-\eqref{k2fluv}.
The sought-after new phase of cascading gauge theory on $S^3$ with spontaneous breaking of chirally symmetry is
obtained in the limit $\frac{m}{\Lambda}\to 0$, provided the dimension-3 and dimension-7 condensates do not vanish in 
in this limit. 
We omit further details associated with discussion of the equations of motion, the appropriate 
boundary conditions for the holographic dual of mass-deformed cascading gauge theory on $S^3$ and present only the 
results\footnote{See sections \ref{kseoms}-\ref{ksir} and \cite{ksbh} for a related detailed discussion.}.

Figure \ref{figure6} shows expectation values of dimension-3 operators $\calo_{3,i}\propto \{f_{3,0},k_{1,3,0}\}$
and a dimension-7 operator $\calo_7\propto f_{7,0}$ of cascading gauge theory on $S^3$ with compactification 
scale $\mu_*<\mu_c$ (see \eqref{mustardef}) as a function of gaugino mass deformation parameter 
$f_{1,0}\propto \frac{m}{\Lambda}$. Notice that all the curves are odd with respect to $f_{1,0}$ --- for instance, 
for the range\footnote{It is difficult numerically to reach larger values of $|f_{1,0}|$ reliably.} $f_{1,0}\in [-0.01,0.01]$,
\begin{equation}
\bigg|\frac{f_{3,0}(f_{1,0})}{f_{3,0}(-f_{1,0})}+1\bigg|\sim (10^{-2}\cdots 5)\times 10^{-7}\,,
\eqlabel{oddness}
\end{equation}
and likewise for the remaining parameters. All these suggest that in the chiral limit, \ie $\frac{m}{\Lambda}\to 0$, 
all the $\csb$ condensates vanish, and the only state we find is that of (perturbatively unstable) chirally symmetric phase.
Thus, condensation of the $\csb$ tachyons discussed in section \ref{tachyon} is not a signature of the second-order
(spontaneous) $\csb$ phase transition --- in other words, the end point of tachyon condensation for $\mu<\mu_c$ 
describes a state that can not be continuously connected to a chirally symmetric state of cascading gauge theory on 
$S^3$.  

\section{Cascading gauge theory on $S^3$ with spontaneously broken chiral symmetry}
In section \ref{tachyon} we showed that $SO(4)$-invariant states of cascading gauge theory on $S^3$ with unbroken chiral symmetry 
are perturbatively unstable once the $S^3$ compactification scale $\mu<\mu_c$, see  \eqref{mucres}.
In section \eqref{endpoint} we argued that there is no $SO(4)$-invariant phase of cascading gauge theory on $S^3$ with 
spontaneously broken chiral symmetry that is continuously connected to above chirally symmetric phase at $\mu=\mu_c$. 
One possibility (that we will not pursue here) is that the end point of the tachyon condensation is some $SO(4)$ non-invariant state 
of the theory. Another, more likely, outcome is that while the condensation end point is $SO(4)$ invariant, it is not connected 
via a second-order phase transition to a chirally symmetric phase. In this section we construct such a  candidate state and show
that it is connected via the first order phase transition to a chirally symmetric phase 
at $\mu=\mu_{\csb}>\mu_c$.

\subsection{``Continuous'' $R^{3}\to S^{3}$ compactification of Klebanov-Strassler state of cascading gauge theory}\label{kscompact}
$\caln=1$ supersymmetric ground state of cascading gauge theory on $R^{3,1}$ --- referred to as Klebanov-Strassler state --- 
spontaneously breaks chiral symmetry \cite{ks}. 
A natural route to construct a $\csb$ state of the theory on $S^3$ is to ``compactify''  Klebanov-Strassler state:
$R^3\to S^3$. We explain now how to achieve this in a ``continuous'' fashion.
  
Consider the five-dimensional metric of the type:
\begin{equation}
ds_5^2=g_{\mu\nu}(y)dy^\mu dy^{\nu}=-c_1^2\ dt^2+c_2^2\ (d\calm_3)^2+c_3^2\ (d\r)^2\,,
\eqlabel{5dks}
\end{equation} 
where $c_i=c_i(\r)$, $\calm_3$ is either $R^3$ or $S^3$ and $(d\calm_3)^2$ is a standard metric on it.
We will be interested in $\csb$ states of cascading gauge theory on $\calm_3$. 
One can derive equations of motion from \eqref{ea1}-\eqref{ea8}. Alternatively, we can construct an effective 1-dimensional 
action\footnote{Effectively, in obtaining $S_1$ we perform Kaluza-Klein-like reduction of $S_5$ on 
$R\times \calm_3$.} from \eqref{5action}, by restricting to the metric ansatz \eqref{5dks}, and the $\r$-only dependence 
of the scalar fields $\{\Phi, h_i, \om_i\}$: 
\begin{equation}
S_5\left[g_{\mu\nu},\om_i,h_i,\Phi\right]\ \Longrightarrow\ S_1\left[c_i,\om_i,h_i,\Phi\right]\,.
\eqlabel{1action}
\end{equation}  
It can be verified that equations of motion obtained from $S_1$ coincide with those obtained from 
 \eqref{ea1}-\eqref{ea8}, provided 
we vary\footnote{This produces the first order constraint similar to \eqref{eq7}.} 
$S_1$ with respect to $c_3$, treating it as an unconstrained field. The 1-dimensional effective action approach 
makes it clear that the only place where the information about $\calm_3$ enters is through the evaluation of $R_5$ in \eqref{ric5}:
\begin{equation}
\begin{split}
R_5=-\frac{6 c_2''}{c_3^2 c_2}-\frac{2 c_1''}{c_3^2 c_1}+\frac{2 c_1' c_3'}{c_3^3 c_1}
-\frac{6 c_1' c_2'}{c_3^2 c_1 c_2}+\frac{6 c_2' c_3'}{c_3^3 c_2}-\frac{6 (c_2')^2}{c_3^2 c_2^2}
+\frac{6 \k}{c_2^2}\,,
\end{split}
\eqlabel{r5cuv}
\end{equation}
where derivatives are with respect to $\r$, and 
\begin{equation}
\k=\begin{cases} 0, &\ \text{if}\ \calm_3=R^3 
\\
1, &\ \text{if}\ \calm_3=S^3
\end{cases}\,.
\eqlabel{kappa}
\end{equation}
Even though $\k$ takes on  discrete values in \eqref{kappa}, there is no obstruction in treating $\k\in [0,1]$ as a continuous 
parameter  in the effective action $S_1$, thus providing a smooth, "continuous'', interpolation between $R^3$ and $S^3$.
For a general value $\k$ we denote cascading gauge theory compactification manifold $\calm_3^{(\k)}$.

\subsubsection{Equations of motion}\label{kseoms}
As in \eqref{metricaby} and \eqref{redef} we denote
\begin{equation}
\begin{split}
c_1=&h^{-1/4}\r^{-1}\,,\qquad  c_2=h^{-1/4}\r^{-1} f_1\,,\qquad c_3=h^{1/4}\r^{-1}\,,\qquad \Phi=\ln g\,,\\
h_1=&\frac 1P\left(\frac{K_1}{12}-36\Om_0\right)\,,\qquad h_2=\frac{P}{18}\ K_2\,,\qquad 
h_3=\frac 1P\left(\frac{K_3}{12}-36\Om_0\right)\,,\\
\Om_1=&\frac 13 f_c^{1/2} h^{1/4}\,,\qquad \Om_2=\frac {1}{\sqrt{6}} f_a^{1/2} h^{1/4}\,,\qquad 
\Om_3=\frac {1}{\sqrt{6}} f_b^{1/2} h^{1/4} \,.
\end{split}
\eqlabel{redef2}
\end{equation}
The equations of motion obtained from $S_1\left[c_i,\om_i,h_i,\Phi\right]$ are 
\begin{equation}
\begin{split}
&0=f_1''+\frac{f_1 (K_1')^2}{16h g P^2 f_b^2}+\frac{f_1 (K_3')^2}{16h g P^2 f_a^2}
+\frac{g P^2 f_1 (K_2')^2}{18h f_a f_b}-\frac{f_1 (f_a')^2}{4f_a^2}
-\frac{f_1 (f_b')^2}{4f_b^2}+\frac{f_1 (h')^2}{4h^2}\\
&+\frac{f_1 (g')^2}{4g^2}
-\frac{f_c' f_1 f_a'}{2f_c f_a}-\frac{f_c' f_1 f_b'}{2f_c f_b}-\frac{2 f_1' f_b'}{f_b}-\frac{f_1' f_c'}{f_c}
- \frac{3f_1' h'}{2h}-\frac{2 f_1' f_a'}{f_a}-\frac{f_1 f_a' f_b'}{f_a f_b}-\frac{(f_1')^2}{f_1}\\
&+\frac{2 f_1 f_c'}{f_c \r}
+\frac{4 f_1 f_a'}{f_a \r}+\frac{2 f_1 h'}{h \r}+\frac{6 f_1'}{\r}+\frac{4 f_1 f_b'}{f_b \r}
-\frac{2 f_1 f_c}{f_a f_b \r^2}- \frac{9f_1 f_a}{8f_c f_b \r^2}-\frac{9f_1 f_b}{8f_c f_a \r^2}\\
&-\frac{9f_1 K_3^2}{32g f_c P^2 f_a h f_b \r^2}
-\frac{9f_1 K_1^2}{32g f_c P^2 f_a h f_b \r^2}+\frac{f_1 g P^2 K_2}{f_c f_a^2 h \r^2}
-\frac{f_1 g P^2 K_2^2}{4f_c f_a^2 h \r^2}-\frac{f_1 g P^2 K_2^2}{4f_c h f_b^2 \r^2}
\\
&+\frac{9f_1 K_1 K_3}{16g f_c P^2 f_a h f_b \r^2}
-\frac{f_1 K_1^2}{4f_c f_a^2 h^2 f_b^2 \r^2}
-\frac{f_1 g P^2}{f_c f_a^2 h \r^2}-\frac{16f_1 K_2^2 K_1^2}{f_c f_a^2 h^2 f_b^2 \r^2}
+ \frac{f_1 K_2 K_1^2}{4f_c f_a^2 h^2 f_b^2 \r^2}\\
&- \frac{f_1 K_2^2 K_3^2}{16f_c f_a^2 h^2 f_b^2 \r^2}
+\frac{f_1 K_2^2 K_1 K_3}{8f_c f_a^2 h^2 f_b^2 \r^2}
- \frac{f_1 K_2 K_3 K_1}{4f_c f_a^2 h^2 f_b^2 \r^2}-\frac{6 f_1}{\r^2}+\frac{6 f_1}{f_a \r^2}+\frac{6 f_1}{f_b \r^2}
+\frac{9f_1}{4f_c \r^2}+\k\ \frac{h }{f_1}\,,
\end{split}
\eqlabel{kseq1}
\end{equation}
\begin{equation}
\begin{split}
&0=f_c''+\frac{3 f_c (f_1')^2}{f_1^2}+\frac{f_c f_a' f_b'}{f_a f_b}+\frac{3 f_c f_1' f_a'}{f_a f_1}
- \frac{f_c (h')^2}{4h^2}+\frac{3 f_c f_1' f_b'}{f_b f_1}+\frac{f_c (f_b')^2}{4f_b^2}-\frac{f_c (g')^2}{4g^2}
\\
&+\frac{f_c (f_a')^2}{4f_a^2}-\frac{(f_c')^2}{2f_c}+ \frac{3f_c' f_a'}{2f_a}-\frac{3f_c (K_1')^2}{16h f_b^2 g P^2}
-\frac{3f_c (K_3')^2}{16f_a^2 h g P^2}+ \frac{3f_c f_1' h'}{2h f_1}+\frac{3f_c' f_b'}{2f_b}
+\frac{9f_1' f_c'}{2f_1}\\
&-\frac{g P^2 f_c (K_2')^2}{6f_a h f_b}-\frac{6 f_c f_a'}{f_a \r}
-\frac{6 f_c f_b'}{f_b \r}-\frac{15 f_c f_1'}{f_1 \r}-\frac{6 f_c'}{\r}-\frac{2 f_c h'}{h \r}+\frac{K_2^2 K_1^2}{16f_a^2 h^2 f_b^2 \r^2}
- \frac{K_2^2 K_1 K_3}{8f_a^2 h^2 f_b^2 \r^2}\\
&-\frac{K_2 K_1^2}{4f_a^2 h^2 f_b^2 \r^2}
+\frac{K_2^2 K_3^2}{16f_a^2 h^2 f_b^2 \r^2}+\frac{3 g P^2}{f_a^2 h \r^2}+\frac{K_1^2}{4f_a^2 h^2 f_b^2 \r^2}
+ \frac{27K_1^2}{32f_a h f_b g P^2 \r^2}+ \frac{3g P^2 K_2^2}{4h f_b^2 \r^2}\\
&+ \frac{3g P^2 K_2^2}{4f_a^2 h \r^2}
-\frac{3 g P^2 K_2}{f_a^2 h \r^2}+\frac{27K_3^2}{32f_a h f_b g P^2 \r^2}+ \frac{K_2 K_3 K_1}{4f_a^2 h^2 f_b^2 \r^2}
+\frac{45f_b}{8f_a \r^2}-\frac{45}{4 \r^2}-\frac{6 f_c}{f_a \r^2}+\frac{45 f_a}{8f_b \r^2}\\
&-\frac{6 f_c}{f_b \r^2}
-\frac{6 f_c^2}{f_a f_b \r^2}+\frac{14 f_c}{\r^2}- \frac{27K_1 K_3}{16f_a h f_b g P^2 \r^2}-\k\ \frac{3 h f_c }{f_1^2}\,,
\end{split}
\eqlabel{kseq2}
\end{equation}
\begin{equation}
\begin{split}
&0=f_a''+\frac{f_c' f_a'}{f_c}+\frac{3f_a f_1' h'}{2h f_1}-\frac{3f_a (K_1')^2}{16h f_b^2 g P^2}
+\frac{f_a (f_b')^2}{4f_b^2}+\frac{6 f_1' f_a'}{f_1}+\frac{3 f_a (f_1')^2}{f_1^2}+\frac{3 f_a f_1' f_b'}{f_b f_1}
\\
&+\frac{2 f_a' f_b'}{f_b}-\frac{g P^2 (K_2')^2}{18h f_b}+\frac{(K_3')^2}{16f_a h g P^2}
-\frac{f_a (g')^2}{4g^2}+\frac{(f_a')^2}{4f_a}+\frac{3f_a f_c' f_1'}{2f_1 f_c}+\frac{f_a f_c' f_b'}{2f_b f_c}
-\frac{f_a (h')^2}{4h^2}\\
&-\frac{9 f_a'}{\r}-\frac{3 f_a f_c'}{f_c \r}-\frac{6 f_a f_b'}{f_b \r}-\frac{15 f_a f_1'}{f_1 \r}
-\frac{2 f_a h'}{h \r}-\frac{f_a g P^2 K_2^2}{4h f_b^2 f_c \r^2}-\frac{3 g P^2 K_2}{f_a h f_c \r^2}+\frac{3g P^2 K_2^2}{4f_a h f_c \r^2}
\\
&-\frac{K_2^2 K_1 K_3}{8f_a h^2 f_b^2 f_c \r^2}+\frac{K_2 K_3 K_1}{4f_a h^2 f_b^2 f_c \r^2}
+\frac{9K_3^2}{32h f_b g P^2 f_c \r^2}+\frac{27f_b}{8f_c \r^2}- \frac{9f_a^2}{8f_b f_c \r^2}
-\frac{9f_a}{4f_c \r^2}\\
&+\frac{9K_1^2}{32h f_b g P^2 f_c \r^2}
+\frac{K_1^2}{4f_a h^2 f_b^2 f_c \r^2}-\frac{K_2 K_1^2}{4f_a h^2 f_b^2 f_c \r^2}
+\frac{K_2^2 K_1^2}{16f_a h^2 f_b^2 f_c \r^2}+
\frac{K_2^2 K_3^2}{16f_a h^2 f_b^2 f_c \r^2}\\
&+\frac{3 g P^2}{f_a h f_c \r^2}-\frac{18}{\r^2}+\frac{6 f_c}{f_b \r^2}-\frac{6 f_a}{f_b \r^2}
+\frac{14 f_a}{\r^2}- \frac{9K_1 K_3}{16h f_b g P^2 f_c \r^2}-\k\ \frac{3 f_a h}{f_1^2}\,,
\end{split}
\eqlabel{kseq3}
\end{equation}
\begin{equation}
\begin{split}
&0=f_b''-\frac{3f_b (K_3')^2}{16h g f_a^2 P^2}-\frac{f_b (h')^2}{4h^2}
+ \frac{3f_b f_1' h'}{2h f_1}+\frac{3f_b f_c' f_1'}{2f_1 f_c}-\frac{f_b (g')^2}{4g^2}
+\frac{f_c' f_b'}{f_c}+\frac{(f_b')^2}{4f_b}\\
&+\frac{(K_1')^2}{16h g f_b P^2}
+\frac{f_b f_c' f_a'}{2f_a f_c}+\frac{2 f_a' f_b'}{f_a}+\frac{3 f_b f_1' f_a'}{f_1 f_a}+\frac{3 f_b (f_1')^2}{f_1^2}
+\frac{f_b (f_a')^2}{4f_a^2}-\frac{g P^2 (K_2')^2}{18h f_a}+\frac{6 f_1' f_b'}{f_1}\\
&-\frac{2 f_b h'}{h \r}
-\frac{15 f_b f_1'}{f_1 \r}-\frac{6 f_b f_a'}{f_a \r}-\frac{3 f_b f_c'}{f_c \r}-\frac{9 f_b'}{\r}
-\frac{K_2 K_1^2}{4h^2 f_a^2 f_b f_c \r^2}+\frac{K_2^2 K_1^2}{16h^2 f_a^2 f_b f_c \r^2}\\&
+\frac{K_2^2 K_3^2}{16h^2 f_a^2 f_b f_c \r^2}-\frac{g f_b P^2}{h f_a^2 f_c \r^2}
-\frac{K_2^2 K_1 K_3}{8h^2 f_a^2 f_b f_c \r^2}+\frac{K_2 K_3 K_1}{4h^2 f_a^2 f_b f_c \r^2}
-\frac{9f_b^2}{8f_a f_c \r^2}+ \frac{27f_a}{8f_c \r^2}-\frac{9f_b}{4f_c \r^2}\\&
+\frac{9K_1^2}{32h g f_a P^2 f_c \r^2}+\frac{3g P^2 K_2^2}{4h f_b f_c \r^2}-\frac{g f_b P^2 K_2^2}{4h f_a^2 f_c \r^2}
+\frac{g f_b P^2 K_2}{h f_a^2 f_c \r^2}+\frac{9K_3^2}{32h g f_a P^2 f_c \r^2}+\frac{K_1^2}{4h^2 f_a^2 f_b f_c \r^2}
\\
&-\frac{6 f_b}{f_a \r^2}+\frac{6 f_c}{f_a \r^2}-\frac{18}{\r^2}+\frac{14 f_b}{\r^2}
-\frac{9K_1 K_3}{16h g f_a P^2 f_c \r^2}-\k\ \frac{3 h f_b }{f_1^2}\,,
\end{split}
\eqlabel{kseq4}
\end{equation}
\begin{equation}
\begin{split}
&0=h''-\frac{(h')^2}{4h}-\frac{9h f_c' f_1'}{2f_c f_1}-\frac{3h f_c' f_a'}{2f_c f_a}
-\frac{3h f_c' f_b'}{2f_c f_b}-\frac{9 h f_1' f_b'}{f_b f_1}+\frac{3h (g')^2}{4g^2}-\frac{3h (f_b')^2}{4f_b^2}
\\
&+ \frac{5g P^2 (K_2')^2}{18f_a f_b}-\frac{3f_1' h'}{2f_1}+\frac{f_a' h'}{f_a}+\frac{f_b' h'}{f_b}
-\frac{9 h (f_1')^2}{f_1^2}-\frac{9 h f_1' f_a'}{f_a f_1}-\frac{3h (f_a')^2}{4f_a^2}-\frac{3 h f_a' f_b'}{f_a f_b}
+\frac{h' f_c'}{2f_c}\\
&+\frac{5(K_1')^2}{16f_b^2 g P^2}+\frac{5(K_3')^2}{16f_a^2 g P^2}+\frac{8 h f_c'}{f_c \r}
+\frac{3 h'}{\r}+\frac{16 h f_a'}{f_a \r}+\frac{16 h f_b'}{f_b \r}+\frac{39 h f_1'}{f_1 \r}
+\frac{K_1^2}{4f_c f_a^2 h f_b^2 \r^2}\\
&+\frac{K_2^2 K_1^2}{16f_c f_a^2 h f_b^2 \r^2}
+\frac{K_2 K_3 K_1}{4f_c f_a^2 h f_b^2 \r^2}+\frac{g P^2 K_2}{f_c f_a^2 \r^2}
-\frac{K_2 K_1^2}{4f_c f_a^2 h f_b^2 \r^2}- \frac{g P^2 K_2^2}{4f_c f_b^2 \r^2}
+\frac{K_2^2 K_3^2}{16f_c f_a^2 h f_b^2 \r^2}\\
&-\frac{g P^2}{f_c f_a^2 \r^2}-\frac{g P^2 K_2^2}{4f_c f_a^2 \r^2}
-\frac{9K_3^2}{32f_c f_a f_b g P^2 \r^2}-\frac{27h f_b}{8f_c f_a \r^2}- \frac{27f_a h}{8f_c f_b \r^2}
+\frac{27 h}{4f_c \r^2}-\frac{9K_1^2}{32f_c f_a f_b g P^2 \r^2}\\
&-\frac{K_2^2 K_1 K_3}{8f_c f_a^2 h f_b^2 \r^2}
+\frac{18 h}{f_a \r^2}-\frac{6 f_c h}{f_a f_b \r^2}+\frac{18 h}{f_b \r^2}-\frac{34 h}{\r^2}
+ \frac{9K_1 K_3}{16f_c f_a f_b g P^2 \r^2}+\k\ \frac{9 h^2} {f_1^2}\,,
\end{split}
\eqlabel{kseq5}
\end{equation}
\begin{equation}
\begin{split}
&0=K_1''-\frac{K_1' f_b'}{f_b}+\frac{K_1' f_c'}{2f_c}-\frac{K_1' h'}{h}+\frac{3 K_1' f_1'}{f_1}
-\frac{K_1' g'}{g}+\frac{K_1' f_a'}{f_a}-\frac{3 K_1'}{\r}+\frac{9f_b K_3}{2f_c f_a \r^2}
\\
&+\frac{g P^2 K_2^2 K_3}{f_c f_a^2 h \r^2}- \frac{9f_b K_1}{2f_c f_a \r^2}-\frac{4 g P^2 K_1}{f_c f_a^2 h \r^2}
+\frac{4 g P^2 K_2 K_1}{f_c f_a^2 h \r^2}-\frac{g P^2 K_2^2 K_1}{f_c f_a^2 h \r^2}-\frac{2 g P^2 K_2 K_3}{f_c f_a^2 h \r^2}\,,
\end{split}
\eqlabel{kseq6}
\end{equation}
\begin{equation}
\begin{split}
&0=K_2''+\frac{3 K_2' f_1'}{f_1}+\frac{K_2' f_c'}{2f_c}+\frac{g' K_2'}{g}-\frac{K_2' h'}{h}
-\frac{3 K_2'}{\r}-\frac{9K_3 K_1}{4f_c g P^2 f_a h f_b \r^2}+\frac{9K_2 K_1 K_3}{4f_c g P^2 f_a h f_b \r^2}
\\
&- \frac{9f_b K_2}{2f_c f_a \r^2}-\frac{9f_a K_2}{2f_c f_b \r^2}+\frac{9 f_b}{f_c f_a \r^2}
-\frac{9K_2 K_1^2}{8f_c g P^2 f_a h f_b \r^2}- \frac{9K_2 K_3^2}{8f_c g P^2 f_a h f_b \r^2}
+\frac{9K_1^2}{4f_c g P^2 f_a h f_b \r^2}\,,
\end{split}
\eqlabel{kseq7}
\end{equation}
\begin{equation}
\begin{split}
&0=K_3''-\frac{K_3' h'}{h}+\frac{3 K_3' f_1'}{f_1}-\frac{K_3' g'}{g}+ \frac{K_3' f_c'}{2f_c}
+\frac{K_3' f_b'}{f_b}-\frac{K_3' f_a'}{f_a}-\frac{3 K_3'}{\r}-\frac{g P^2 K_2^2 K_3}{f_c f_b^2 h \r^2}
\\&- \frac{9f_a K_3}{2f_c f_b \r^2}+\frac{g P^2 K_2^2 K_1}{f_c f_b^2 h \r^2}+ \frac{9f_a K_1}
{2f_c f_b \r^2}-\frac{2 g P^2 K_2 K_1}{f_c f_b^2 h \r^2}\,,
\end{split}
\eqlabel{kseq8}
\end{equation}
\begin{equation}
\begin{split}
&0=g''-\frac{g^2 P^2 (K_2')^2}{9f_a f_b h}+\frac{g' f_b'}{f_b}-\frac{(g')^2}{g}+\frac{3 g' f_1'}{f_1}
+\frac{(K_1')^2}{8f_b^2 h P^2}+\frac{(K_3')^2}{8f_a^2 h P^2}+\frac{g' f_c'}{2f_c}+\frac{g' f_a'}{f_a}
\\
&-\frac{3 g'}{\r}-\frac{g^2 P^2 K_2^2}{2f_c f_b^2 h \r^2}+ \frac{9K_3^2}{16f_c f_a f_b h \r^2 P^2}
-\frac{2 g^2 P^2}{f_c f_a^2 h \r^2}-\frac{9K_1 K_3}{8f_c f_a f_b h \r^2 P^2}+\frac{9K_1^2}{16f_c f_a f_b h \r^2 P^2}
\\
&-\frac{g^2 P^2 K_2^2}{2f_c f_a^2 h \r^2}+\frac{2 g^2 P^2 K_2}{f_c f_a^2 h \r^2}\,.
\end{split}
\eqlabel{kseq9}
\end{equation}
Additionally, we have the first order constraint
\begin{equation}
\begin{split}
&0=(K_1')^2 f_a^2+(K_3')^2 f_b^2-4 h g f_a^2 P^2 (f_b')^2+\frac{4}{h} g f_a^2 f_b^2 P^2 (h')^2
-\frac{24}{f_1f_c} h g f_a^2 f_b^2 P^2 f_c' f_1'\\
&-\frac{48}{f_1} h g f_a^2 f_b P^2 f_1' f_b'
+\frac{4}{g} h (g')^2 f_a^2 f_b^2 P^2-4 h g f_b^2 P^2 (f_a')^2-16 h g f_a f_b P^2 f_a' f_b'
\\&
-\frac{48}{f_1} h g f_a f_b^2 P^2 f_1' f_a'-\frac{48}{f_1^2} h g f_a^2 f_b^2 P^2 (f_1')^2
+\frac89 g^2 P^4 (K_2')^2 f_a f_b-\frac{8}{f_c} h g f_a f_b^2 P^2 f_c' f_a'\\
&-\frac{8}{f_c} h g f_a^2 f_b P^2 f_c' f_b'
-\frac{24}{f_1} g f_a^2 f_b^2 P^2 f_1' h'+\frac{32}{f_c\r} h g f_a^2 f_b^2 P^2 f_c'
+\frac{64}{\r} h g f_a^2 f_b P^2 f_b'\\
&+\frac{144}{f_1\r} h g f_a^2 f_b^2 P^2 f_1'
+\frac{32}{\r} g f_a^2 f_b^2 P^2 h'+\frac{64}{\r} h g f_a f_b^2 P^2 f_a'
+\frac{16}{f_c\r^2} g^2 P^4 f_b^2 K_2\\
&+\frac{9}{f_c\r^2} f_a f_b K_1 K_3-\frac{4}{f_c\r^2} g^2 P^4 f_b^2 K_2^2
-\frac{18}{f_c\r^2} h g f_a^3 f_b P^2+\frac{36}{f_c\r^2} h g f_a^2 f_b^2 P^2+\frac{96}{\r^2} h g f_a f_b^2 P^2
\\
&+\frac{96}{\r^2} h g f_a^2 f_b P^2-\frac{96}{\r^2} h g f_a^2 f_b^2 P^2-\frac{4 g P^2 K_1^2}{h f_c \r^2}
-\frac{4}{f_c\r^2} g^2 P^4 K_2^2 f_a^2-\frac{18}{f_c\r^2} h g f_a f_b^3 P^2
\\
&+\frac{2 g P^2 K_2^2 K_1 K_3}{h f_c \r^2}-\frac{32}{\r^2} h g f_c f_a f_b P^2-\frac{g P^2 K_2^2 K_1^2}{h f_c \r^2}
+\frac{4 g P^2 K_2 K_1^2}{h f_c \r^2}-\frac{4 g P^2 K_2 K_3 K_1}{h f_c \r^2}\\
&-\frac{9}{2f_c\r^2} f_a f_b K_1^2
-\frac{9}{2f_c\r^2} f_a f_b K_3^2
-\frac{16}{f_c\r^2} g^2 P^4 f_b^2-\frac{g P^2 K_2^2 K_3^2}{h f_c \r^2}+\k\ \frac{48}{f_1^2} h^2 g f_a^2 f_b^2 P^2 \,.
\end{split}
\eqlabel{kseq10}
\end{equation}
We explicitly verified that for any value $\k$ the constraint \eqref{kseq10} is 
consistent with \eqref{kseq1}-\eqref{kseq9}. Moreover, 
with 
\begin{equation}
\k=1\,,\qquad f_c=f_2\,,\qquad f_a=f_b=f_3\,,\qquad K_1=K_3=K\,,\qquad K_2=1\,,
\eqlabel{chirallimit}
\end{equation}
equations \eqref{kseq1}-\eqref{kseq10} are equivalent to \eqref{eq1}-\eqref{eq7}.

\subsubsection{UV asymptotics}\label{ksuv}
The general UV (as $\r\to 0$) asymptotic solution of \eqref{kseq1}-\eqref{kseq10} describing the phase of cascading
 gauge theory with spontaneously broken chiral symmetry takes form
\begin{equation}
\begin{split}
f_1=&f_0\biggl(1+\left(-\frac{\k}{8}  K_0-\frac{\k}{16}P^2 g_0 +\frac{\k}{4}P^2 g_0  \ln\r\right) \frac{\r^2}{f_0^2}
+\biggl(\frac{\k}{16} \a_{1,0} P^2g_0-\frac{\k}{8} \a_{1,0} K_0\\
&+\frac{\k}{4} \a_{1,0} P^2 g_0 
 \ln\r\biggr) \frac{\r^3}{f_0^3}+\sum_{n=4}^\infty\sum_k f_{1,n,k}\ \frac{\r^n}{f_0^n}\ln^k\r\biggr)\,,
\end{split}
\eqlabel{ksf1}
\end{equation}
\begin{equation}
\begin{split}
f_c=&1-\a_{1,0} \frac{\r}{f_0}+ \biggl(\frac{\k}{4} K_0+\frac{\a_{1,0}^2}{4}+\frac{3\k}{8} P^2g_0 
-\frac\k2 P^2g_0 \ln\r
\biggr)\frac{\r^2}{f_0^2}-\frac\k4 \a_{1,0} P^2 g_0  \ \frac{\r^3}{f_0^3}\\
&+\sum_{n=4}^\infty\sum_k f_{c,n,k}\ \frac{\r^n}{f_0^n}\ln^k\r\,,
\end{split}
\eqlabel{ksfc}
\end{equation}
\begin{equation}
\begin{split}
f_a=&1-\a_{1,0} \frac{\r}{f_0}+ \biggl(\frac\k4 K_0+\frac{\a_{1,0}^2}{4}+\frac{5\k}{16}P^2 g_0 -\frac\k2 P^2g_0  \ln\r
\biggr)\frac{\r^2}{f_0^2}\\
&+ \biggl(-\frac\k4 \a_{1,0} P^2g_0 +f_{a,3,0}\biggr)\frac{\r^3}{f_0^3}+\sum_{n=4}^\infty\sum_k f_{a,n,k}\ \frac{\r^n}{f_0^n}\ln^k\r\,,
\end{split}
\eqlabel{ksfa}
\end{equation}
\begin{equation}
\begin{split}
f_b=&1-\a_{1,0} \frac{\r}{f_0}+ \biggl(\frac\k4 K_0+ \frac{\a_{1,0}^2}{4}+\frac{5\k}{16}P^2 g_0 -\frac\k2 P^2g_0  \ln\r
\biggr)\frac{\r^2}{f_0^2}\\
&+\biggl(-\frac\k4 \a_{1,0} P^2g_0 -f_{a,3,0}\biggr)\frac{\r^3}{f_0^3}+\sum_{n=4}^\infty\sum_k f_{b,n,k}\ \frac{\r^n}{f_0^n}\ln^k\r\,,
\end{split}
\eqlabel{ksfb}
\end{equation}
\begin{equation}
\begin{split}
h=&\frac18 P^2 g_0+\frac14 K_0-\frac12 P^2 g_0 \ln\r+\a_{1,0} \left(\frac 12 K_0- P^2 g_0 \ln\r\right) \frac{\r}{f_0}
+ \biggl(\frac{23\k}{288} P^4 g_0^2-\frac\k8 K_0^2\\&
 -\frac\k6 P^2 g_0  K_0+\frac{\a_{1,0}^2}{8}(5K_0-2 P^2 g_0)
+\frac16 P^2 g_0\biggl(3\k K_0+2\k P^2g_0 -\frac{15}{2}  \a_{1,0}^2\biggr) \ln\r
\\&-\frac\k2 P^4 g_0^2 \ln^2\r\biggr)\frac{\r^2}{f_0^2}
+\biggl(\frac{13\k}{32} \a_{1,0} P^4 g_0^2 -\frac{11}{24} P^2 g_0 \a_{1,0}^3-\frac \k4 \a_{1,0}  K_0 P^2 g_0
+\frac58 K_0 \a_{1,0}^3\\
&-\frac{3\k}{8} \a_{1,0} K_0^2+\biggl(\frac\k2 \a_{1,0} P^4 g_0^2 -\frac54 P^2 g_0 \a_{1,0}^3
+\frac{3\k}{2} \a_{1,0}  K_0 P^2 g_0\biggr) \ln\r\\
&-\frac{3\k}{2} \a_{1,0} P^4 g_0^2  \ln^2\r\biggr)
\frac{\r^3}{f_0^3}+\sum_{n=4}^\infty\sum_k h_{n,k}\ \frac{\r^n}{f_0^n}\ln^k\r\,,
\end{split}
\eqlabel{ksh}
\end{equation}
\begin{equation}
\begin{split}
K_1=&K_0-2 P^2 g_0 \ln\r-P^2 g_0 \a_{1,0} \frac{\r}{f_0}+ \biggl(
\frac14 P^2 g_0 (\k K_0+3 P^2 g_0  \k-\a_{1,0}^2)-\frac\k2 P^4 g_0^2 \ln\r\biggr)\frac{\r^2}{f_0^2}\\
&+\biggl(
\frac{1}{12} P^2 g_0 (6 \a_{1,0} P^2 g_0  \k+12 k_{1,3,0}-\a_{1,0}^3+3 \a_{1,0} \k K_0)
+\frac12 P^2 g_0 (4 f_{a,3,0}\\
&-\a_{1,0} P^2 g_0  \k) 
\ln\r\biggr)\frac{\r^3}{f_0^3}+\sum_{n=4}^\infty\sum_k k_{1,n,k}\ \frac{\r^n}{f_0^n}\ln^k\r\,,
\end{split}
\eqlabel{ksK1}
\end{equation}
\begin{equation}
\begin{split}
K_2=&1+ \left(-f_{a,3,0}+\frac32 k_{1,3,0}+3 f_{a,3,0} \ln\r\right)
\frac{\r^3}{f_0^3}+\sum_{n=4}^\infty\sum_k k_{2,n,k}\ \frac{\r^n}{f_0^n}\ln^k\r\,,
\end{split}
\eqlabel{ksK2}
\end{equation}
\begin{equation}
\begin{split}
K_3=&K_0-2 P^2 g_0 \ln\r-P^2 g_0 \a_{1,0} \frac{\r}{f_0}+ \biggl(
\frac14 P^2 g_0 (\k K_0+3 P^2 g_0  \k-\a_{1,0}^2)-\frac \k2 P^4  g_0^2 \ln\r\biggr)
\frac{\r^2}{f_0^2}\\
&+ \biggl(
\frac{1}{12} P^2 g_0 (6 \a_{1,0} P^2 g_0  \k-12 k_{1,3,0}-\a_{1,0}^3+3 \a_{1,0} \k K_0)+\frac 12 P^2 g_0 (
-4 f_{a,3,0}\\
&-\a_{1,0} P^2 g_0  \k) 
\ln\r\biggr)\frac{\r^3}{f_0^3}+\sum_{n=4}^\infty\sum_k k_{3,n,k}\ \frac{\r^n}{f_0^n}\ln^k\r\,,
\end{split}
\eqlabel{ksK3}
\end{equation}
\begin{equation}
\begin{split}
g=&g_0\biggl(1-\frac\k4 P^2g_0   \frac{\r^2}{f_0^2}-\frac\k4  \a_{1,0} P^2g_0 \frac{\r^3}{f_0^3}
+\sum_{n=4}^\infty\sum_k g_{n,k}\ \frac{\r^n}{f_0^n}\ln^k\r\biggr)\,.
\end{split}
\eqlabel{ksg}
\end{equation}
It is characterized by 12 parameters:
\begin{equation}
\{K_0\,,\ f_0\,,\ g_0\,,\  \a_{1,0}\,,\ k_{1,3,0}\,,\ f_{c,4,0}\,,\ f_{a,3,0}\,,\ f_{a,4,0}\,,\ f_{a,6,0}\,,\ f_{a,7,0}\,,\ f_{a,8,0}\,,\ g_{4,0}\}\,.
\eqlabel{uvparks}
\end{equation}
In what follows we developed the UV expansion to order $\calo(\r^{10})$ inclusive.

\subsubsection{IR asymptotics}\label{ksir}
As in section \ref{ircond}, we use a radial coordinate $\r$ that extends to infinity, see \eqref{extend}. 
The crucial difference between the IR boundary conditions for a chirally symmetric phase discussed in section \ref{ircond}
and the IR boundary conditions for a $\csb$ phase discussed here is that in the former case the manifold $\calm_5$ geodesically
completes with (a smooth) shrinking to zero size of  $S^3\subset \calm_5$, while in the latter case, much like in supersymmetric 
Klebanov-Strassler state of cascading gauge theory \cite{ks}, the 10-dimensional uplift of $\calm_5$,
\begin{equation}
\calm_5\ \to\ \calm_{10}=\calm_5\times X_5\,,
\eqlabel{uplift}
\end{equation}
 (with the metric given by 
\eqref{10dmetric}), geodesically completes with (a smooth) shrinking of a 2-cycle in the compact manifold $X_5$ \cite{ks}.        
Introducing 
\begin{equation}
y\equiv \frac 1\r\,,\qquad h^h\equiv y^{-4}\ h\,,\qquad  f^h_{a,c}\equiv y^2\ f_{a,c}\,,
\eqlabel{kshorfunc}
\end{equation}
the general IR (as $y\to 0$) asymptotic solution of  \eqref{kseq1}-\eqref{kseq10} describing the $\csb$ phase of cascading 
gauge theory takes form
\begin{equation}
\begin{split}
f_1=f_{1,0}^h+\frac{h_0^h \k}{3f_{1,0}^h} y^2+\sum_{n=2}^\infty f_{1,n}^h\ y^{2n}\,,
\end{split}
\eqlabel{f1hks}
\end{equation}
\begin{equation}
\begin{split}
f_c^h=&\frac34 f_{a,0}^h+\biggl(-\frac{3f_{a,0}^h k_{2,4}^h}{2k_{2,2}^h}+\frac{f_{a,0}^h (k_{1,3}^h)^2}{64h_0^h P^2 g_0^h}
-\frac{13P^2 g_0^h}{15(f_{a,0}^h)^2 h_0^h}+\frac{19(k_{3,1}^h)^2}{320f_{a,0}^h h_0^h P^2 g_0^h}
\\&-\frac{19(k_{2,2}^h)^2 P^2 g_0^h}{540h_0^h}+\frac65
-\frac{f_{a,0}^h h_0^h \k}{2(f_{1,0}^h)^2}-\frac{27}{5 k_{2,2}^h f_{a,0}^h}+\frac{3k_{3,1}^h k_{1,3}^h}{20
k_{2,2}^h f_{a,0}^h h_0^h P^2 g_0^h}
\biggr) y^2\\&+\sum_{n=2}^\infty f_{c,n}^h\ y^{2n}\,,
\end{split}
\eqlabel{fchks}
\end{equation}
\begin{equation}
\begin{split}
f_a^h=&f_{a,0}^h+\biggr(\frac{f_{a,0}^h (k_{1,3}^h)^2}{48h_0^h P^2 g_0^h}-\frac{4P^2 g_0^h}{45(f_{a,0}^h)^2 h_0^h}
-\frac{17(k_{3,1}^h)^2}{240f_{a,0}^h h_0^h P^2 g_0^h}+ \frac{17(k_{2,2}^h)^2 P^2 g_0^h}{405h_0^h}
+\frac{11}{5}\\
&+\frac{f_{a,0}^h k_{2,4}^h}{k_{2,2}^h}
+ \frac{f_{a,0}^h h_0^h \k}{3(f_{1,0}^h)^2}
+\frac{18}{5 k_{2,2}^h f_{a,0}^h}-
\frac{k_{3,1}^h k_{1,3}^h}{10k_{2,2}^h f_{a,0}^h h_0^h P^2 g_0^h}\biggr)y^2+\sum_{n=2}^\infty f_{a,n}^h\ y^{2n}\,,
\end{split}
\eqlabel{fahks}
\end{equation}
\begin{equation}
\begin{split}
f_b=&3+\biggl(-\frac{(k_{1,3}^h)^2}{16P^2 g_0^h h_0^h}+\frac{4P^2 g_0^h}{3(f_{a,0}^h)^3 h_0^h}
+\frac{(k_{3,1}^h)^2}{16(f_{a,0}^h)^2 P^2 g_0^h h_0^h}- \frac{P^2 g_0^h (k_{2,2}^h)^2}{27f_{a,0}^h h_0^h}
-\frac{h_0^h \k}{(f_{1,0}^h)^2}\\&-\frac{3}{f_{a,0}^h}\biggr)y^2+\sum_{n=2}^\infty f_{b,n}^h\ y^{2n}\,,
\end{split}
\eqlabel{fbhks}
\end{equation}
\begin{equation}
h^h=h_0^h+\biggl(- \frac{(k_{1,3}^h)^2}{48P^2 g_0^h}-\frac{4P^2 g_0^h}{9(f_{a,0}^h)^3}
-\frac{(k_{3,1}^h)^2}{16(f_{a,0}^h)^2 P^2 g_0^h}- \frac{P^2 g_0^h (k_{2,2}^h)^2}{27f_{a,0}^h}
\biggr)y^2+\sum_{n=2}^\infty h_{n}^h\ y^{2n}\,,
\eqlabel{hhks}
\end{equation}
\begin{equation}
\begin{split}
K_1=&k_{1,3}^h y^3+\biggl(- \frac{P^2 g_0^h k_{1,3}^h (k_{2,2}^h)^2}{54f_{a,0}^h h_0^h}
- \frac{9(k_{1,3}^h)^3}{160P^2 g_0^h h_0^h}
+ \frac{6P^2 g_0^h k_{1,3}^h}{5(f_{a,0}^h)^3 h_0^h}- \frac{7k_{1,3}^h (k_{3,1}^h)^2}{160(f_{a,0}^h)^2 P^2 g_0^h h_0^h}
\\&-\frac{4 h_0^h k_{1,3}^h \k}{5(f_{1,0}^h)^2}- \frac{12k_{1,3}^h}{5f_{a,0}^h}
-\frac{9k_{3,1}^h}{5(f_{a,0}^h)^2}+ \frac{4P^2 g_0^h k_{2,2}^h 
k_{3,1}^h}{15(f_{a,0}^h)^3 h_0^h}\biggr)y^5+\sum_{n=2}^\infty k_{1,n}^h\ y^{2n+1}\,,
\end{split}
\eqlabel{k1hks}
\end{equation}
\begin{equation}
K_2=k_{2,2}^h y^2+k_{2,4}^h y^4+\sum_{n=3}^\infty k_{2,n}^h\ y^{2n}\,,
\eqlabel{k2hks}
\end{equation}
\begin{equation}
\begin{split}
K_3=&k_{3,1}^h y+\biggl(\frac{18k_{3,1}^h}{5(f_{a,0}^h)^2 k_{2,2}^h}
+ \frac{k_{3,1}^h (k_{1,3}^h)^2}{480h_0^h P^2 g_0^h}
+ \frac{41P^2 g_0^h (k_{2,2}^h)^2 k_{3,1}^h}{810f_{a,0}^h h_0^h}
+ \frac{4P^2 g_0^h k_{2,2}^h k_{1,3}^h}{135 f_{a,0}^h h_0^h}
\\&+\frac{k_{3,1}^h k_{2,4}^h}{k_{2,2}^h}
- \frac{k_{1,3}^h (k_{3,1}^h)^2}{10(f_{a,0}^h)^2 h_0^h P^2 g_0^h k_{2,2}^h}
+ \frac{2P^2 g_0^h k_{3,1}^h}{15(f_{a,0}^h)^3 h_0^h}
- \frac{41(k_{3,1}^h)^3}{480(f_{a,0}^h)^2 h_0^h P^2 g_0^h}
+ \frac{4k_{3,1}^h}{5f_{a,0}^h}\\&+\frac{4 h_0^h k_{3,1}^h \k}{15(f_{1,0}^h)^2}-\frac 15 k_{1,3}^h\biggr)y^3
+\sum_{n=2}^\infty k_{3,n}^h\ y^{2n+1}\,,
\end{split}
\eqlabel{k3hks}
\end{equation}
\begin{equation}
\begin{split}
g=&g^h_0\biggl(1+\biggl(-\frac{ (k_{3,1}^h)^2}
{16(f_{a,0}^h)^2 h_0^hP^2g_0^h}
+\frac{P^2 g_0^h (k_{2,2}^h)^2}{27f_{a,0}^h h_0^h}- \frac{(k_{1,3}^h)^2}{48h_0^hP^2 g_0^h}+\frac {4
P^2 g_0^h}{9(f_{a,0}^h)^3 h_0^h}\biggr)y^2\\&+\sum_{n=2}^\infty g_{n}^h\ y^{2n}\biggr)\,.
\end{split}
\eqlabel{ghks}
\end{equation}
Notice that the prescribed IR boundary conditions imply 
\begin{equation}
\lim_{y\to 0}\ \om_3^2 = \lim_{y\to 0}\ \frac 16\ f_b\ h^{1/2}=\lim_{y\to 0}\ \frac {y^2}{6}\ f_b\ (h^h)^{1/2}=0\,,
\eqlabel{2cycle}
\end{equation}  
with all the other warp factors in \eqref{10dmetric} being finite. Moreover, see \eqref{10dmetric},
\begin{equation}
\begin{split}
\lim_{y\to 0}\biggl(\om_1^2\ g_5^2+\om_2^2\ [g_3^2+g_4^2]\biggr)=\frac 16 f_{a,0}^h (h_0^h)^{1/2}\left(\frac 12 g_5^2
+g_3^2+g_4^2\right)\,,
\end{split}
\eqlabel{ids3}
\end{equation}
which is the metric of the round $S^3$ which stays of finite size in the deep infrared as the 2-cycle 
fibered over it (smoothly) shrinks to zero size \eqref{2cycle}.
Asymptotic solution \eqref{f1hks}-\eqref{ghks} is characterized by 8 parameters:
\begin{equation}
\{f_{1,0}^h\,,\ f_{a,0}^h\,,\ h_{0}^h\,,\ k_{1,3}^h\,,\ k_{2,2}^h\,,\ k_{2,4}^h\,,\ k_{3,1}^h\,, g_0^h\}\,.
\eqlabel{ksirpar}
\end{equation}
In what follows we developed the IR expansion to order $\calo(y^{10})$ inclusive.

\subsubsection{Symmetries and numerical procedure}
The background geometry \eqref{redef2} dual to a phase of cascading gauge theory with 
spontaneously broken chiral symmetry on $\calm_3^{(\k)}$ enjoys all the symmetries\footnote{We assume that $\k\in (0,1]$.}, 
properly generalized,
discussed in section \ref{symmetries}:
\nxt 
\begin{equation}
P\to \lambda P\,,\ g\to \frac 1\l g\,,\ \{\r,f_{1,a,b,c},h,K_{1,2,3}\}\to \{\r,f_{1,a,b,c},h,K_{1,2,3}\}\,,
\eqlabel{kssym1}
\end{equation} 
\nxt 
\begin{equation}
P\to \lambda P\,,\ \r\to \frac 1\l \r\,,\ \{h,K_{1,3}\}\to \l^2\{h,K_{1,3}\}\,,\ 
\{f_{1,a,b,c},K_{2},g\}\to \{f_{1,a,b,c},K_{2},g\}\,,
\eqlabel{kssym2}
\end{equation} 
\nxt 
\begin{equation}
\r\to \lambda \r\,,\ f_1\to \l f_1\,,\ 
\{P,f_{a,b,c},h,K_{1,2,3},g\}\to \{P,f_{a,b,c},h,K_{1,2,3},g\}\,,
\eqlabel{kssym3}
\end{equation} 
\nxt \begin{equation}
\left( \begin{array}{c}
P\\
\r  \\
h  \\
f_1\\
f_{a,b,c}\\
K_{1,2,3}\\
g   
\end{array} \right)\
\Longrightarrow \left( \begin{array}{c}
\hat{P}\\
\hr  \\
\hh  \\
\hf_1\\
\hf_{a,b,c}\\
\hK_{1,2,3}\\
\hat{g}   \end{array} \right)
=
\left( \begin{array}{c}
P\\
{\r}/{(1+\a\ \r)}  \\
(1+\a\ \r)^4\ h \\
f_1\\
(1+\a\ \r)^{-2}\ f_{a,b,c}\\
K_{1,2,3}\\
{g}   \end{array} \right)\,,\qquad \a={\rm const}\,.
\eqlabel{kssym4}
\end{equation}
Thus, much like in section \ref{symmetries}, we can set 
\begin{equation}
g_0=1\,,\qquad f_0=1\,,\qquad \frac{K_0}{P^2}=\ln\ \frac{\mu^2}{\Lambda^2 P^2}\equiv \frac 1\dd\,,
\eqlabel{ksset}
\end{equation}
where $\mu\equiv \frac {1}{f_0}$ is the compactification scale.
The residual diffeomorphisms \eqref{kssym4} are actually completely fixed once we insist on the IR asymptotics as 
in \eqref{f1hks}-\eqref{ghks}.

The numerical procedure for solving the background equations \eqref{kseq1}-\eqref{kseq10}, subject to the 
boundary conditions \eqref{ksf1}-\eqref{ksg} and \eqref{f1hks}-\eqref{ghks}  is identical to the one 
described earlier, see section \ref{numerical}. Given \eqref{ksset}, for a fixed $\dd$,  
the gravitational solution is characterized by 9 parameters in the UV 
and 8 parameters in the IR:
\begin{equation}
\begin{split}
&{\rm UV}:\qquad \{\a_{1,0}\,,\ k_{1,3,0}\,,\ f_{c,4,0}\,,\ f_{a,3,0}\,,\ f_{a,4,0}\,,\ f_{a,6,0}\,,\ f_{a,7,0}\,,\ f_{a,8,0}\,,\ g_{4,0}\}\,,
\\
&{\rm IR}:\qquad 
\{f_{1,0}^h\,,\ f_{a,0}^h\,,\ h_{0}^h\,,\ k_{1,3}^h\,,\ k_{2,2}^h\,,\ k_{2,4}^h\,,\ k_{3,1}^h\,, g_0^h\}\,.
\end{split}
\eqlabel{ksuvirfinal}
\end{equation} 
Notice that $9+8=17$ is precisely the number of integration constants needed to specify a solution to \eqref{kseq1}
-\eqref{kseq10} ---
we have 9 second order differential equations and a single first order differential constraint: $2\times 9-1=17$.

In practice, we replace the second-order differential equation for $f_c$ \eqref{kseq2} with the constraint equation \eqref{kseq10},
which we use to  algebraically eliminate $f_c'$ from \eqref{kseq1}, \eqref{kseq3}-\eqref{kseq9}. The solution is found using the 
``shooting'' method as detailed in \cite{abk}. 

Ultimately, we are interested in the solution at $\k=1$. 
Finding such a ``shooting'' solution in 17-dimensional parameter space  \eqref{ksuvirfinal} is quite challenging. 
Thus, we start with analytic result for $\k=0$ (the Klebanov-Strassler state of cascading gauge theory),
and a fixed value of $\dd$, and  
slowly increase 
$\k$, \ie continuously deform $\calm_3^{(\k)}$ from $R^3$ to $S^3$.  
We further use the obtained solution as a starting point to explore other values of $\dd$.

\subsubsection{$\k$-deformation of Klebanov-Strassler state}\label{ksk}

\begin{figure}[t]
\begin{center}
\psfrag{a10}{{$\a_{1,0}$}}
\psfrag{fa30}{{$f_{a,3,0}$}}
\psfrag{k130}{{$k_{1,3,0}$}}
\psfrag{K0}{{$K_0$}}
\includegraphics[width=2in]{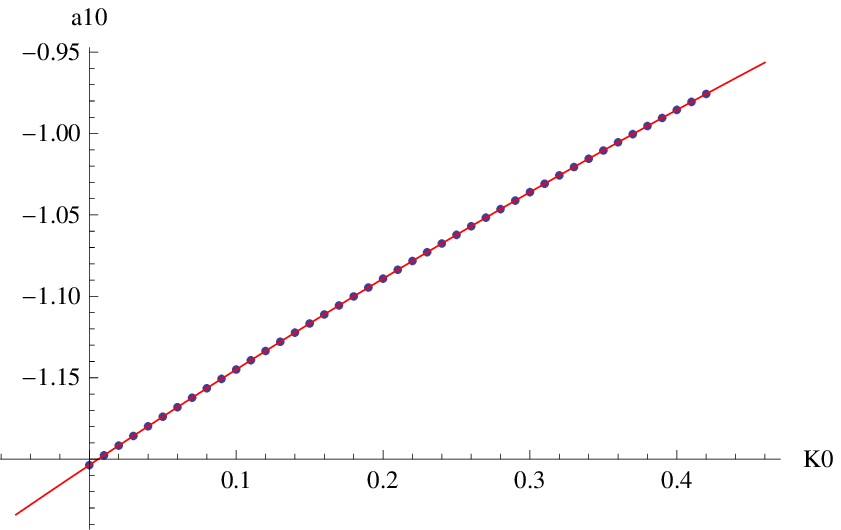}
\includegraphics[width=2in]{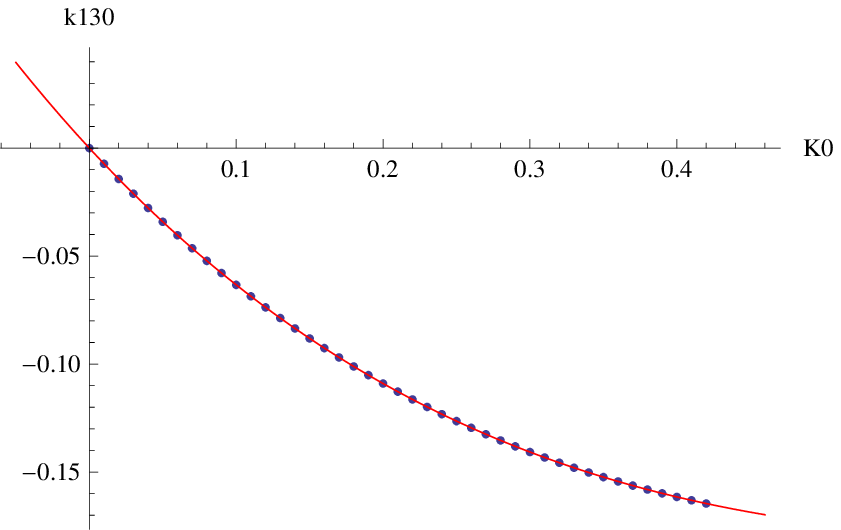}
\includegraphics[width=2in]{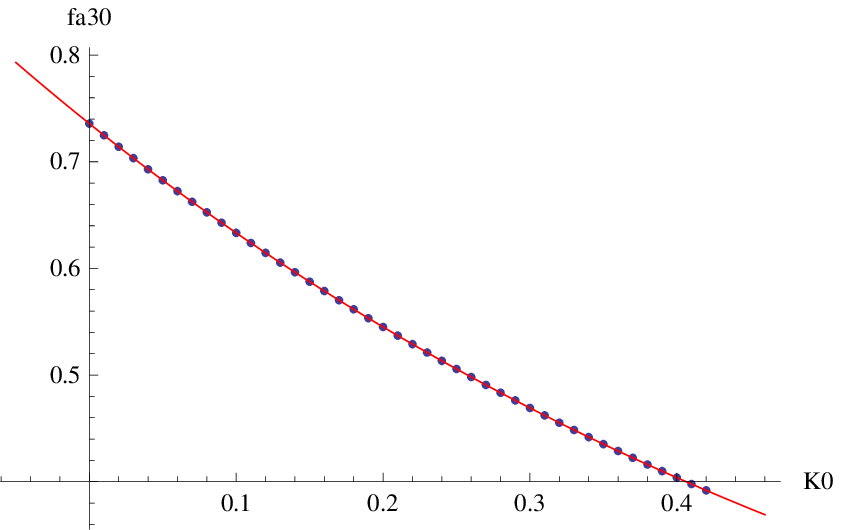}
\end{center}
  \caption{(Colour online) Comparison of values of  select UV  parameters  $\{\a_{1,0},k_{1,3,0},f_{a,3,0}\}$  
of Klebanov-Strassler state obtained numerically (blue dots) with the analytic prediction (red curves), see \eqref{susyuv}.
} \label{figure7}
\end{figure}

\begin{figure}[t]
\begin{center}
\psfrag{fah0}{{$f_{a,0}^h$}}
\psfrag{hh0}{{$h_{0}^h$}}
\psfrag{K1h3}{{$k_{1,3}^h$}}
\psfrag{K0}{{$K_0$}}
\includegraphics[width=2in]{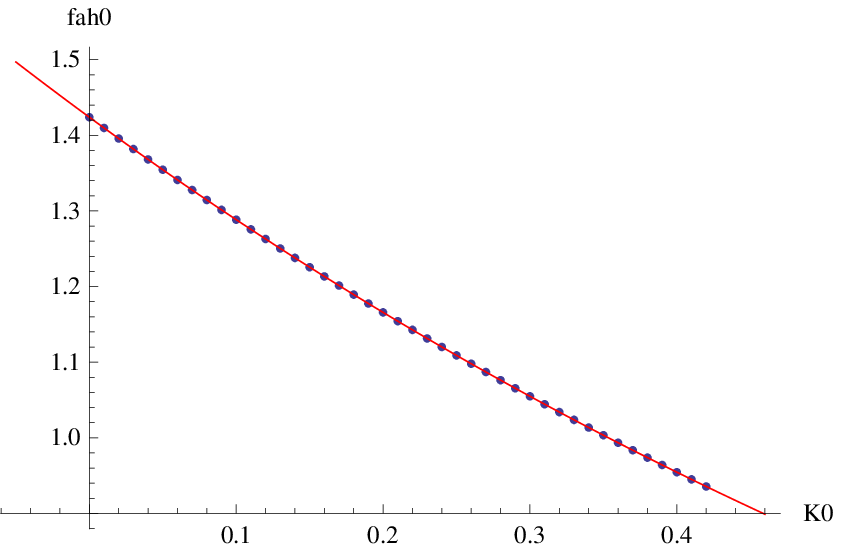}
\includegraphics[width=2in]{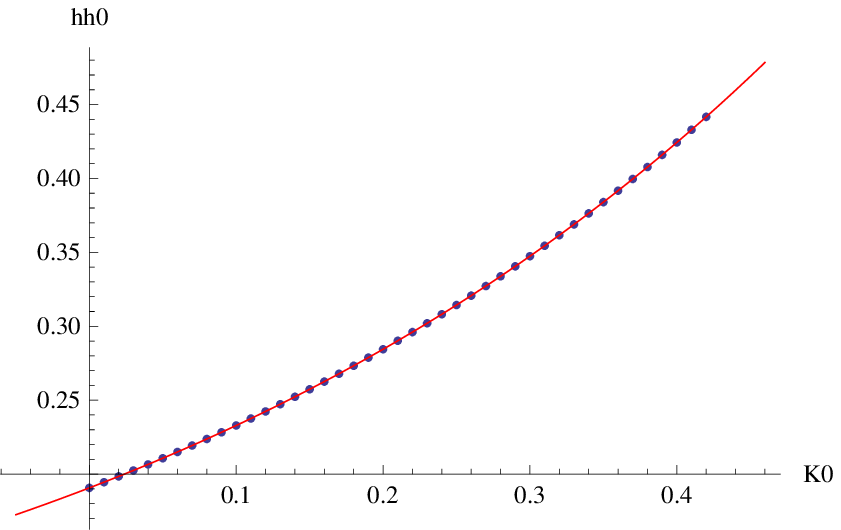}
\includegraphics[width=2in]{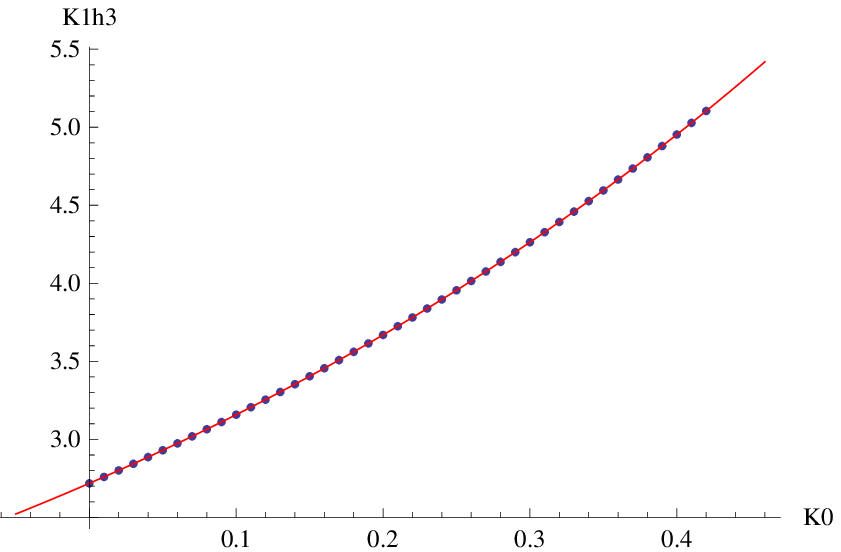}
\end{center}
  \caption{(Colour online) Comparison of values of  select IR  parameters  $\{f_{a,0}^h,h_0^h, k_{1,3}^h\}$  
of Klebanov-Strassler state obtained numerically (blue dots) with the analytic prediction (red curves), see \eqref{susyir}.
} \label{figure8}
\end{figure}

We begin with mapping the Klebanov-Strassler solution \cite{ks} to a $\k=0$ solution of \eqref{kseq1}-\eqref{kseq10}. 
We set
\begin{equation}
g_0=1\,,\qquad P=1\,.
\eqlabel{setkappa0}
\end{equation}
$\caln=1$ supersymmetric Klebanov-Strassler solution takes form\footnote{See eqs. (2.22) and (2.34) in \cite{ksbh}.}:
\begin{equation}
\begin{split}
&ds_5^2=H_{KS}^{-1/2}\ \left(-dt^2+dx_1^2+dx_2^2+dx_3^2\right)+H_{KS}^{1/2}\ \w_{1,KS}^2\ dr^2\,,\\
&\om_i=\w_{i,KS}\ H^{1/2}_{KS}\,,\qquad h_i=h_{i,KS}\,,
\end{split}
\eqlabel{ks1}
\end{equation}
\begin{equation}
\begin{split}
&h_{1,KS}=\frac{\cosh r-1}{18\sinh r}
\left(\frac{r\cosh r}{\sinh r}-1\right)\,,\qquad 
h_{2,KS}=\frac{1}{18}\left(1-\frac {r}{\sinh r}\right)\,,\\
&h_{3,KS}=\frac{\cosh r+1}{18\sinh r}
\left(\frac{r\cosh r}{\sinh r}-1\right)\,,
\qquad g=1\,,\qquad f_1=1\,,\\
&\w_{1,KS}=\frac{\epsilon^{2/3}}{\sqrt{6}{\hat K_{KS}}}\,,\qquad 
\w_{2,KS}=\frac{\epsilon^{2/3}{\hat K_{KS}}^{1/2}}{\sqrt{2}}\cosh\frac r2\,,\qquad  \w_{3,KS}=\frac{\epsilon^{2/3}
{\hat K_{KS}}^{1/2}}{\sqrt{2}}
\sinh\frac r2\,,
\end{split}
\eqlabel{ksks}
\end{equation}
with 
\begin{equation}
{\hat K_{KS}}=\frac{(\sinh (2r)-2r)^{1/3} }{2^{1/3}\sinh r}\,,\ H'_{KS}=\frac{16((9 h_{2,KS}-1)h_{1,KS}-9 h_{3,KS} h_{2,KS})}
{9\epsilon^{8/3}{\hat K_{KS}}^2\sinh^2 r }\,,\ 
\Om_0=0\,,
\eqlabel{kk}
\end{equation}
where now $r\to \infty$ is the boundary and $r\to 0$ is the IR. Above solution is parametrized by a single 
constant $\e$ which will be mapped to $K_0$, and which in turn will determine all the parameters in \eqref{ksuvirfinal}
once $\k=0$.

Comparing the metric ansatz in \eqref{ks1} and \eqref{5dks}, \eqref{redef2} we identify
\begin{equation}
\frac{(d\r)^2}{\r^4}=(w_{1,KS}(r))^2 (dr)^2\,.
\eqlabel{rrho}
\end{equation}
Introducing 
\begin{equation}
z\equiv e^{-r/3}\,,
\eqlabel{defz}
\end{equation}
we find from \eqref{rrho}
\begin{equation}
\frac1\r=\frac {\sqrt{6}\ (2\e)^{2/3}}{4}\ \int_1^z\ du\  \frac{u^6-1}{u^2(1-u^{12}+12u^6 \ln u)^{1/3}}\,.
\eqlabel{solverho}
\end{equation}
In the UV, $r\to \infty$, $z\to 0$ and $\r\to 0$ we have
\begin{equation}
\begin{split}
&e^{-r/3}\equiv z=
\frac{\sqrt{6}\ (2\e)^{2/3}}{4} \r \biggl(1+\calq \r+\calq^2 \r^2+\calq^3 \r^3+\calq^4 \r^4+\calq^5 \r^5
+\biggl(\frac{27}{80} \e^4 \ln 3+\calq^6\\
&+\frac{27}{800} \e^4-\frac{9}{16} \e^4 \ln 2+\frac{9}{20}
 \e^4 \ln\e+\frac{27}{40} \e^4 \ln\r\biggr) \r^6+\biggl(
-\frac{63}{16} \e^4 \calq \ln 2+\frac{189}{80} \e^4 \calq \ln 3+\calq^7\\
&+\frac{729}{800} \calq \e^4+\frac{63}{20} \e^4 \calq \ln\e
+\frac{189}{40} \calq \e^4 \ln\r\biggr) \r^7+\biggl(\frac{2403}{400} \e^4 \calq^2
-\frac{63}{4} \e^4 \calq^2 \ln 2+\frac{189}{20} \e^4 \calq^2 \ln 3\\
&+\frac{63}{5} \e^4 \calq^2 \ln\e+\calq^8
+\frac{189}{10} \e^4 \calq^2 \ln\r\biggr) \r^8+\biggl(\frac{189}{5} \e^4 \calq^3 \ln\e+\frac{9729}{400}
 \e^4 \calq^3-\frac{189}{4} \e^4 \calq^3 \ln 2\\
&+\frac{567}{20} \e^4 \calq^3 \ln 3+\calq^9+\frac{567}{10} \e^4 \calq^3 
\ln\r\biggr) 
\r^9+\calo(\r^{10}\ln\r)\biggr)\,,
\end{split}
\eqlabel{rrhouv}
\end{equation} 
where 
\begin{equation}
\begin{split}
\calq=&\frac{\sqrt{6}\ (2\e)^{2/3}}{4}\ \biggl\{
\int_0^1\ du\  \biggl(\frac{1-u^6}{u^2(1-u^{12}+12u^6 \ln u)^{1/3}}-\frac{1}{u^2}\biggr)-1\biggr\}\\
=&-\frac{\sqrt{6}\ (2\e)^{2/3}}{4}\ \times\ 0.839917(9)\,.
\end{split}
\eqlabel{qdef}
\end{equation}
In the IR, $r\to 0$, $z\to 1_-$ and $\frac1\r\to 0$ we have
\begin{equation}
\begin{split}
r=\frac{\sqrt 6\ 2^{1/3}}{3^{1/3}\ \e^{2/3}}\ y\ \biggl(1-\frac{2^{2/3}\ 3^{1/3}}{15\ \e^{4/3}}\  y^2+
\frac{71\ 3^{2/3}\ 2^{1/3}}{2625\ \e^{8/3}}\  y^4
+\calo(y^6)\biggr)\,.
\end{split}
\eqlabel{rrhoir}
\end{equation}
Using \eqref{rrhouv} and \eqref{rrhoir}, and the exact analytic solution describing the Klebanov-Strassler state of 
cascading gauge theory \eqref{ksks}, \eqref{kk} we can identify parameters\footnote{We matched the asymptotic
expansions \eqref{ksf1}-\eqref{ksg} and \eqref{f1hks}-\eqref{ghks} with the exact solution \eqref{ksks}
to the order we developed them: $\calo(\r^{10})$ and $\calo(y^{10})$ correspondingly.} \eqref{ksuvirfinal} 
\begin{equation}
\begin{split}
&K_0=-\ln 3+\frac53\ \ln2-\frac43\ \ln\e-\frac23\,,\\
&a_{1,0}=2 \calq\,,\qquad k_{1,3,0}=\frac{\e^2\sqrt{6}}{4}(-5\ln 2+3\ln 3+4\ln\e+2)\,,\qquad f_{c,4,0}=0\,,\\
&f_{a,3,0}=\frac{3\sqrt{6}}{4}\ \e^2\,,\qquad f_{a,4,0}=\frac{3\sqrt{6}}{4}\ \e^2\ \calq\\
&f_{a,6,0}=\frac{3\e^2}{400} (-225\e^2\ln 2+180\e^2\ln\e+216\e^2+135\e^2\ln3+100\sqrt{6}\calq^3)\,,
\\ &f_{a,7,0}=\frac{3\sqrt{6}}{4}\ \e^2\ \calq^4\,,\\
&f_{a,8,0}=\frac{3\e^2}{16}\calq^2(4\sqrt{6}\calq^3+135\e^2-90\e^2\ln2+54\e^2\ln3+72\e^2\ln\e)\,,\qquad g_{4,0}=0\,,
\end{split}
\eqlabel{susyuv}
\end{equation} 
in the UV, and 
\begin{equation}
\begin{split}
&f_{1,0}^h=1\,,\qquad f_{a,0}^h=2^{1/3}\ 3^{2/3}\ \e^{4/3}\,,\qquad h_0^h=\e^{-8/3}\ \times\ 0.056288(0)\,,\\
&k_{1,3}^h=\frac{4\sqrt{6}}{9\ \e^2}\,,\qquad k_{2,2}^h=\frac{2^{2/3}}{3^{2/3}\ \e^{4/3}}\,,\qquad 
k_{2,4}^h=-\frac{11\ 2^{1/3}\ 3^{2/3}}{45\ \e^{8/3}} \,,\\
&k_{3,1}^h=\frac{4\sqrt{6}\ 2^{1/3}\ 3^{2/3}}{27\ \e^{2/3}}\,,\qquad g_0^h=1\,,
\end{split}
\eqlabel{susyir}
\end{equation}
in the IR. Notice that inverting the first identification in \eqref{susyuv}, $\e=\e(K_0)$, we obtain a prediction 
for all the parameters \eqref{ksuvirfinal} as a function of $K_0$. 

Figures \ref{figure7} and \ref{figure8} compares the results of select UV and IR parameters in \eqref{ksuvirfinal} 
obtained numerically (blue dots) with analytic predictions (red curves) \eqref{susyuv} and \eqref{susyir} for the 
supersymmetric Klebanov-Strassler state. In this numerical computation we must set $\k=0\,,\ f_1(\r)\equiv 1$, \ie we remove 
the differential equation \eqref{kseq1}. Correspondingly, we have to remove (fix) 2 parameters in \eqref{ksuvirfinal}
for the numerical shooting procedure to be well-posed. Requiring that $f_1\equiv 1$ (for $\k=0$) both in the 
UV asymptotic solution \eqref{ksf1} and the IR asymptotic solution  \eqref{f1hks} implies
\begin{equation}
f_{a,4,0}=\frac12 \a_{1,0} f_{a,3,0}\,,\qquad f_{1,0}^h=1\,.
\eqlabel{fix2}
\end{equation}   
Notice that in Klebanov-Strassler state the string coupling is identically constant, \ie $g=1$. The latter in particular
implies that $g_{4,0}= 0$ and $g_0^h=1$. Numerically, over the range of values $K_0$ in figure \ref{figure7}, we find 
\begin{equation}
g_{4,0}\sim 10^{-6}\cdots 10^{-5}\,,\qquad |1-g_0^h|\sim 10^{-9}\cdots 10^{-8} \,.
\eqlabel{ksnumerics}
\end{equation}

As we mentioned earlier, we are after the states of cascading gauge theory with broken chiral symmetry 
on $S^3$, \ie the deformations of  Klebanov-Strassler states at $\k=1$. In practice we start with 
numerical Klebanov-Strassler state at $K_0=0.25$ ($P=1$) and increase $\k$ in increments of $\dd\k=10^{-3}$
up to $\k=1$.  
The resulting state is then used as a starting point to explore the states of cascading gauge theory 
on $S^3$ with $\csb$ for other values of $K_0\ne 0.25$.

\subsection{The first-order $\csb$ phase transition in $S^3$-compactified cascading gauge theory}\label{transition}
In section \ref{ksk} we numerically constructed states of cascading gauge theory on $S^3$ with spontaneously broken chiral 
symmetry over a range of $\ln\frac{\mu}{\Lambda}$. To determine whether (and when in terms of $\ln\frac{\mu}{\Lambda}$) 
these states represent the true ground state\footnote{As opposite to the states of cascading gauge theory on $S^3$ with 
unbroken chiral symmetry discussed in section \ref{symmetric}.} of $S^3$-compactified cascading gauge theory one has 
to compute their energies. The energy density of a chirally symmetric state was computed in \eqref{epc} using the full 
holographic renormalization of cascading gauge theory  implemented in \cite{aby}. To compute the energy density of the 
state of cascading gauge theory with spontaneously broken chiral symmetry one has to properly refine the holographic renormalization 
of \cite{aby}. We explain the main features of such refinement here.

For a static $S^3$-invariant states described by the effective action \eqref{5action} the  energy density is given 
\begin{equation}
\cale=\int_{\r_{UV}}^{\infty}d\r\ \call_E\,,
\eqlabel{energydef}
\end{equation}    
where $\call_E$ is the Euclidean one-dimensional Lagrangian density corresponding to the state, and $\r_{UV}$ is the 
UV cut-off, regularizing the Euclidean gravitational action in \eqref{energydef}. Briefly, holographic renormalization 
of the theory modifies the energy density 
\begin{equation}
\int_{\r_{UV}}^{\infty}d\r\ \call_E \to \int_{\r_{UV}}^{\infty}d\r\ \call_E+ S_{GH}^{\r_{UV}}+S_{counterterms}^{\r_{UV}}\,,
\eqlabel{holrenac}
\end{equation}
to include the Gibbons-Hawking and the local counterterms at the cut-off boundary $\r=\r_{UV}$ in a way that would render 
the  renormalized energy density finite in the limit $\r_{UV}\to 0$. 

Using the equations of motion \eqref{kseq1}-\eqref{kseq10}, it is possible to show that the on-shell gravitational 
action \eqref{5action} for static, $S^3$-invariant states of cascading gauge theory is a total derivative. Specifically,
we find\footnote{See  \eqref{5dks} and \eqref{redef2} for the  background metric.} 
\begin{equation}
\call_{E}^b=\frac{108}{16\pi G_5}\times \frac{d}{d\r}\biggl(\frac{2c_2^3c_1'\Omega_1\Omega_2^2\Omega_3^2}{c_3}\biggr)
=-\frac{108}{16\pi G_5}\times 
\frac{d}{d\r}\bigg(\frac{f_1^3f_c^{1/2} f_af_b (\r h'+4 h)}{216 h \r^4}\biggr)\,.
\eqlabel{lb}
\end{equation}
The integral in \eqref{energydef} now becomes the boundary values of the expression in \eqref{lb}.
Note that 
\begin{equation}
\lim_{\r\to \infty}\ \frac{f_1^3f_c^{1/2} f_af_b (\r h'+4 h)}{216 h \r^4}=-\lim_{y\to 0}
\frac{f_1^3 (f_c^h)^{1/2} f_a^h f_b y^2(h^h)'}{216 h^h}=0\,,
\eqlabel{ircontributions}
\end{equation}
where in the last equality we used \eqref{f1hks}-\eqref{hhks}.
 Thus,
\begin{equation}
\begin{split}
&\frac{16\pi G_5}{108}\ \cale^b=\biggl\{
\cale^b_{-4}\ \frac{1}{\r^4}+\cale^b_{-3}\ \frac{1}{\r^3}+\cale^b_{-2}\ \frac{1}{\r^2}
+\cale^b_{-1}\ \frac{1}{\r}+\cale^b_{0} +\calo(\r)
\biggr\}\bigg|_{\r=\r_{UV}}\,,
\end{split}
\eqlabel{caleb}
\end{equation} 
with 
\begin{equation}
\begin{split}
\cale_{-4}^b=&\frac{K_0-2 \ln\r}{27(2 K_0+1-4 \ln\r)}\,,
\end{split}
\eqlabel{ebm4}
\end{equation}
\begin{equation}
\begin{split}
\cale_{-3}^b=&- \frac{\a_{1,0}^{b}}{27(2 K_0+1-4 \ln\r)^2} \biggl(2 K_0+4 K_0^2+1-(16 K_0+4) \ln\r+16 \ln^2\r\biggr)\,,
\end{split}
\eqlabel{ebm3}
\end{equation}
\begin{equation}
\begin{split}
\cale_{-2}^b=&-\frac{1}{3888(2 K_0+1-4 \ln\r)^3} \biggl(-720 (\a_{1,0}^{b})^2 K_0-117-864 (\a_{1,0}^{b})^2 K_0^3-394 K_0\\
&-312 K_0^3+36 (\a_{1,0}^{b})^2-864 (\a_{1,0}^{b})^2 K_0^2-476 K_0^2+(3456 (\a_{1,0}^{b})^2 K_0+788\\
&+1440 (\a_{1,0}^{b})^2+1904 K_0+5184 (\a_{1,0}^{b})^2 K_0^2+1872 K_0^2) \ln\r+(-3744 K_0\\
&-3456 (\a_{1,0}^{b})^2-1904-10368 (\a_{1,0}^{b})^2 K_0) \ln^2\r+(2496+6912 (\a_{1,0}^{b})^2) \ln^3\r\biggr)\,,
\end{split}
\eqlabel{ebm2}
\end{equation}
\begin{equation}
\begin{split}
\cale_{-1}^b=&-\frac{1}{3888(2 K_0+1-4 \ln\r)^4} \a_{1,0}^{b} \biggl(-191+300 (\a_{1,0}^{b})^2+1248 K_0^2+1264 K_0^3\\
&+624 K_0^4+4 K_0+864 (\a_{1,0}^{b})^2 K_0^3+1056 (\a_{1,0}^{b})^2 K_0^2+576 (\a_{1,0}^{b})^2 K_0^4-168 (\a_{1,0}^{b})^2 K_0
\\&(336 (\a_{1,0}^{b})^2-8-4992 K_0-4608 (\a_{1,0}^{b})^2 K_0^3-5184 (\a_{1,0}^{b})^2 K_0^2-4992 K_0^3\\
&-4224 (\a_{1,0}^{b})^2 K_0-7584 K_0^2) \ln\r+(4992+4224 (\a_{1,0}^{b})^2+15168 K_0\\
&+13824 (\a_{1,0}^{b})^2 K_0^2+14976 K_0^2+10368 (\a_{1,0}^{b})^2 K_0) \ln^2\r+(-6912 (\a_{1,0}^{b})^2\\
&-18432 (\a_{1,0}^{b})^2 K_0-19968 K_0-10112) \ln^3\r+(9216 (\a_{1,0}^{b})^2+9984) \ln^4\r\biggr)\,,
\end{split}
\eqlabel{ebm1}
\end{equation}
\begin{equation}
\begin{split}
\cale_{0}^b=&-\frac{1}{648} \ln^2\r+\left(\frac{5}{2592}+\frac{1}{648} K_0\right) \ln\r
+\frac{23}{9216}+\frac{1}{5184} K_0+\frac{1}{864} (\a_{1,0}^{b})^4\\
&+\frac{1}{54} \a_{1,0}^{b} 
f_{a,3,0}+\frac{7}{5184} (\a_{1,0}^{b})^2-\frac{1}{27} f_{a,4,0}+\frac{1}{54} f_{c,4,0}+\calo(\ln^{-1}\r)\,,
\end{split}
\eqlabel{ebm0}
\end{equation}
where we set $P=1$, $g_0=1$, $f_0=1$, and used \eqref{ksf1}-\eqref{ksh}. The superscript $\ ^b$ in the UV parameter 
$\a_{1,0}$ is used to indicate that it is computed in the phase with broken chiral symmetry. 

Clearly, the expression \eqref{caleb} is divergent in the limit $\r_{UV}\to 0$. Turns out that all the divergences are removed 
once we include the generalized\footnote{``Generalized" five-dimensional Gibbons-Hawking term is just a dimensional reduction of the 
10-dimensional Gibbons-Hawking term corresponding to \eqref{10dmetric}.} Gibbons-Hawking term, see \cite{aby}, 
\begin{equation}
\begin{split}
S_{GH}^{\r_{UV}}=\frac{108}{8\pi G_5}\ \frac{1}{c_3}\left(c_1 c_2^3 \Om_1\Om_2^2\Om_3^2\right)'\bigg|_{\r=\r_{UV}}=
\frac{1}{8\pi G_5}\ \frac{\r}{h^{1/4}}\left(\frac{h^{1/4}f_1^3f_c^{1/2}f_af_b}{\r^4}\right)'\bigg|_{\r=\r_{UV}}\,,
\end{split}
\eqlabel{ghterm}
\end{equation}
and the local counter-terms obtained in \cite{aby} with the following obvious modifications:
\begin{equation}
\begin{split}
&K^{KT}=\frac 12 K_1+\frac 12 K_3\,,\qquad \Om_1^{KT}=3\Om_1\,,\qquad \Om_2^{KT}=\frac{\sqrt{6}}{2} \left(\Om_2+\Om_3\right) \,.
\end{split}
\eqlabel{modifications}
\end{equation}
We find
\begin{equation}
\cale^b=\frac{1}{8\pi G_5}\ \left(\frac{403}{1920}+\frac{1}{32}K_0^2+\frac{3}{32}K_0-3f_{a,4,0}
+\frac32 f_{c,4,0}+\frac32 a_{1,0}^b f_{a,3,0}-\frac{3}{32}(\a_{1,0}^b)^2\right)\,.
\eqlabel{calebfin}
\end{equation}
Notice that \eqref{calebfin} coincides with \eqref{epc} once restricted to chirally symmetric states:
\begin{equation}
f_{a,3,0}=0\,,\qquad f_{c,4,0}=a_{4,0}\,,\qquad f_{a,4,0}=b_{4,0}\,.
\eqlabel{srestric}
\end{equation}
We can now compare the energy densities of a chirally symmetric state and a state spontaneously  breaking chiral symmetry
for cascading gauge theory on $S^3$ (we restored the full $\{P,g_0,f_0\}$ dependence)
\begin{equation}
\begin{split}
\cale^b-\cale^s=&\frac{1}{8\pi G_5}\frac{1}{f_0^4}\biggl(3(b_{4,0}-f_{a,4,0})
+\frac32 (f_{c,4,0}-a_{4,0})+\frac32 a_{1,0}^b f_{a,3,0}\\
&+\frac{3}{32}\left((\a_{1,0}^s)^2-(\a_{1,0}^b)^2\right)\biggr)\,.
\end{split}
\eqlabel{finaldiff}
\end{equation} 

\begin{figure}[t]
\begin{center}
\psfrag{ebmes}{{$\frac{8\pi G_5}{P^4g_0^2}\times\frac{\cale^b-\cale^s}{\mu^4}$}}
\psfrag{lnmul2}{{$\ln\frac{\mu^2}{\Lambda^2}$}}
\includegraphics[width=4in]{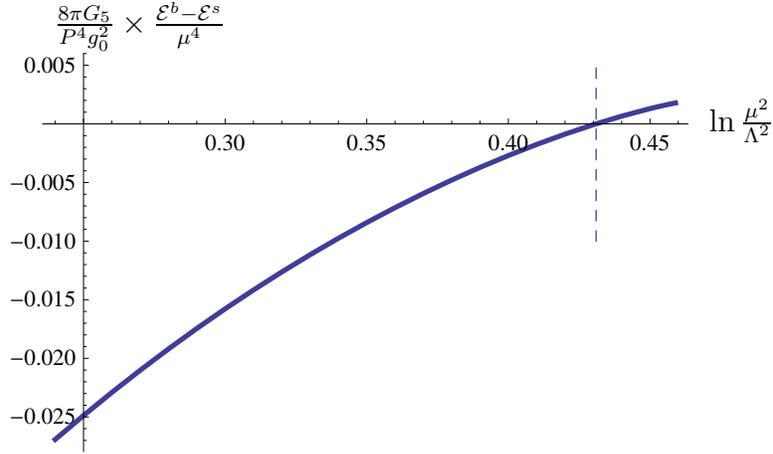}
\end{center}
  \caption{Energy densities difference between the  state with spontaneously broken chiral 
symmetry, $\cale^b$, and the chirally symmetric state, $\cale^s$, of cascading gauge theory on
$S^3$ as a function of compactification scale $\mu$ and the strong coupling scale $\Lambda$ of the theory.
The vertical dashed line, see \eqref{muscb}, indicates the location of the first-order chiral symmetry 
breaking QPT in cascading gauge theory. 
} \label{figure9}
\end{figure}

Figure \ref{figure9} presents the energy densities difference between the  state with spontaneously broken chiral 
symmetry, $\cale^b$, and the chirally symmetric state, $\cale^s$, of cascading gauge theory on
$S^3$ as a function of $\ln\frac{\mu^2}{\Lambda^2}$. Notice that the for $\mu>\mu_{\csb}$ (indicated by 
a vertical dashed line),
\begin{equation}
\mu_{\csb}=1.240467(8)\ \Lambda\,,
\eqlabel{muscb}
\end{equation} 
$\cale^b>\cale^s$, \ie the true ground state of cascading gauge theory on $S^3$ is chirally symmetric.
Since 
\begin{equation}
\frac{d(\cale^b-\cale^s)}{d\ln\mu}\bigg|_{\mu=\mu_{\csb}}\ne 0\,,
\eqlabel{ordertake2}
\end{equation}
at  $\mu=\mu_{\csb}$ the theory undergoes the first-order quantum phase transition associated with spontaneous breaking 
of chiral symmetry.  Finally, notice that $\mu_{\csb}>\mu_c$ (see \eqref{mucres}) associated with the condensation of $\csb$
tachyons in a chirally symmetric phase of the  theory.

\section*{Acknowledgments}
I would like to thank Ofer Aharony, Philip Argyres, Cliff Burgess,
Rob Myers and Andrei Starinets for valuable discussions.  I would like to thank Aspen
Center for Physics, Galileo Galilei Institute for Theoretical Physics, 
International Centre for Mathematical Sciences
and  Centro de Ciencias de Benasque Pedro Pascual for hospitality
during the various stages of this project.  Research at Perimeter
Institute is supported by the Government of Canada through Industry
Canada and by the Province of Ontario through the Ministry of
Research \& Innovation. I gratefully acknowledge further support by an
NSERC Discovery grant.


\begin{thebibliography}{99}

\bibitem{kt}
  I.~R.~Klebanov and A.~A.~Tseytlin,
  ``Gravity duals of supersymmetric SU(N) x SU(N+M) gauge theories,''
  Nucl.\ Phys.\  B {\bf 578}, 123 (2000)
  [arXiv:hep-th/0002159].


\bibitem{ks}
  I.~R.~Klebanov and M.~J.~Strassler,
  JHEP {\bf 0008}, 052 (2000)
  [arXiv:hep-th/0007191].


\bibitem{kw}
  I.~R.~Klebanov and E.~Witten,
  Nucl.\ Phys.\  B {\bf 536}, 199 (1998)
  [arXiv:hep-th/9807080].




\bibitem{sd}
  N.~Seiberg,
  Nucl.\ Phys.\  B {\bf 435}, 129 (1995)
  [arXiv:hep-th/9411149].


\bibitem{b}
  A.~Buchel,
  Nucl.\ Phys.\  B {\bf 600}, 219 (2001)
  [arXiv:hep-th/0011146].

\bibitem{k}
  M.~Krasnitz,
  JHEP {\bf 0212}, 048 (2002)
  [arXiv:hep-th/0209163].



\bibitem{aby2}
  O.~Aharony, A.~Buchel and A.~Yarom,
  JHEP {\bf 0611}, 069 (2006)
  [arXiv:hep-th/0608209].


\bibitem{dks}
  A.~Dymarsky, I.~R.~Klebanov and N.~Seiberg,
  JHEP {\bf 0601}, 155 (2006)
  [arXiv:hep-th/0511254].


\bibitem
{juan} J.~M.~Maldacena,
Adv.\ Theor.\ Math.\ Phys.\  {\bf 2}, 231 (1998)
[Int.\ J.\ Theor.\ Phys.\  {\bf 38}, 1113 (1999)]
[arXiv:hep-th/9711200].

\bibitem
{adscft} O.~Aharony, S.~S.~Gubser, J.~M.~Maldacena, H.~Ooguri and
Y.~Oz,
  Phys.\ Rept.\  {\bf 323}, 183 (2000)
  [arXiv:hep-th/9905111].




\bibitem{aby}
  O.~Aharony, A.~Buchel and A.~Yarom,
  Phys.\ Rev.\  D {\bf 72}, 066003 (2005)
  [arXiv:hep-th/0506002].


\bibitem{bt}
  A.~Buchel and A.~A.~Tseytlin,
  Phys.\ Rev.\  D {\bf 65}, 085019 (2002)
  [arXiv:hep-th/0111017].



\bibitem{kbh1}
  A.~Buchel, C.~P.~Herzog, I.~R.~Klebanov, L.~A.~Pando Zayas and A.~A.~Tseytlin,
  JHEP {\bf 0104}, 033 (2001)
  [arXiv:hep-th/0102105].

\bibitem{kbh2}
  S.~S.~Gubser, C.~P.~Herzog, I.~R.~Klebanov and A.~A.~Tseytlin,
  JHEP {\bf 0105}, 028 (2001)
  [arXiv:hep-th/0102172].


\bibitem{abk}
  O.~Aharony, A.~Buchel and P.~Kerner,
  Phys.\ Rev.\  D {\bf 76}, 086005 (2007)
  [arXiv:0706.1768 [hep-th]].

\bibitem{bp1}
  A.~Buchel and C.~Pagnutti,
  Nucl.\ Phys.\  B {\bf 834}, 222 (2010)
  [arXiv:0912.3212 [hep-th]].


\bibitem{ksbh}
  A.~Buchel,
  Nucl.\ Phys.\  B {\bf 847}, 297 (2011)
  [arXiv:1012.2404 [hep-th]].



\bibitem{hyd1}
  A.~Buchel and J.~T.~Liu,
  Phys.\ Rev.\ Lett.\  {\bf 93}, 090602 (2004)
  [arXiv:hep-th/0311175].

\bibitem{hyd2}
  A.~Buchel,
  Phys.\ Rev.\  D {\bf 72}, 106002 (2005)
  [arXiv:hep-th/0509083].

\bibitem{hyd3}
  A.~Buchel,
  Nucl.\ Phys.\  B {\bf 820}, 385 (2009)
  [arXiv:0903.3605 [hep-th]].

\bibitem{oa}
  O.~Aharony, private communication.




\end{thebibliography}
\end{document}